\documentclass[preprintnumbers,amsmath,11pt,amssymb,floatfix,superscriptaddress,nofootinbib]{revtex4}
\usepackage{graphicx}
\usepackage{epsfig}
\usepackage{bm}
\usepackage{amsfonts}

\input epsf

\begin{document}

\title{A Higher-Derivative Hubble Parameter Dark Energy Model: Cosmological Analysis and Scalar Field Correspondence}
\author{Antonio Pasqua}
\email{toto.pasqua@gmail.com}

\date{\today}
\newpage


\begin{abstract}
In this work, we study a Dark Energy (DE) energy density model which depends on the Hubble parameter squared $H^2$ and on its first, second and third time derivatives $\dot{H}$, $\ddot{H}$ and $\dddot{H}$. Considering a scale factor $a$ with a power-law dependence on the time (with $n$ indicating the power-law index), we obtain some important cosmological quantities as function of the , like the energy densities of Matter $\rho_m$ and of DE $\rho_D$, the fractional energy densities of DM $\Omega_m$ and of DE $\Omega_D$, the Hubble parameter squared $H^2$, the deceleration parameter $q$, the evolutionary form of the fractional energy density of DE $\Omega'_D$, the pressure of DE $p_D$ and the Equation of State (EoS) parameter of DE $\omega_D$, for both non interacting and interacting cases. For the interacting case, we consider 9 different interacting term $Q$, all functions of the Hubble parameter $H$ and/or of $\rho_m$ and $\rho_D$. Finally, we establish a correspondence between the DE model we study and some scalar field theories, including tachyon, k-essence, quintessence, Yang-Mills (YM) and Nonlinear Electrodynamics (NLED) fields.
\end{abstract}

\maketitle
\tableofcontents

\section{Introduction}
Observational data collected from a variety of independent probes, including the Planck satellite mission \cite{planck}, the Supernova Cosmology Project \cite{sn2,sn4}, the Wilkinson Microwave Anisotropy Probe (WMAP) \cite{cmb1,cmb2}, the Sloan Digital Sky Survey (SDSS) \cite{sds1,sds3,sds4}  
and X-ray observations of galaxy clusters \cite{xray}, 
consistently indicate that the Universe is experiencing a phase of accelerated expansion. 
This unexpected behavior cannot be explained within the framework of standard matter and radiation alone, 
and thus points to the existence of a dominant energy component with negative pressure, 
commonly referred to as dark energy (DE). 
Among the various theoretical candidates proposed, the cosmological constant  $\Lambda_{CC}$ represents the most straightforward and successful description at the phenomenological level. 
However, its interpretation faces deep conceptual challenges, notably the cosmological constant problem and the coincidence problem,  which have motivated the development of numerous alternative dynamical models of dark energy \cite{cosm3,cosm4,cosm5}.

The Cosmological Constant $ \Lambda_{CC} $, often introduced as an additional geometric term in Einstein’s field equations, is commonly interpreted as the energy density associated with the quantum vacuum.  Nonetheless, when Quantum Field Theory (QFT) is applied with natural cut-offs at the Planck or electroweak scales, the predicted vacuum energy density exceeds the observed value by an enormous margin —  approximately $10^{123}$ or $10^{55}$ times larger, respectively. 
This vast discrepancy, in the absence of any symmetry or physical mechanism capable of naturally fine-tuning 
$ \Lambda_{CC} $ to such a small value, constitutes the essence of the so-called cosmological constant problem. 
A related puzzle, known as the \emph{coincidence problem}, concerns the apparent timing of cosmic evolution: 
why are the energy densities of dark energy and matter of comparable magnitude precisely today? 
Extensive discussions of these long-standing issues can be found in \cite{cosm3,cosm4,cosm5},  whereas a comprehensive overview of the inflationary dynamics in the early Universe is presented in \cite{rev2}.

Within the context of General Relativity, dark energy (DE) is estimated to account for approximately two-thirds of the total energy density of the Universe, 
$\rho_{\text{tot}}$ \cite{twothirds}, with the remaining portion primarily composed of dark matter (DM) and a smaller fraction of baryonic matter. 
Despite numerous observational campaigns and theoretical investigations, the fundamental properties of DE remain elusive.

As an alternative to the cosmological constant, several dynamical DE models have been proposed, 
characterized by a time-dependent equation of state parameter $\omega_D$. 
Observational analyses, particularly of Type Ia Supernovae, suggest that these models may provide a better fit to the data compared to a simple constant-$\Lambda$ scenario. 
Among the theoretical candidates are a variety of scalar field frameworks, including quintessence \cite{quint1,quint4,quint5}, k-essence \cite{kess3,kess4,kess5}, tachyon fields \cite{tac2,tac3,tac4}, phantom energy \cite{pha2,pha5,pha6}, dilaton fields \cite{dil1,dil2}, and quintom models \cite{qui3,qui8,qui10,qui12}. 
Additionally, interacting DE scenarios have been extensively studied, encompassing models based on Chaplygin gas \cite{cgas1,cgas2,cgas3}, as well as Agegraphic Dark Energy (ADE) and its extension, the New ADE (NADE) \cite{ade1,ade2}. 
For a comprehensive discussion of multifluid cosmologies with complex forms of the equation of state parameter, see \cite{rev2}.

Beyond scalar field and Chaplygin gas models, a distinct class of dynamical dark energy (DE) theories emerges from the holographic principle \cite{holo1,holo2,holo5}. 
Among these, the Holographic Dark Energy (HDE) model, initially proposed by Li \cite{li}, has attracted significant attention and has been extensively investigated in numerous studies \cite{nood1,nood2,nood3,nood4,nood5,nood6,nood7}. 
The holographic principle, which plays a central role in black hole thermodynamics and string theory, posits that the entropy of a physical system scales with the area $A$ of its boundary rather than its volume $V$, i.e., $S \sim A \sim L^2$, where $L$ is a characteristic length scale.

In Li's formulation, the DE energy density is expressed as
\begin{equation}
\rho_\Lambda = 3\alpha M_p^2 L^{-2},
\end{equation}
where $\alpha$ is a dimensionless constant, and $M_p = \left(8\pi G_N\right)^{-1/2}$ denotes the reduced Planck mass, with $G_N$ representing Newton's gravitational constant. 
The concept of holographic DE was first introduced by Cohen et al.~\cite{coh1}, who argued that the DE energy density should be bounded by the size of the Universe to prevent black hole formation. 
Their initial proposal, with $\rho_\Lambda \propto H^2$, however, failed to produce cosmic acceleration, resulting in an effective equation of state $\omega = 0$.

To overcome this limitation, Li suggested using the future event horizon as the infrared (IR) cutoff, leading to a model compatible with late-time cosmic acceleration \cite{li}. 
Subsequent generalizations include the Holographic Ricci Dark Energy model, in which the IR cutoff scales as $L \propto R^{-1/2}$ with $R$ being the Ricci scalar \cite{gaoprimo}, 
and the Granda–Oliveros proposal, where the DE density depends on both the Hubble parameter $H$ and its first derivative $\dot{H}$ \cite{go1,go4}. 
These holographic DE models have been rigorously tested against Type Ia supernovae, cosmic microwave background observations, and baryon acoustic oscillation data \cite{cons1,cons2,cons3,cons4,cons5,cons6,cons7,cons8,cons9}. 
For comprehensive reviews and broader discussions on holographic DE, see \cite{hde1,hde2,hde7,hde10,hde12,hde13,hde17,hde18,hde19,
hde22,hde23,hde24,hde26,hde28,hde30,hde32,hde33,hde34,hde35,saridakis11,saridakis22,sc1,sc2,sc3}.

In this work, we focus on a recently proposed model introduced in Pasqua \cite{mio!-}. The main features and implications of this model are discussed in the following sections.

The paper is structured as follows.\\
In Section 2, we present the cosmological framework adopted in this study. Section 3 introduces the HDE model considered in our analysis. In Section 4, we investigate several cosmological quantities by assuming a power-law form for the scale factor, both for non-interacting and interacting dark sectors.
Section 5 establishes a correspondence between the proposed DE model and various scalar field theories.
Finally, Section 6 summarizes our main results and conclusions.

\section{Holographic Dark Energy Model in a Non-Flat Universe}
In this section, we present the key characteristics of the Dark Energy (DE) model considered in this work and derive the corresponding fundamental cosmological quantities.\\
The geometry of a homogeneous and isotropic Universe is described by the Friedmann–Lemaître–Robertson–Walker (FLRW) metric, given by:
\begin{eqnarray}
    ds^2 &=&-dt^2 + a^2(t)\left[\frac{dr^2}{1 - kr^2} + r^2d\Omega^2\right]\nonumber \\
    &=&-dt^2 + a^2(t)\left[\frac{dr^2}{1 - kr^2} + r^2 (d\theta^2 + \sin^2\theta\, d\varphi^2)\right], \label{6}
\end{eqnarray}
where $t $ denotes the cosmic time, $a(t) $ is the scale factor describing the expansion of the Universe, $r $ represents the comoving radial coordinate, and $\theta $ and $\varphi $ are the usual angular coordinates in spherical symmetry, taking values $0 \leq \theta \leq \pi $ and $0 \leq \varphi < 2\pi $, respectively. \\
The parameter $k $ characterizes the spatial curvature and can assume the values $-1$, $0$, or $+1$, corresponding to open, flat, and closed Universes, respectively. \\
The evolution of a homogeneous and isotropic Universe in the framework of General Relativity is governed by the Friedmann equations, which, in the presence of both Dark Energy (DE) and Dark Matter (DM), take the form:
\begin{eqnarray}
    H^2  &=& \frac{1}{3M^2_p}\left( \rho_D + \rho_m \right), \label{7} \\
    \dot{H} + 2H^2 &=& \left(\frac{8\pi G}{6}\right)\, p_D, \label{7fri2}
\end{eqnarray}
where $H = \dot{a}/a $ denotes the Hubble parameter, $\rho_D $ is the energy density of Dark Energy (DE), and $p_D $ represents its corresponding pressure. Moreover, $\rho_m $ indicates the energy density of Dark Matter (DM). \\
We define the fractional energy densities of DE and DM as follows:
\begin{eqnarray}
    \Omega_D &=& \frac{\rho_D}{\rho_{\text{cr}}} = \frac{\rho_D}{3M^2_p H^2}, \label{10} \\
    \Omega_m &=& \frac{\rho_m}{\rho_{\text{cr}}} = \frac{\rho_m}{3M^2_p H^2}, \label{8}
\end{eqnarray}
The quantity $\rho_{\text{cr}} $ denotes the critical energy density required for a spatially flat Universe, and it is expressed as:
\begin{eqnarray}
    \rho_{\text{cr}} = 3M^2_p H^2.
\end{eqnarray}
Using the expressions for $\Omega_D $ and $\Omega_m $ given in Eqs.~(\ref{10}) and (\ref{8}), the Friedmann equation in Eq.~(\ref{7}) can be rewritten in the following form:
\begin{eqnarray}
    \Omega_D + \Omega_m = 1. \label{11}
\end{eqnarray}
To ensure the fulfillment of the Bianchi identity, or equivalently the local conservation of energy-momentum, the total energy density $\rho_{\text{tot}} = \rho_D + \rho_m $ must satisfy the continuity equation:
\begin{eqnarray}
    \dot{\rho}_{\text{tot}} + 3H\left( \rho_{\text{tot}} + p_{\text{tot}} \right) = 0, \label{12old}
\end{eqnarray}
where the quantities $\rho_{\text{tot}} $ and $p_{\text{tot}} $ represent the total energy density and total pressure of the cosmic fluid, respectively, and are given by the following relations:
\begin{eqnarray}
    \rho_{\text{tot}} &=& \rho_m + \rho_D, \\
    p_{\text{tot}} &=& p_D,
\end{eqnarray}
since we assume that Dark Matter (DM) is pressureless, i.e., $p_m = 0 $. \\
The continuity equation obtained in Eq.~(\ref{12old}) can also be written in terms of the total equation-of-state (EoS) parameter $\omega_{\text{tot}} = p_{\text{tot}} / \rho_{\text{tot}} $ as follows:
\begin{eqnarray}
    \dot{\rho}_{\text{tot}} + 3H\left(1 + \omega_{\text{tot}} \right) \rho_{\text{tot}} = 0. \label{12}
\end{eqnarray}
Since the energy densities of Dark Matter (DM) and Dark Energy (DE) are assumed to be conserved separately, Eq.~(\ref{12}) can be decomposed into two independent continuity equations. In the non-interacting case, these can be written as:
\begin{eqnarray}
    \dot{\rho}_D + 3H\left( 1 + \omega_D \right)\rho_D &=& 0, \label{12deold} \\
    \dot{\rho}_m + 3H\rho_m &=& 0. \label{12dm}
\end{eqnarray}
Starting from the general expression of the equation of state (EoS) parameter for Dark Energy (DE), defined as
\begin{eqnarray}
    \omega_D = \frac{p_D}{\rho_D}, \label{mammud}
\end{eqnarray}
the continuity equation for the DE component, Eq.~(\ref{12deold}), can be recast in the following form:
\begin{eqnarray}
    \dot{\rho}_D + 3H\left(p_D + \rho_D\right) = 0. \label{12de}
\end{eqnarray}
Currently, Dark Energy (DE) accounts for nearly two-thirds of the total energy content of the Universe, while its contribution was negligible during the early cosmic epochs. This fact implies that DE is not a static component but evolves alongside the cosmic expansion. Within this context, it becomes natural to explore models in which the DE density is expressed as a function of the Hubble parameter $H$ and its time derivatives, which encapsulate the dynamical behavior of the expanding Universe.

\section{HDE model with third time derivative of Hubble parameter $H$}
We now consider a model for the dark energy (DE) density that incorporates higher-order time derivatives of the Hubble parameter $H$, and is defined as follows \cite{mio!-}:
\begin{eqnarray}
\rho_D = 3 \left[ \alpha \left(\frac{\dddot{H}}{H^2}\right)  +\beta \left(\frac{\ddot{H}}{H}\right)  + \gamma \dot{H}+\delta H^2\right] ,\label{cz1}
\end{eqnarray}
where $\alpha$, $\beta$, $\gamma$, and $\delta$ are dimensionless parameters. For mathematical convenience, we adopt natural units by setting the reduced Planck mass to unity, $M_p = 1$. It is worth noting that the inclusion of the inverse powers $H^{-2}$ and $H^{-1}$ in the first and second terms, respectively, ensures dimensional consistency among all components of the expression.

The cosmological evolution and the main characteristics of this DE model are strongly determined by the four free parameters introduced in its definition. The energy density in Eq.~(\ref{cz1}) can be interpreted as a generalized form encompassing several DE models previously proposed in the literature. For instance, by setting $\alpha = 0$, one recovers the energy density proposed by Chen and Jing~\cite{modelhigher}, later extended in subsequent studies. In the limiting case $\alpha = \beta = 0$, Eq.~(\ref{cz1}) reduces to the well-known Granda–Oliveros (GO) DE model~\cite{gohnde}. Moreover, for the particular parameter choice $\alpha=\beta=0$, $\gamma=1$, and $\delta=2$, the expression reproduces the DE density with an infrared (IR) cutoff determined by the average radius associated with the Ricci scalar, valid for a spatially flat Universe ($k=0$). Since the present formulation introduces an additional degree of freedom, it provides a more general framework than the standard Ricci Dark Energy (RDE) model. Related generalized DE scenarios have been investigated in~\cite{altri3,altri1,altri2}.

\subsection{Non Interacting case}
We now consider a power-law form of the scale factor $a(t)$ given by the following relation:
\begin{eqnarray}
a(t) = a_0 t^n, \label{scale}
\end{eqnarray}
where $a_0$ denotes the present-day value of the scale factor $a$, and $n$ (assumed to be positive) represents the power-law index characterizing the cosmic expansion. \\
We now proceed to compute some relevant cosmological quantities using the expression of the scale factor given in Eq.~(\ref{scale}). Substituting Eq.~(\ref{scale}) into the definition of the DE energy density provided in Eq.~(\ref{cz1}), we obtain the following expression for $\rho_D$ as a function of cosmic time:
\begin{eqnarray}
\rho_{D}(t) 
&=& \frac{3}{t^2} \left( -\frac{6\alpha}{n} + 2\beta - \gamma n + \delta n^2 \right)\nonumber \\
&=& 3 \left( -\frac{6\alpha}{n} + 2\beta - \gamma n + \delta n^2 \right)t^{-2}.\label{cz3}
\end{eqnarray}
By substituting the chosen form of the scale factor $a(t)$ into the continuity equation for dark matter (DM), we can express it as follows:
\begin{eqnarray}
\dot{\rho}_m +\left(\frac{3n}{t}\right)\rho_m=0 .   \label{annamia1}
\end{eqnarray}
The general solution of Eq. (\ref{annamia1}) is given by:
\begin{eqnarray}
    \rho_m&=& \rho_{m0}t^{-3n}\nonumber \\
    &=& 3H_0^2\Omega_{m0}t^{-3n},\label{annamia2}
\end{eqnarray}
where $\rho_{m0}$ denotes the present-day value of the matter energy density $\rho_m$, and we used the relation $\rho_{m0} =3H_0^2  \Omega_{m0}$.\\
By combining the expression of $\rho_m$ obtained in Eq.~(\ref{annamia2}) with the expression of $\rho_D$ derived in Eq.~(\ref{cz3}), and substituting them into the Friedmann equation given in Eq.~(\ref{7}), we obtain the following expression for $H^2$:
\begin{eqnarray}
    H^2(t)= H_0^2\left\{ \Omega_{m0}t^{-3n} +\frac{1}{H_0^2} \left( -\frac{6\alpha}{n} + 2\beta - \gamma n + \delta n^2 \right)t^{-2}\right\}. \label{annarella}
\end{eqnarray}
The general definition of the deceleration parameter $q$ is given by:
\begin{eqnarray}
    q= -1-\frac{\dot{H}}{H^2}.  \label{qgen}
\end{eqnarray}
Using the expression of $H^2$ given in Eq. (\ref{annarella}), we obtain:
\begin{eqnarray}
q(t) = -1 +  
\frac{
3H_0^2n \, \Omega_{m0} \, t^{-3n -1} + 2  \left(-\frac{6\alpha}{n} + 2\beta - \gamma n + \delta n^2\right) \, t^{-3}
}{2H_0^2
\Omega_{m0} \, t^{-3n} +2\left( -\frac{6\alpha}{n} + 2\beta - \gamma n + \delta n^2\right) \, t^{-2}
}.
\end{eqnarray}

We now calculate the expressions of the fractional energy densities of DM and DE in two different ways.\\
Using in the general definition of $\Omega_m$ given in Eq.(\ref{8}) the expression of $\rho_m$ given in Eq. (\ref{annamia2}) along with the definition of $H$ given in Eq. (\ref{acca}), we obtain:
\begin{eqnarray}
    \Omega_m= \Omega_{m0}t^{2-3n}
\end{eqnarray}
Therefore, using the result of the Friedmann equation given in Eq. (\ref{11}), we can write:
\begin{eqnarray}
    \Omega_D=1-\Omega_m=1- \Omega_{m0}t^{2-3n}
\end{eqnarray}
We now want to find the evolutionary form of the fractional energy density $\Omega_D'$, where usually define $'=\frac{d}{dx}$ with $x=\ln a$.\\
We underline there is the following relation between derivative with respect to $x$ and derivative with respect to $t$:
\begin{eqnarray}
    \frac{d}{dx}=\frac{t}{n}\frac{d}{dt}
\end{eqnarray}
In our case, we obtain:
\begin{eqnarray}
    \Omega_D'&=&\left(3-\frac{2}{n}\right)\Omega_m\nonumber \\
    &=& \left(3-\frac{2}{n}\right)(1-\Omega_D)
\end{eqnarray}

 Instead, using in the general definition of $\Omega_m$ given in Eq.(\ref{8}) the expression of $\rho_m$ given in Eq. (\ref{annamia2}) along with the definition of $H^2$ given in Eq. (\ref{annarella}), we obtain:
\begin{equation}
\Omega_m(t) = \frac{\rho_m(t)}{3 H^2(t)} 
= \frac{H_0^2 \Omega_{m0} \, t^{-3n}}
{H_0^2 \Omega_{m0} \, t^{-3n} + \left(-\frac{6\alpha}{n} + 2\beta - \gamma n + \delta n^2\right)  t^{-2}}.
\end{equation}
Therefore, using the result of the Friedmann equation given in Eq. (\ref{11}), we can write:
\begin{equation}
\Omega_D(t) =1-
 \frac{H_0^2 \Omega_{m0} \, t^{-3n}}
{H_0^2 \Omega_{m0} \, t^{-3n} + \left(-\frac{6\alpha}{n} + 2\beta - \gamma n + \delta n^2\right)  t^{-2}}.
\end{equation}

The evolutionary form of the fractional energy density of DE is given by:
\begin{equation}
\Omega_D'(x)
= -\left(\frac{2 - 3n}{n}\right)
\frac{
H_0^2 \Omega_{m0}
\left(-\frac{6\alpha}{n} + 2\beta - \gamma n + \delta n^2\right)
t^{-3n - 2}
}{
\left[
H_0^2 \Omega_{m0} t^{-3n}
+ \left(-\frac{6\alpha}{n} + 2\beta - \gamma n + \delta n^2\right)t^{-2}
\right]^2
}.
\end{equation}

We now derive the expression of the pressure of DE $p_{D}$.\\
From the continuity equation for DE, we obtain the following general expression for $ p_{D}$:
\begin{eqnarray}
    p_{D}= -\rho_{D} - \frac{\dot{\rho}_{D}}{3H}.
\end{eqnarray}
From Eq. (\ref{cz3}), we obtain:
\begin{eqnarray}
\dot{\rho}_{D}(t) 
&=& -\frac{6}{t^3} \left( -\frac{6\alpha}{n} + 2\beta - \gamma n + \delta n^2 \right)\nonumber \\
&=&-6\left( -\frac{6\alpha}{n} + 2\beta - \gamma n + \delta n^2 \right)t^{-3}.
\end{eqnarray}
Moreover, we obtain that the Hubble parameter as function of the time is given by:
\begin{eqnarray}
    H(t) = \frac{\dot{a}}{a}=\frac{n}{t}. \label{acca}
\end{eqnarray}
Using the expressions of $\rho_{D}$, $\dot{\rho}_{D}$ and $H$, we can write: 
\begin{eqnarray}
    p_{D}(t) &=& \left( \frac{2}{3n} -1 \right)\rho_{D}(t)\nonumber \\
    &=& \left( \frac{2}{n} -3 \right)\left( -\frac{6\alpha}{n} + 2\beta - \gamma n + \delta n^2 \right)t^{-2}.
\end{eqnarray}
We now want to derive the expression of the EoS parameter of DE $\omega_D$.\\
From the continuity equation for DE given in Eq. (\ref{12deold}), we find the following general expression for $\omega_D$:
\begin{eqnarray}
\omega_{D} = -1 - \frac{\dot{\rho}_{D}}{3H\rho_{D}}  .  
\end{eqnarray}
Using the expressions of $\rho_{D}$, $\dot{\rho}_{D}$ and $H$, we obtain the following relation for $\omega_D$:
\begin{eqnarray}
    \omega_{D_1} = -1+\frac{2}{3n} .\label{eos}
\end{eqnarray}
This expression implies that $\omega_{D}$ is always greater than $-1$, 
indicating a quintessence-like behavior. In particular, for small values of $n$,  the EoS parameter deviates significantly from the cosmological constant limit,  reflecting a less dominant dark energy component. 
For $n=1$, the equation of state (EoS) parameter of DE is 
\begin{eqnarray}
\omega_{D} = -\frac{1}{3}.
\end{eqnarray}
This value corresponds to the critical limit between decelerated and accelerated expansion of the Universe:
\begin{itemize}
    \item If $\omega_{D} > -1/3$ (i.e. $0< n <1$), the expansion is decelerating.
    \item If $\omega_{D} < -1/3$, (i.e. ($n>1$)) the expansion is accelerating.
\end{itemize}
Thus, for $n=1$, the model predicts a marginal expansion at the boundary between acceleration and deceleration.

As $n$ increases, $\omega_{D}$ approaches $-1$, effectively mimicking a cosmological constant. 
Therefore, the parameter $n$ controls the deviation of dark energy from a pure  $\Lambda$-like behavior, with higher values of $n$ corresponding to an accelerated  expansion that closely resembles a $\Lambda$CDM scenario. 
Overall, the model predicts a non-phantom, accelerating universe driven by  quintessence-like dark energy.\\

\subsection{First Interacting Case}
We now consider the presence of interaction between Dark Sectors.\\
We now extend our analysis by considering the possibility of an interaction between the dark sectors. This idea refers to scenarios where dark matter (DM) and dark energy (DE) are not entirely independent but may exchange energy or momentum. Such a coupling is often motivated by attempts to address the so-called coincidence problem, namely why the energy densities of DM and DE are of the same order today despite their different evolutionary histories. Allowing for an interaction modifies the standard cosmological dynamics and can leave distinctive observational imprints, such as changes in the expansion history, deviations in structure formation, or shifts in the cosmic microwave background (CMB) anisotropies. These models have therefore been widely studied as possible alternatives or extensions to the concordance $\Lambda$CDM framework.

In the presence of such a coupling, the conservation equations for DE and DM are modified as follows:
\begin{eqnarray}
\dot{\rho}_{D} + 3H \rho_{D} (1+\omega_{D}) &=& -Q, \label{eq:DE_cons}\\
\dot{\rho}_{m} + 3H \rho_{m} &=& Q, \label{eq:DM_cons}
\end{eqnarray}
where $Q$ specifies the rate of energy transfer between the two sectors. In general, $Q$ may be a function of several cosmological quantities, including the Hubble parameter $H$, the deceleration parameter $q$, and the energy densities $\rho_{m}$ and $\rho_{D}$, i.e. $Q=Q(\rho_{m},\rho_{D},H,q)$. A variety of choices for this function have been considered in the literature. We now adopt the phenomenological form
\begin{equation}
Q_1 = 3 d^{2} H \rho_{m}, \label{Q}
\end{equation}
where $d^{2}$ is a dimensionless constant quantifying the strength of the interaction, often called the transfer rate or coupling parameter~\cite{ref144d,ref145d,ref146d}.

Observational analyses combining different cosmological probes — such as the Gold SNe~Ia sample, CMB data from WMAP, and BAO measurements from SDSS — suggest that $d^{2}$ should be positive defined and it should assume a small value. This outcome is in agreement with the requirements imposed by the cosmic coincidence problem as well as with thermodynamical considerations~\cite{ref147d}. Additional constraints from CMB anisotropy studies and galaxy cluster observations further indicate the range $0 < d^{2} < 0.025$~\cite{ref148d}. More generally, the parameter is usually considered within $[0,1]$, with the special case $d^{2}=0$ reducing to the standard non-interacting FLRW cosmology. It is worth stressing that many other functional forms of $Q$ have been proposed in the literature, each leading to different phenomenological consequences.

The expression of the energy density of DE is the same as in the non-interacting case.\\ 
Inserting the expression of $Q_1$ along with the expression of $H$ given in Eq. (\ref{acca}) in the continuity equation for $\rho_m$ defined in Eq. (\ref{eq:DM_cons}), we obtain:
\begin{eqnarray}
\dot{\rho}_{m} + \left(\frac{3n}{t} \right)\rho_{m} &=& \left(\frac{3nd^2}{t}\right)\rho_m \label{rhom1}
\end{eqnarray}
The solution of Eq. (\ref{rhom1}) is given by:
\begin{eqnarray}
    \rho_{m,I}(t)&=& \rho_{m0}t^{-3(1-d^2)n}\\
    &=& 3H_0^2\Omega_{m0}t^{-3(1-d^2)n}  \label{rhom1sol}.
\end{eqnarray}
Using the expression of $\rho_m$ obtained in Eq. (\ref{rhom1sol}) along with the expression of $\rho_D$ obtained in Eq. (\ref{cz3}), we can obtain the expression of $H^2$ from the Friedmann equation given in Eq. (\ref{7}), which leads to
\begin{eqnarray}
    H^2(t)= H_0^2\left[ \Omega_{m0}t^{-3(1-d^2)n} +\frac{1}{H_0^2} \left( -\frac{6\alpha}{n} + 2\beta - \gamma n + \delta n^2 \right)t^{-2}\right].  \label{acca1sol}
\end{eqnarray}
We now want to obtain the expression of the deceleration parameter  $q$ using the general definition given in Eq. (\ref{qgen}).\\
Using the expression of $H^2$ given in Eq. (\ref{acca1sol}), we obtain:
\begin{eqnarray}
q(t) = -1
- \frac{ H_0^2\Omega_{m0}\,[-3(1-d^2)n]\,t^{-3(1-d^2)n-1}
-2\left(-\dfrac{6\alpha}{n}+2\beta-\gamma n+\delta n^2\right)\,t^{-3} }
{2\left[ H_0^2\Omega_{m0}\,t^{-3(1-d^2)n} + \left(-\dfrac{6\alpha}{n}+2\beta-\gamma n+\delta n^2\right)\,t^{-2} \right]^{3/2} }.
\end{eqnarray}

We now calculate the expressions of the fractional energy densities of DM and DE in two different ways.\\
Using in the general definition of $\Omega_m$ given in Eq. (\ref{8}) the expression of $\rho_m$ given in Eq. (\ref{acca1sol}) along with the definition of $H$ given in Eq. (\ref{acca}), we obtain:
\begin{eqnarray}
    \Omega_m= \Omega_{m0}t^{2-3(1-d^2)n}.\label{Omegam1sol1}
\end{eqnarray}
Therefore, using the result of the Friedmann equation given in Eq. (\ref{11}), we can write:
\begin{eqnarray}
    \Omega_D=1-\Omega_m=1- \Omega_{m0}t^{2-3(1-d^2)n}.
\end{eqnarray}
The evolutionary form of the fractional energy density of DE is given by
\begin{eqnarray}
    \Omega'_D&=&\left[ 3(1-d^2) -\frac{2}{n} \right] \Omega_m\nonumber \\
    &=&  \left[ 3(1-d^2) -\frac{2}{n} \right] (1-\Omega_D).
\end{eqnarray}
 Instead,using in the general definition of $\Omega_m$ given in Eq. (\ref{8}) the expression of $\rho_m$ given in Eq. (\ref{rhom1sol})  along with the definition of $H^2$ given in Eq. (\ref{acca1sol}), we obtain:
\begin{equation}
\Omega_m(t) = 
\frac{
\Omega_{m0}\, t^{-3(1-d^2)n}
}{
\Omega_{m0}\, t^{-3(1-d^2)n}
+ \dfrac{1}{H_0^2}
\left( -\dfrac{6\alpha}{n} + 2\beta - \gamma n + \delta n^2 \right)
t^{-2}
}.
\label{Omegam1sol2}
\end{equation}
Therefore, using the result of the Friedmann equation given in Eq. (\ref{11}), we can write:
\begin{equation}
\Omega_D(t) =1- 
\frac{
\Omega_{m0}\, t^{-3(1-d^2)n}
}{
\Omega_{m0}\, t^{-3(1-d^2)n}
+ \dfrac{1}{H_0^2}
\left( -\dfrac{6\alpha}{n} + 2\beta - \gamma n + \delta n^2 \right)
t^{-2}
}.
\label{}
\end{equation}
The evolutionary form of the fractional energy density of DE is given by
\begin{eqnarray}
\Omega'_D
= -\left(\frac{A}{nH_0^2}\right)
\frac{\left[2-3(1-d^2)n\right]\Omega_{m0}t^{-3(1-d^2)n-2}}
{\left[\Omega_{m0}\,t^{-3(1-d^2)n}+\left(\frac{A}{H_0^2}\right)\,t^{-2}\right]^2},
\end{eqnarray}
where $A$ is defined as:
\begin{eqnarray}
A\equiv -\frac{6\alpha}{n}+2\beta-\gamma n+\delta n^2.
\end{eqnarray}
We now want to obtain the expression of the pressure of DE $p_D$.\\
The general expression of the pressure $p_D$ is given by:
\begin{eqnarray}
    p_{D}= -\rho_{D} - \frac{\dot{\rho}_{D}}{3H} - \frac{Q_1}{3H}.
\end{eqnarray}
 Using the expression of $Q_1$ given in Eq. (\ref{Q}), we can write:
\begin{eqnarray}
    \frac{Q_1}{3H} =    d^{2}  \rho_{m} = 3d^2H_0^2\Omega_{m0}t^{-3(1-d^2)n}.
\end{eqnarray}
Therefore, we obtain that:
\begin{eqnarray}
    p_{D}(t) 
    &=& \left( \frac{2}{n} -3 \right)\left[ -\frac{6\alpha}{n} + 2\beta - \gamma n + \delta n^2 \right]t^{-2}-3d^2H_0^2\Omega_{m0}t^{-3(1-d^2)n}.
\end{eqnarray}
We now want to calculate the expression of the EoS parameter of DE $\omega_D$.\\
From the continuity equation given in Eq. (\ref{eq:DE_cons}), we obtain the following expression for the EoS parameter:
\begin{eqnarray}
    \omega_{D,1,I} = -1 - \frac{\dot\rho_{D}}{3H\rho_{D}}- \frac{Q_1}{3H\rho_{D}}.
\end{eqnarray}
Using the expression of $Q_1$ given in Eq. (\ref{Q}), we can write
\begin{eqnarray}
  \frac{Q_1}{3H\rho_{D}} &=& d^2\left( \frac{\rho_{m,I} }{\rho_D}    \right)\nonumber \\
  &=&d^2 \left( \frac{ \Omega_{m,I}  }{\Omega_D}  \right).
\end{eqnarray}
Using the expression of $\Omega_m$ obtained in Eq. (\ref{Omegam1sol1}), we can write:
\begin{eqnarray}
    \omega_{D_{1,I-1}} = -1+\frac{2}{3n} - d^2 \left[\frac{\Omega_{m0}t^{2-3(1-d^2)n}}{1- \Omega_{m0}t^{2-3(1-d^2)n}}  \right].
\end{eqnarray}
Instead, using the expression of $\Omega_{m}$ given in Eq. (\ref{Omegam1sol2}), we can write:
\begin{eqnarray}
    \omega_{D_{1,I-2}} = -1+\frac{2}{3n} -  d^2\left[\frac{H_0^2\Omega_{m0} t^{2 - 3(1 - d^2)n}}{
-\frac{6\alpha}{n} + 2\beta - \gamma n + \delta n^2
}\right].
\end{eqnarray}

\subsection{Second Interacting Case}
We now consider the second interaction term given by:
\begin{eqnarray}
    Q_2 = 3Hd^2\rho_D\label{Q2}.
\end{eqnarray}
Inserting the expression of $Q_2$ along with the expression of $H$ given in Eq. (\ref{acca}) in the continuity equation for $\rho_m$ defined in Eq. (\ref{eq:DM_cons}), we obtain:
\begin{eqnarray}
\dot{\rho}_{m} + \left(\frac{3n}{t} \right)\rho_{m} &=& \left(\frac{3nd^2}{t}\right)\rho_D .
\end{eqnarray}
Therefore, using the expression of $\rho_D$ given in Eq. (\ref{cz3}), we obtain:
\begin{eqnarray}
\dot{\rho}_{m} + \left(\frac{3n}{t}\right) \rho_{m} &=& \frac{9d^2n}{t^3}\left( -\frac{6\alpha}{n} + 2\beta - \gamma n + \delta n^2 \right) \label{rhom2}.
\end{eqnarray}
The solution of Eq. (\ref{rhom2}) is given by
\begin{eqnarray}
\rho_m(t) = K t^{-3n}+
\frac{9 d^2 n}{3n - 2}\left(
-\frac{6\alpha}{n} + 2\beta - \gamma n + \delta n^2
\right)t^{-2} ,\label{rhom2sol}
\end{eqnarray}
where $K$ is a constant.\\
We can relate the constant $K$ to the present day of $\rho_m$, indicated with $\rho_{m0}$, and to the other parameters of the model we consider.\\
At present time, i.e. for $t=t_0$ (with $t_0$ being the present day age of the Universe), we have that $\rho_m(t=t_0)=\rho_{m0}$, therefore from Eq. (\ref{rhom2sol}) we obtain:
\begin{eqnarray}
K = t_0^{3n} \left[ \rho_{m0} - \frac{9 d^2 n}{3n - 2} \left(-\frac{6\alpha}{n} + 2\beta - \gamma n + \delta n^2\right) t_0^{-2} \right].
\end{eqnarray}
Using the expression of $\rho_m$ obtained in Eq. (\ref{rhom2sol}) along with the expression of $\rho_D$ obtained in Eq. (\ref{cz3}), we can obtain the expression of $H^2$ from the Friedmann equation given in Eq. (\ref{7}), which leads to
\begin{eqnarray}
 H^2(t)
&=&\left\{ \left(\frac{K}{3}\right) t^{-3n}  +\left[ 1+\frac{3 d^2 n}{(3n - 2)}  \right] \left( -\frac{6\alpha}{n} + 2\beta - \gamma n + \delta n^2 \right)t^{-2}\right\}. \label{acca2sol}
\end{eqnarray}
We can also relate the constant $K$ to the present day value of the Hubble parameter, i.e. $H_0$.\\
Considering $H(t=t_0)=H_0$ in Eq. (\ref{acca2sol}), we obtain:
\begin{eqnarray}
K &=& 3\, t_0^{3n} \left[
H_0^2 -
\left( 1 + \frac{3 d^2 n}{3n - 2} \right)
\left( -\frac{6\alpha}{n} + 2\beta - \gamma n + \delta n^2 \right)
t_0^{-2}
\right]. \label{Kresult}
\end{eqnarray}
We now want to obtain the expression of the deceleration parameter  $q$ using the general definition given in Eq. (\ref{qgen}).\\
Using the expression of $H^2$ given in Eq. (\ref{acca2sol}), we obtain:
\begin{equation}
q(t) = -1 + \frac{3 n A \, t^{-3n-1} + 2 B \, t^{-3}}{2 \left( A \, t^{-3n} + B  t^{-2} \right)^{3/2}},
\end{equation}
where $A$ and $B$ are given by:
\begin{eqnarray}
A &=& \frac{K}{3}, \\
B &=& \left( 1 + \frac{3 d^2 n}{3 n - 2} \right) \left( -\frac{6\alpha}{n} + 2 \beta - \gamma n + \delta n^2 \right).
\end{eqnarray}

We now calculate the expressions of the fractional energy densities of DM and DE in two different ways.\\
Using in the general definition of $\Omega_m$ given in Eq. (\ref{8}) the expression of $\rho_m$ given in Eq. (\ref{rhom2sol}) along with the definition of $H$ given in Eq. (\ref{acca}), we obtain:
\begin{equation}
\Omega_m(t) = \frac{1}{3 n^2} \left[  K \, t^{-(3n-2)}+ \frac{9 d^2 n}{3n - 2} \left( -\frac{6\alpha}{n} + 2\beta - \gamma n + \delta n^2 \right)  \right].
\end{equation}
Therefore, using the result of the Friedmann equation given in Eq. (\ref{11}), we can write:
\begin{equation}
\Omega_D(t) = 1-\frac{1}{3 n^2} \left[ K \, t^{-(3n-2)}+ \frac{9 d^2 n}{3n - 2} \left( -\frac{6\alpha}{n} + 2\beta - \gamma n + \delta n^2 \right)  \right].
\end{equation}
The evolutionary form of the fractional energy density of DE is given by
\begin{equation}
\Omega_D' 
= \left[\frac{K (3n-2)}{3 n^3} \right]\, t^{-(3n-2)} \,.
\end{equation}
 Instead, using in the general definition of $\Omega_m$ given in Eq. (\ref{8}) the expression of $\rho_m$ given in Eq. (\ref{rhom2sol}) along with the definition of $H^2$ given in Eq. (\ref{acca2sol}), we obtain:
\begin{eqnarray}
\Omega_m(t)&=&
\frac{K\,t^{-3n} + \dfrac{9 d^2 n}{3n-2}\left(-\dfrac{6\alpha}{n}+2\beta-\gamma n+\delta n^2\right)\,t^{-2}}
{K\,t^{-3n} + 3\left(1+\dfrac{3 d^2 n}{3n-2}\right)\left(-\dfrac{6\alpha}{n}+2\beta-\gamma n+\delta n^2\right)\,t^{-2}}\nonumber \\
&=&\frac{K + \dfrac{9 d^2 n}{3n-2}\left(-\dfrac{6\alpha}{n}+2\beta-\gamma n+\delta n^2\right)\,t^{3n-2}}
{K + 3\left(1+\dfrac{3 d^2 n}{3n-2}\right)\left(-\dfrac{6\alpha}{n}+2\beta-\gamma n+\delta n^2\right)\,t^{3n-2}}.
\end{eqnarray}
Therefore, using the result of the Friedmann equation given in Eq. (\ref{11}), we can write:
\begin{eqnarray}
\Omega_D(t)&=&1-
\frac{K\,t^{-3n} + \dfrac{9 d^2 n}{3n-2}\left(-\dfrac{6\alpha}{n}+2\beta-\gamma n+\delta n^2\right)\,t^{-2}}
{K\,t^{-3n} + 3\left(1+\dfrac{3 d^2 n}{3n-2}\right)\left(-\dfrac{6\alpha}{n}+2\beta-\gamma n+\delta n^2\right)\,t^{-2}}\nonumber \\
&=&1-\frac{K + \dfrac{9 d^2 n}{3n-2}\left(-\dfrac{6\alpha}{n}+2\beta-\gamma n+\delta n^2\right)\,t^{3n-2}}
{K + 3\left(1+\dfrac{3 d^2 n}{3n-2}\right)\left(-\dfrac{6\alpha}{n}+2\beta-\gamma n+\delta n^2\right)\,t^{3n-2}}.
\end{eqnarray}
The evolutionary form of the fractional energy density of DE is given by
\begin{eqnarray}
\Omega_D'(t) = \frac{3(3n-2)\,K\,St^{3n-2}}{n\left(K + Z S \, t^{3n-2}\right)^2},
\end{eqnarray}
where we defined:
\begin{eqnarray}
    S &=& -\frac{6\alpha}{n} + 2\beta - \gamma n + \delta n^2, \\
Z &=& 3 + \frac{9 d^2 n}{3n-2}.
\end{eqnarray}
We now want to obtain the expression of the pressure of DE $p_D$.\\ 
Using the expression of $Q_2$ given in Eq. (\ref{Q2}), we can write:
\begin{eqnarray}
     \frac{Q_2}{3H} = d^2\rho_D. \label{frac2}
\end{eqnarray}
Therefore, using the expressions of $\rho_{D}$ and $H$ we derived in Eqs. (\ref{cz3}) and (\ref{acca}) along with the result of Eq. (\ref{frac2}), we can write: 
\begin{eqnarray}
    p_{D}(t) 
    &=& \left( \frac{2}{n} -3 -d^2\right)\left[ -\frac{6\alpha}{n} + 2\beta - \gamma n + \delta n^2 \right]t^{-2}.
\end{eqnarray}
We now want to calculate the expression of the EoS parameter of DE $\omega_D$. \\
The general expression of $\omega_D$ for this case is given by
\begin{eqnarray}
    \omega_{D} = -1 - \frac{\dot\rho_{D}}{3H\rho_{D}}- \frac{Q_2}{3H\rho_{D}}.
\end{eqnarray}
Using  the expression of $Q_2$ given in Eq. (\ref{Q2}), we can write
\begin{eqnarray}
    \frac{Q_2}{3H\rho_{D}}  =  d^2.
\end{eqnarray}
Therefore, we obtain the following relation for $\omega_{D_2}$:
\begin{eqnarray}
 \omega_{D_{2,I}} = -1+\frac{2}{3n}-d^2
\end{eqnarray}
which is a constant depending on the power law index of the scale factor $n$ and on the interaction term $d^2$.\\
The positive term $2/(3n)$ partially offsets the negative contributions of $-1$ and $-d^2$, so that for sufficiently small $n$ satisfying $2/(3n) > d^2$, the dark energy behaves as quintessence ($\omega_D > -1$). Conversely, for large $n$ such that $2/(3n) < d^2$, the equation of state becomes phantom-like ($\omega_D < -1$). This demonstrates that the model can naturally accommodate both quintessence and phantom regimes depending on the choice of parameters.

\subsubsection{Limiting Case of $n = 2/3$}
In the limiting case of $n = 2/3$, the solution of the continuity equation for $\rho_m$ is given by:
\begin{eqnarray}
\rho_m(t) = K\,t^{-2} + 9 d^2 n \left( -\frac{6\alpha}{n} + 2\beta - \gamma n + \delta n^2  \right)\, t^{-2}\ln t,
\end{eqnarray}
where $K$ is an integration constant. \\
Following the same procedure of the general case, we obtain the following relation for $K$:
\begin{eqnarray}
K = t_0^{2} \rho_{m0} - 9 d^2 n \left( -\frac{6\alpha}{n} + 2\beta - \gamma n + \delta n^2 \right) \ln t_0. \label{rhom2sol2}
\end{eqnarray}
Using the expression of $\rho_m$ obtained in Eq. (\ref{rhom2sol2}) along with the expression of $\rho_D$ obtained in Eq. (\ref{cz3}), we can obtain the expression of $H^2$ from the Friedmann equation given in Eq. (\ref{7}), which leads to
\begin{eqnarray}
H^2(t) = t^{-2} \left[ \frac{K}{3} + \Bigl( -\frac{6\alpha}{n} + 2\beta - \gamma n + \delta n^2 \Bigr) \Bigl( 1 + 3 d^2 n \ln t \Bigr) \right]. \label{acca2sol2}
\end{eqnarray}
In this case, the expression of $K$ as a function of $H_0$ is given by:
\begin{eqnarray}
K &=& 3 \left[
t_0^{2} H_0^2 -
\left( -\frac{6\alpha}{n} + 2\beta - \gamma n + \delta n^2 \right)
\left( 1 + 3 d^2 n \ln t_0 \right)
\right]. \label{Kresult2}
\end{eqnarray}
Using the expression of $H^2$ obtained in Eq. (\ref{acca2sol2}), we can write the deceleration parameter as follow:
\begin{equation}
q(t) = -2 + \frac{3\, A\, d^2 n}{2 \Bigl[ \frac{K}{3} + A \bigl( 1 + 3 d^2 n \ln t \bigr) \Bigr]},
\end{equation}
where we defined $A$ as:
\begin{eqnarray}
    A = -\frac{6\alpha}{n} + 2\beta - \gamma n + \delta n^2.
\end{eqnarray}

We now calculate the expressions of the fractional energy densities of DM and DE in two different ways.\\
Using in the general definition of $\Omega_m$ given in Eq. (\ref{8}) the expression of $\rho_m$ given in Eq. (\ref{rhom2sol2}) along with the definition of $H$ given in Eq. (\ref{acca}), we obtain:
\begin{eqnarray}
\Omega_m(t) =
\frac{K + 9 d^2 n \left(-\frac{6\alpha}{n} + 2\beta - \gamma n + \delta n^2 \right) \ln t}{3 n^2}.
\end{eqnarray}
Therefore, using the result of the Friedmann equation given in Eq. (\ref{11}), we can write:
\begin{eqnarray}
\Omega_D(t) =1-
\frac{K + 9 d^2 n \left(-\frac{6\alpha}{n} + 2\beta - \gamma n + \delta n^2 \right) \ln t}{3 n^2}.
\end{eqnarray}
The evolutionary form of the fractional energy density of DE is given by
\begin{eqnarray}
\Omega_D' = -\frac{3 d^2}{n^2}
\left(-\frac{6\alpha}{n} + 2\beta - \gamma n + \delta n^2 \right),
\end{eqnarray}
which is a constant depending on the parameters of the model we consider and on the value of the power law index of the scale factor.\\
 Instead, using in the general definition of $\Omega_m$ given in Eq. (\ref{8}) the expression of $\rho_m$ given in Eq. (\ref{rhom2sol2}) along with the definition of $H^2$ given in Eq. (\ref{acca2sol2}), we obtain:
\begin{eqnarray}
\Omega_m(t) =
\frac{
K + 9 d^2 n \left(-\frac{6\alpha}{n} + 2\beta - \gamma n + \delta n^2 \right) \ln t
}{
K + 3 \left(-\frac{6\alpha}{n} + 2\beta - \gamma n + \delta n^2 \right)
\left( 1 + 3 d^2 n \ln t \right)
}.
\end{eqnarray}
Therefore, using the result of the Friedmann equation given in Eq. (\ref{11}), we can write:
\begin{eqnarray}
\Omega_D(t) =1-
\frac{
K + 9 d^2 n \left(-\frac{6\alpha}{n} + 2\beta - \gamma n + \delta n^2 \right) \ln t
}{
K + 3 \left(-\frac{6\alpha}{n} + 2\beta - \gamma n + \delta n^2 \right)
\left( 1 + 3 d^2 n \ln t \right)
}.
\end{eqnarray}
The evolutionary form of the fractional energy density of DE is given by
\begin{eqnarray}
\Omega_D'(t) =
-\,\frac{
27\,d^2
\left(
-\dfrac{6\alpha}{n} + 2\beta - \gamma n + \delta n^2
\right)^2
}{
\left[
K
+ 3\left(-\dfrac{6\alpha}{n} + 2\beta - \gamma n + \delta n^2\right)
+ 9 d^2 n \left(-\dfrac{6\alpha}{n} + 2\beta - \gamma n + \delta n^2\right) \ln t
\right]^2
}.
\end{eqnarray}
The expressions of the pressure of DE $p_D$ and of the EoS parameter of DE $\omega_D$ are the same of the general case but for $n=2/3$.

\subsection{Third Interacting Case}
We now consider the third interaction term given by:
\begin{eqnarray}
    Q_3 = 3Hd^2(\rho_m + \rho_D)\label{Q3}.
\end{eqnarray}
Inserting the expression of $Q_3$ given in Eq. (\ref{Q3}) along with the expression of $H$ given in Eq. (\ref{acca}) in the continuity equation for $\rho_m$ defined in Eq. (\ref{eq:DM_cons}), we obtain:
\begin{eqnarray}
\dot{\rho}_{m} + \left(\frac{3n}{t}\right) \rho_{m} &=& \left(\frac{3d^2n}{t}\right)\left(\rho_m+\rho_D\right) ,
\end{eqnarray}
or equivalently
\begin{eqnarray}
\dot{\rho}_{m} + \left[\frac{3n(1-d^2)}{t}\right] \rho_{m} &=& \left(\frac{3d^2n}{t}\right)\rho_D.
\end{eqnarray}
Therefore, using the expression of $\rho_D$ given in Eq. (\ref{cz3}), we obtain:
\begin{eqnarray}
\dot{\rho}_{m} + \left[\frac{3n(1-d^2)}{t}\right]  \rho_{m} &=&  \frac{9d^2n}{t^3}\left( -\frac{6\alpha}{n} + 2\beta - \gamma n + \delta n^2 \right). \label{rhomgen3}
\end{eqnarray}
The solution of Eq. (\ref{rhomgen3}) is given by:
\begin{eqnarray}
\rho_m(t) = K\,t^{-3n(1-d^2)}+
\frac{9d^2 n}{3n(1-d^2) - 2}
\left(-\frac{6\alpha}{n} + 2\beta - \gamma n + \delta n^2\right)t^{-2}, \label{rhom3sol}
\end{eqnarray}
where $K$ is an integration constant.\\
We can relate the constant $K$ to the present day of $\rho_m$, indicated with $\rho_{m0}$, and to the other parameters of the model we consider.\\
At present time, i.e. for $t=t_0$ (with $t_0$ being the present day age of the Universe), we have that $\rho_m(t=t_0)=\rho_{m0}$, therefore from Eq. (\ref{rhom3sol}) we obtain:
\begin{eqnarray}
K &=& t_0^{3n(1-d^2)}
\left[
\rho_{m0} -
\frac{9d^2 n}{3n(1-d^2) - 2}
\left(-\frac{6\alpha}{n} + 2\beta - \gamma n + \delta n^2\right)
t_0^{-2}
\right].
\label{K_rhom3}
\end{eqnarray}
Using the expression of $\rho_m$ obtained in Eq. (\ref{rhom3sol}) along with the expression of $\rho_D$ obtained in Eq. (\ref{cz3}), we can obtain the expression of $H^2$ from the Friedmann equation given in Eq. (\ref{7}), which leads to
\begin{equation}
H^2(t) = \left(\frac{K}{3}\right) \, t^{-3n(1-d^2)} 
+ \left[ 1 + \frac{3 d^2 n}{3n(1-d^2)-2} \right] 
\left(-\frac{6\alpha}{n} + 2\beta - \gamma n + \delta n^2 \right) t^{-2}. \label{acca3sol}
\end{equation}
We can also relate the constant $K$ to the present day value of the Hubble parameter, i.e. $H_0$.\\
Considering $H(t=t_0)=H_0$ in Eq. (\ref{acca2sol}), we obtain:
\begin{eqnarray}
K &=& 3\, t_0^{3n(1-d^2)} \left\{ H_0^2 - \left[ 1 + \frac{3 d^2 n}{3n(1-d^2)-2} \right] \left(-\frac{6\alpha}{n} + 2\beta - \gamma n + \delta n^2 \right) t_0^{-2} \right\}.
\end{eqnarray}
We now want to obtain the expression of the deceleration parameter  $q$ using the general definition given in Eq. (\ref{qgen}).\\
Using the expression of $H^2$ given in Eq. (\ref{acca3sol}), we obtain:
\begin{equation}
q(t) 
= -1 + \frac{
K\, 3n(1-d^2) \, t^{-3n(1-d^2)-1} + 2 
\left[ 1 + \frac{3 d^2 n}{3n(1-d^2)-2} \right] 
\left(-\frac{6\alpha}{n} + 2\beta - \gamma n + \delta n^2 \right) t^{-3}
}{2 \left[ \frac{K}{3} \, t^{-3n(1-d^2)} 
+ \left[ 1 + \frac{3 d^2 n}{3n(1-d^2)-2} \right] 
\left(-\frac{6\alpha}{n} + 2\beta - \gamma n + \delta n^2 \right) t^{-2} \right]^{3/2}} .
\end{equation}
We now calculate the expressions of the fractional energy densities of DM and DE in two different ways.\\
Using in the general definition of $\Omega_m$ given in Eq. (\ref{8}) the expression of $\rho_m$ given in Eq. (\ref{rhom3sol}) along with the definition of $H$ given in Eq. (\ref{acca}), we obtain:
\begin{equation}
\Omega_m(t) = \left(\frac{K}{3 n^2}\right) t^{2 - 3 n (1-d^2)}+\frac{3 d^2}{\,n \,[3 n(1-d^2) - 2]\,} \left(-\frac{6\alpha}{n} + 2\beta - \gamma n + \delta n^2 \right).
\end{equation}
Therefore, using the result of the Friedmann equation given in Eq. (\ref{11}), we can write:
\begin{equation}
\Omega_D(t) = 1- \left(\frac{K}{3 n^2}\right) t^{2 - 3 n (1-d^2)}-\frac{3 d^2}{\,n \,[3 n(1-d^2) - 2]\,} \left(-\frac{6\alpha}{n} + 2\beta - \gamma n + \delta n^2 \right).\label{Omegad3sol1}
\end{equation}
The evolutionary form of the fractional energy density of DE is given by
\begin{equation}
\Omega_D' = - \left(\frac{K}{3 n^3}\right) \left[2 - 3 n (1-d^2)\right]\, t^{2 - 3 n (1-d^2)}.
\end{equation}
 Instead, using in the general definition of $\Omega_m$ given in Eq. (\ref{8}) the expression of $\rho_m$ given in Eq. (\ref{rhom3sol}) along with the definition of $H^2$ given in Eq. (\ref{acca3sol}), we obtain:
\begin{eqnarray}
\Omega_m(t) = 
\frac{
K\, t^{-3n(1-d^2)} + \frac{9 d^2 n}{3n(1-d^2) - 2} \left(-\frac{6\alpha}{n} + 2\beta - \gamma n + \delta n^2\right) t^{-2}
}{
K\, t^{-3n(1-d^2)} + 3 \left( 1 + \frac{3 d^2 n}{3n(1-d^2)-2} \right) \left(-\frac{6\alpha}{n} + 2\beta - \gamma n + \delta n^2 \right) t^{-2}
}.
\end{eqnarray}
Therefore, using the result of the Friedmann equation given in Eq. (\ref{11}), we can write:
\begin{eqnarray}
\Omega_D(t) =1- 
\frac{
K\, t^{-3n(1-d^2)} + \frac{9 d^2 n}{3n(1-d^2) - 2} \left(-\frac{6\alpha}{n} + 2\beta - \gamma n + \delta n^2\right) t^{-2}
}{
K\, t^{-3n(1-d^2)} + 3 \left[ 1 + \frac{3 d^2 n}{3n(1-d^2)-2} \right] \left(-\frac{6\alpha}{n} + 2\beta - \gamma n + \delta n^2 \right) t^{-2}
}.\label{Omegad3sol2}
\end{eqnarray}
The evolutionary form of the fractional energy density of DE is given by:
\begin{eqnarray}
\Omega_D'(t) &=&
- \left\{\left[3n(1-d^2)-2\right] \left\{
\frac{9 d^2 n}{3n(1-d^2)-2} \left(-\frac{6\alpha}{n}+2\beta - \gamma n + \delta n^2\right) \right. \right. \nonumber \\
&&\left.
\left. - 3\left[1+\frac{3 d^2 n}{3n(1-d^2)-2}\right]\left(-\frac{6\alpha}{n}+2\beta - \gamma n + \delta n^2\right)
\right\} t^{-3n(1-d^2)-2}\right\}\times\nonumber \\
&&\left\{
n \left\{
K t^{-3n(1-d^2)} + 3 \left[1+\frac{3 d^2 n}{3n(1-d^2)-2}\right] \times \right. \right. \nonumber \\
&& \left. \left.\left(-\frac{6\alpha}{n}+2\beta - \gamma n + \delta n^2\right) t^{-2}
\right\}^2\right\}^{-1}.
\end{eqnarray}

We now want to obtain the expression of the pressure of DE $p_D$.\\ 
The general expression of the pressure is given by:
\begin{eqnarray}
    p_{D}= -\rho_{D} - \frac{\dot{\rho}_{D}}{3H} - \frac{Q_3}{3H}.
\end{eqnarray}
Using  the expression of $Q_3$ given in Eq. (\ref{Q3}), we can write:
\begin{eqnarray}
   \frac{Q_3}{3H} =  d^2(\rho_m + \rho_D). \label{Q3frac}
\end{eqnarray}
Therefore, using the expressions of $\rho_{D}$ and $H$ we derived in Eqs. (\ref{cz3}) and (\ref{acca}) along with the result of Eq. (\ref{Q3frac}), we can write:
\begin{eqnarray}
    p_{D}(t) 
&=&\left[ \frac{2}{n} -3-d^2 - \frac{9d^4 n}{3n(1-d^2) - 2}  \right]\left(-\frac{6\alpha}{n} + 2\beta - \gamma n + \delta n^2\right)t^{-2}
- d^2K\,t^{-3n(1-d^2)}\nonumber \\
&=& \left[ 
\frac{6 d^{4} n^{2} - 6 d^{2} n^{2} + 4 d^{2} n + 9 n^{2} - 12 n + 4}
{n\,(3 d^{2} n - 3 n + 2)}
  \right]\left(-\frac{6\alpha}{n} + 2\beta - \gamma n + \delta n^2\right)t^{-2}\nonumber \\
  &&
- d^2K\,t^{-3n(1-d^2)}.
\end{eqnarray}

We now want to calculate the expression of the EoS parameter of DE $\omega_D$. \\
The general expression of $\omega_D$ for this case is given by:
\begin{eqnarray}
    \omega_{D} = -1 - \frac{\dot\rho_{D}}{3H\rho_{D}}- \frac{Q_3}{3H\rho_{D}}.
\end{eqnarray}
Using  the expression of $Q_3$ given in Eq. (\ref{Q3}), we can write:
\begin{eqnarray}
  \frac{Q_3}{3H\rho_{D}}&=&    \frac{ 3Hd^2(\rho_m + \rho_D)}{3H\rho_{D}} \nonumber \\
  &=&d^2\left(\frac{\rho_D+\rho_m}{\rho_D}\right)\nonumber \\
  &=&d^2\left(\frac{\Omega_D+\Omega_m}{\Omega_D}\right)\nonumber \\&=& \frac{d^2}{\Omega_D},
\end{eqnarray}
where we used the relation $\Omega_m + \Omega_D=1$.\\
We can write, then:
\begin{eqnarray}
    \omega_{D} = -1 - \frac{\dot\rho_{D}}{3H\rho_{D}}- \frac{d^2}{\Omega_D}.
\end{eqnarray}
Therefore, using the expression of $\Omega_D$ given in Eq. (\ref{Omegad3sol1}), we obtain the following relation for $\omega_D$:
\begin{eqnarray}
    \omega_{D_{3,I-1}} = -1 + \frac{2}{3n }- \frac{d^2}{ 1- \frac{K}{3 n^2}\, t^{2 - 3 n (1-d^2)}-\frac{3 d^2}{\,n \,[3 n(1-d^2) - 2]\,} \left(-\frac{6\alpha}{n} + 2\beta - \gamma n + \delta n^2 \right)}   .
\end{eqnarray}
Instead, using the expression of $\Omega_D$ given in Eq. (\ref{Omegad3sol2}), we obtain the following relation for $\omega_D$:
\begin{eqnarray}
    \omega_{D_{3,I-2}} = -1 + \frac{2}{3n }- \frac{d^2}{1- 
\frac{
K\, t^{-3n(1-d^2)} + \frac{9 d^2 n}{3n(1-d^2) - 2} \left(-\frac{6\alpha}{n} + 2\beta - \gamma n + \delta n^2\right) t^{-2}
}{
K\, t^{-3n(1-d^2)} + 3 \left( 1 + \frac{3 d^2 n}{3n(1-d^2)-2} \right) \left(-\frac{6\alpha}{n} + 2\beta - \gamma n + \delta n^2 \right) t^{-2}
}}.   
\end{eqnarray}

\subsubsection{Limiting Case of $n(1-d^2) = 2/3$}
In the limiting case corresponding to $n(1-d^2) = 2/3$, the continuity equation for $\rho_m$ has the following general solution:
\begin{eqnarray}
\rho_m(t) = K\,t^{-2} + 9d^2 n \left( -\frac{6\alpha}{n} + 2\beta - \gamma n + \delta n^2 \right)\, t^{-2}\ln t. \label{rhom3sol2}
\end{eqnarray}
In this case, the expression of $K$ is given by:
\begin{eqnarray}
K &=& t_0^{2}\rho_{m0} 
- 9d^2 n 
\left( -\frac{6\alpha}{n} + 2\beta - \gamma n + \delta n^2 \right)
\ln t_0.
\label{K_rhom3sol2}
\end{eqnarray}
Using the expression of $\rho_m$ obtained in Eq. (\ref{rhom3sol2}) along with the expression of $\rho_D$ obtained in Eq. (\ref{cz3}), we can obtain the expression of $H^2$ from the Friedmann equation given in Eq. (\ref{7}), which leads to
\begin{equation}
H^2(t) = \left[ \frac{K}{3} + \left(1 + 3d^2 n\ln t\right) \left(-\frac{6\alpha}{n} + 2\beta - \gamma n + \delta n^2 \right) \right] t^{-2} \label{acca3sol2}
\end{equation}
Following the same procedure of the general case, we have:
\begin{eqnarray}
K &=& 3 \left[ H_0^2 t_0^2 - \left(1 + 3 d^2 n \ln t_0 \right) \left(-\frac{6\alpha}{n} + 2\beta - \gamma n + \delta n^2 \right) \right].
\end{eqnarray}

We now want to obtain the expression of the deceleration parameter  $q$ using the general definition given in Eq. (\ref{qgen}).\\
Using the expression of $H^2$ given in Eq. (\ref{acca3sol2}), we obtain:
\begin{eqnarray}
q(t)   
= \frac{C_0 + C_1 (1 + \ln t)}{2 (C_0 + C_1 \ln t)}.
\end{eqnarray}
where $C_0$ and $C_1$ are defined as:
\begin{align}
C_0 &= \frac{K}{3} + \left(-\frac{6\alpha}{n} + 2\beta - \gamma n + \delta n^2 \right), \\
 C_1 &= 3 d^2 n \left(-\frac{6\alpha}{n} + 2\beta - \gamma n + \delta n^2 \right).
\end{align}

We now calculate the expressions of the fractional energy densities of DM and DE in two different ways.\\
Using in the general definition of $\Omega_m$ given in Eq. (\ref{8}) the expression of $\rho_m$ given in Eq. (\ref{rhom3sol2}) along with the definition of $H$ given in Eq. (\ref{acca}), we obtain:
\begin{equation}
\Omega_m(t) = \frac{K + 9 d^2 n \left(-\frac{6\alpha}{n} + 2\beta - \gamma n + \delta n^2 \right) \ln t}{3 n^2}.
\end{equation}
Therefore, using the result of the Friedmann equation given in Eq. (\ref{11}), we can write:
\begin{equation}
\Omega_D(t) = 1-\frac{K + 9 d^2 n \left(-\frac{6\alpha}{n} + 2\beta - \gamma n + \delta n^2 \right) \ln t}{3 n^2}.\label{sergio1}
\end{equation}
The evolutionary form of the fractional energy density of DE is given by
\begin{equation}
\Omega_D' = - \frac{3 d^2 \left(-\frac{6\alpha}{n} + 2\beta - \gamma n + \delta n^2 \right)}{n^2}.
\end{equation}
Therefore we obtain that, in this case, $\Omega_D' $ is a constant.\\
 Instead, using in the general definition of $\Omega_m$ given in Eq. (\ref{8}) the expression of $\rho_m$ given in Eq. (\ref{rhom3sol2}) along with the definition of $H^2$ given in Eq. (\ref{acca3sol2}), we obtain:
\begin{eqnarray}
\Omega_m(t) =
\frac{
K + 9 d^2 n \left(-\frac{6\alpha}{n} + 2\beta - \gamma n + \delta n^2 \right) \ln t
}{
K + 3 \left( 1 + 3 d^2 n \ln t \right) \left(-\frac{6\alpha}{n} + 2\beta - \gamma n + \delta n^2 \right)
}.
\end{eqnarray}
Therefore, using the result of the Friedmann equation given in Eq. (\ref{11}), we can write:
\begin{eqnarray}
\Omega_D(t) =1-
\frac{
K + 9 d^2 n \left(-\frac{6\alpha}{n} + 2\beta - \gamma n + \delta n^2 \right) \ln t
}{
K + 3 \left( 1 + 3 d^2 n \ln t \right) \left(-\frac{6\alpha}{n} + 2\beta - \gamma n + \delta n^2 \right)
}.\label{sergio2}
\end{eqnarray}
The evolutionary form of the fractional energy density of DE is given by:
\begin{eqnarray}
\Omega_D'(t) = - \frac{27 d^2 \left(-\frac{6\alpha}{n} + 2\beta - \gamma n + \delta n^2\right)^2}{\left[ K + 3 \left( 1 + 3 d^2 n \ln t \right) \left(-\frac{6\alpha}{n} + 2\beta - \gamma n + \delta n^2 \right) \right]^2}.
\end{eqnarray}
We now want to obtain the expression of the pressure of DE $p_D$.\\ 
Using the expressions of $\rho_{D}$ and $H$ we derived in Eqs. (\ref{cz3}) and (\ref{acca}) along with the result of Eq. (\ref{rhom3sol2}), we can write: 
\begin{eqnarray}
    p_{D}(t) 
    &=& \left[\left( \frac{2}{n} -3-d^2 -9d^4 n\ln t \right)\left( -\frac{6\alpha}{n} + 2\beta - \gamma n + \delta n^2 \right)-d^2K\right]t^{-2}.
\end{eqnarray}
We now want to calculate the expression of the EoS parameter of DE $\omega_D$. \\
The general expression for this case is given by:
\begin{eqnarray}
     \omega_{D} = -1 - \frac{\dot\rho_{D}}{3H\rho_{D}}- \frac{d^2}{\Omega_D}.
    \end{eqnarray}
Using the expression of $\Omega_D$ given in Eq. (\ref{sergio1}), we obtain the following relation for $\omega_D$:
\begin{eqnarray}
    \omega_{D_{3,I-1,lim}} &=& -1 + \frac{2}{3n }- \frac{d^2}{1-\frac{K + 9 d^2 n \left(-\frac{6\alpha}{n} + 2\beta - \gamma n + \delta n^2 \right) \ln t}{3 n^2}}   \nonumber \\
    &=& -1 + \frac{2}{3n }- \frac{3n^2d^2}{3n^2-K - 9 d^2 n \left(-\frac{6\alpha}{n} + 2\beta - \gamma n + \delta n^2 \right) \ln t}.
\end{eqnarray}
Instead, using the expression of $\Omega_D$ given in Eq. (\ref{sergio2}), we obtain the following relation for $\omega_D$:
\begin{eqnarray}
    \omega_{D_{3,I-2,lim}} = -1 + \frac{2}{3n }- \frac{d^2}{1-
\frac{
K + 9 d^2 n \left(-\frac{6\alpha}{n} + 2\beta - \gamma n + \delta n^2 \right) \ln t
}{
K + 3 \left( 1 + 3 d^2 n \ln t \right) \left(-\frac{6\alpha}{n} + 2\beta - \gamma n + \delta n^2 \right)
}}.   
\end{eqnarray}

\subsection{Fourth Interacting Case}
We now consider the fourth interaction term given by:
\begin{eqnarray}
    Q_4 = 3H \left(d_1^2\rho_m + d_2^2\rho_D\right).\label{Q4}
\end{eqnarray}
Inserting the expression of $Q_4$ along with the expression of $H$ given in Eq. (\ref{acca}) in the continuity equation for $\rho_m$ defined in Eq. (\ref{eq:DM_cons}), we obtain:
\begin{eqnarray}
\dot{\rho}_{m} + \left(\frac{3n}{t}\right)  \rho_{m} &=&\left( \frac{3d_1^2n}{t}\right)\rho_D + \left(\frac{3d_2^2n}{t}\right)\rho_m.\label{rho4gen}
\end{eqnarray}
Using the expression of $\rho_D$ given in Eq. (\ref{cz3}), we obtain:
\begin{eqnarray}
\dot{\rho}_{m} + \left(\frac{3n}{t}\right) \rho_{m} &=& \frac{9d_1^2n}{t^3}\left( -\frac{6\alpha}{n} + 2\beta - \gamma n + \delta n^2 \right) + \left(\frac{3d_2^2n}{t}\right)\rho_m.\label{rho4gen}
\end{eqnarray}
The solution of Eq. (\ref{rho4gen}) is given by:
\begin{eqnarray}
\rho_m(t) = K\,t^{-3n(1-d_2^2)}+
\frac{9d_1^2 n}{3n(1-d_2^2) - 2}
\left(-\frac{6\alpha}{n} + 2\beta - \gamma n + \delta n^2\right)t^{-2}.\label{rho4sol}
\end{eqnarray}
where $K$ is an integration constant.\\
We can relate the constant $K$ to the present day of $\rho_m$, indicated with $\rho_{m0}$, and to the other parameters of the model we consider.\\
At present time, i.e. for $t=t_0$ (with $t_0$ being the present day age of the Universe), we have that $\rho_m(t=t_0)=\rho_{m0}$, therefore from Eq. (\ref{rho4sol}) we obtain:
\begin{eqnarray}
K &=& t_0^{3n(1-d_2^2)}\left[\rho_{m0}
- \frac{9d_1^2 n}{3n(1-d_2^2) - 2}\left( -\frac{6\alpha}{n} + 2\beta - \gamma n + \delta n^2 \right)t_0^{-2}\right].
\label{K_rho4}
\end{eqnarray}

Using the expression of $\rho_m$ obtained in Eq. (\ref{rho4sol}) along with the expression of $\rho_D$ obtained in Eq. (\ref{cz3}), we can obtain the expression of $H^2$ from the Friedmann equation given in Eq. (\ref{7}), which leads to
\begin{equation}
H^2(t) = \frac{K}{3} \, t^{-3n(1-d_2^2)} 
+ \left[ 1 + \frac{3 d_1^2 n}{3n(1-d_2^2)-2} \right] 
\left(-\frac{6\alpha}{n} + 2\beta - \gamma n + \delta n^2 \right) t^{-2}.\label{acca4sol}
\end{equation}
We can also relate the constant $K$ to the present day value of the Hubble parameter, i.e. $H_0$.\\
Considering $H(t=t_0)=H_0$ in Eq. (\ref{acca4sol}), we obtain:
\begin{eqnarray}
K &=& 3\, t_0^{3n(1-d_2^2)} \left\{ H_0^2 - \left[ 1 + \frac{3 d_1^2 n}{3n(1-d_2^2)-2} \right] 
\left(-\frac{6\alpha}{n} + 2\beta - \gamma n + \delta n^2 \right) t_0^{-2} \right\}.
\end{eqnarray}

We now want to obtain the expression of the deceleration parameter  $q$ using the general definition given in Eq. (\ref{qgen}).\\
Using the expression of $H^2$ given in Eq. (\ref{acca4sol}), we obtain:
\begin{equation}
q(t) = -1 + 
\frac{ 3\, n \,(1-d_2^2)\, A \, t^{-3 n (1-d_2^2)-1} + 2B \, t^{-3} }
     { 2\left( A \, t^{-3 n (1-d_2^2)} + B \, t^{-2} \right)^{3/2} },
\end{equation}
where $A$ and $B$ are defined as:
\begin{align}
A &= \frac{K}{3}, \\
B &= \left[ 1 + \frac{3 d_1^2 n}{3n(1-d_2^2)-2} \right] 
\left(-\frac{6\alpha}{n} + 2\beta - \gamma n + \delta n^2 \right).
\end{align}

We now calculate the expressions of the fractional energy densities of DM and DE in two different ways.\\
Using in the general definition of $\Omega_m$ given in Eq. (\ref{8}) the expression of $\rho_m$ given in Eq. (\ref{rho4sol}) along with the definition of $H$ given in Eq. (\ref{acca}), we obtain:
\begin{equation}
\Omega_m(t) =\left(\frac{K}{3 n^2}\right) t^{2 - 3 n (1-d_2^2)}+
\frac{3 d_1^2 }{n\left[3 n (1-d_2^2) - 2\right]}  \left(-\frac{6\alpha}{n} + 2 \beta - \gamma n + \delta n^2 \right).\label{Omegam4sol1} 
\end{equation}
Therefore, using the result of the Friedmann equation given in Eq. (\ref{11}), we can write:
\begin{equation}
\Omega_D(t) =1- \left(\frac{K}{3 n^2}\right) t^{2 - 3 n (1-d_2^2)}-
\frac{3 d_1^2 }{n\left[3 n (1-d_2^2) - 2\right]}  \left(-\frac{6\alpha}{n} + 2 \beta - \gamma n + \delta n^2 \right).
\end{equation}
The evolutionary form of the fractional energy density of DE is given by:
\begin{eqnarray}
\Omega_D'(t) = -\left\{\frac{K\left[2-3n(1-d_2^2)\right]}{3\,n^3}\right\}t^{\,2-3n(1-d_2^2)}.
\end{eqnarray}
 Instead, using in the general definition of $\Omega_m$ given in Eq. (\ref{8}) the expression of $\rho_m$ given in Eq. (\ref{rho4sol}) along with the definition of $H^2$ given in Eq. (\ref{acca4sol}), we obtain:
\begin{eqnarray}
\Omega_m(t) =
\frac{
K\, t^{-3n(1-d_2^2)} + \frac{9 d_1^2 n}{3n(1-d_2^2)-2} \left(-\frac{6\alpha}{n}+2\beta - \gamma n + \delta n^2\right) t^{-2}
}{
K\, t^{-3n(1-d_2^2)} + 3 \left[1 + \frac{3 d_1^2 n}{3n(1-d_2^2)-2} \right] \left(-\frac{6\alpha}{n}+2\beta - \gamma n + \delta n^2\right) t^{-2}\label{Omegam4sol2} 
}.
\end{eqnarray}

Therefore, using the result of the Friedmann equation given in Eq. (\ref{11}), we can write:
\begin{eqnarray}
\Omega_D(t) =1-
\frac{
K\, t^{-3n(1-d_2^2)} + \frac{9 d_1^2 n}{3n(1-d_2^2)-2} \left(-\frac{6\alpha}{n}+2\beta - \gamma n + \delta n^2\right) t^{-2}
}{
K\, t^{-3n(1-d_2^2)} + 3 \left[1 + \frac{3 d_1^2 n}{3n(1-d_2^2)-2} \right] \left(-\frac{6\alpha}{n}+2\beta - \gamma n + \delta n^2\right) t^{-2}
}.
\end{eqnarray}
The evolutionary form of the fractional energy density of DE is given by:
\begin{eqnarray}
\Omega_D'(t)=\frac{3\left[3n(1-d_2^2)-2\right]\!\left(-\dfrac{6\alpha}{n}+2\beta-\gamma n+\delta n^2\right)\!t^{-3n(1-d_2^2)-2}}
{\,n\left\{K\,t^{-3n(1-d_2^2)}+3\left[1+\dfrac{3d_1^2 n}{3n(1-d_2^2)-2}\right]\left(-\dfrac{6\alpha}{n}+2\beta-\gamma n+\delta n^2\right)t^{-2}\right\}^2}.
\end{eqnarray}

We now want to obtain the expression of the pressure of DE $p_D$.\\ 
The general expression of $p_D$ is given by:
\begin{eqnarray}
    p_{D}= -\rho_{D} - \frac{\dot{\rho}_{D}}{3H} - \frac{Q_4}{3H}.
\end{eqnarray}
Using  the expression of $Q_4$ given in Eq. (\ref{Q4}), we can write:
\begin{eqnarray}
   \frac{Q_4}{3H} =  \left(d_1^2\rho_m + d_2^2\rho_D\right).\label{Q4frac}
\end{eqnarray}
Therefore, using the expressions of $\rho_{D}$ and $H$ we derived in Eqs. (\ref{cz3}) and (\ref{acca}) along with the result of Eq. (\ref{Q4frac}), we can write: 
\begin{eqnarray}
    p_{D}(t) 
&=&\left[ \frac{2}{n} -3-d_2^2 --\frac{9d_1^4 n}{3n(1-d_2^2) - 2}    \right]\left(-\frac{6\alpha}{n} + 2\beta - \gamma n + \delta n^2\right)t^{-2}
-d_1^2 K\,t^{-3n(1-d_2^2)}\nonumber \\
&=&\left\{   \frac{n^2 \left(3 d_2^4 + 6 d_2^2 - 9 - 9 d_1^4 \right) + n \left(12 - 4 d_2^2\right) - 4\,}{\,n \left[3 n (1-d_2^2) - 2\right]} \right\}\left(-\frac{6\alpha}{n} + 2\beta - \gamma n + \delta n^2\right)t^{-2}
-d_1^2 Kt^{-3n(1-d_2^2)}.
\end{eqnarray}
We now want to calculate the expression of the EoS parameter of DE $\omega_D$. \\
The general expression of $\omega_D$ for this case is:
\begin{eqnarray}
    \omega_{D} = -1 - \frac{\dot\rho_{D}}{3H\rho_{D}}- \frac{Q_4}{3H\rho_{D}}.
\end{eqnarray}
Using  the expression of $Q_4$ given in Eq. (\ref{Q4}), we can write:
\begin{eqnarray}
  \frac{Q_4}{3H\rho_{D}} = d_1^2 \left(\frac{\rho_m}{\rho_D}\right) + d_2^2= d_1^2 \left(\frac{\Omega_m}{\Omega_D}\right) + d_2^2 = d_1^2 \left(\frac{\Omega_m}{1-\Omega_m}\right) + d_2^2.
\end{eqnarray}
Therefore, using the expression of $\Omega_m$ given in Eq. (\ref{Omegam4sol1}), we obtain the following relation for $\omega_D$:
\begin{eqnarray}
    \omega_{D_{4,I-1}} 
    &=& -1 + \frac{2  }{3n}- d_1^2 \left\{  \frac{\dfrac{Kt^{\,2-3n(1-d_2^2)}}{3n^2}+\dfrac{3 d_1^2F}{n\big[3n(1-d_2^2)-2\big]}}
{1-\dfrac{Kt^{\,2-3n(1-d_2^2)}}{3n^2}-\dfrac{3 d_1^2F}{n\big[3n(1-d_2^2)-2\big]}} \right\},
\end{eqnarray}
where $F$ is defined as:
\begin{eqnarray}
F=-\dfrac{6\alpha}{n}+2\beta-\gamma n+\delta n^2.
\end{eqnarray}
Instead, using the expression of $\Omega_m$ given in Eq. (\ref{Omegam4sol2}), we obtain the following relation for $\omega_D$:
\begin{eqnarray}
    \omega_{D_{4,I-2}} 
    &=& -1 + \frac{2  }{3n}- d_1^2 \left[\frac{Kt^{\,2 - 3n(1 - d_2^2)} }{3\left(-\dfrac{6\alpha}{n} + 2\beta - \gamma n + \delta n^2 \right)}
+ \frac{3 d_1^2 n}{3n(1 - d_2^2) - 2}  \right] - d_2^2.
\end{eqnarray}

\subsubsection{Limiting Case of $n(1-d_2^2) = 2/3$}
In the limiting case of $n(1-d_2^2) = 2/3$, the solution of the continuity equation for $\rho_m$ is given by:
\begin{eqnarray}
\rho_m(t) = K\,t^{-2} + 9d_1^2 n \left( -\frac{6\alpha}{n} + 2\beta - \gamma n + \delta n^2 \right)\, t^{-2}\ln t \label{rhom4sol2}.
\end{eqnarray}
Following the same procedure of the general case, we obtain:
\begin{eqnarray}
K &=& \rho_{m0} t_0^2 - 9 d_1^2 n \left( -\frac{6\alpha}{n} + 2\beta - \gamma n + \delta n^2 \right) \ln t_0.
\end{eqnarray}
Using the expression of $\rho_m$ obtained in Eq. (\ref{rhom4sol2}) along with the expression of $\rho_D$ obtained in Eq. (\ref{cz3}), we can obtain the expression of $H^2$ from the Friedmann equation given in Eq. (\ref{7}), which leads to
\begin{equation}
H^2(t) = \left[ \frac{K}{3} + \left(1 + 3 d_1^2 n\ln t\right) \left(-\frac{6\alpha}{n} + 2\beta - \gamma n + \delta n^2 \right) \right] t^{-2}\label{acca4sol2}.
\end{equation}
Following the same procedure of the general case, we have:
\begin{eqnarray}
K &=& 3 \left[ H_0^2 t_0^2 - \left(1 + 3 d_1^2 n \ln t_0\right) \left(-\frac{6\alpha}{n} + 2\beta - \gamma n + \delta n^2 \right) \right].
\end{eqnarray}

We now want to obtain the expression of the deceleration parameter  $q$ using the general definition given in Eq. (\ref{qgen}).\\
Using the expression of $H^2$ given in Eq. (\ref{acca4sol2}), we obtain:
\begin{eqnarray}
q(t) = \frac{C_0 + C_1 (1 + \ln t)}{2 (C_0 + C_1 \ln t)},
\end{eqnarray}
where $C_0$ and $C_1$ are given by:
\begin{eqnarray}
C_0 &=& \frac{K}{3} + \left(-\frac{6\alpha}{n} + 2\beta - \gamma n + \delta n^2 \right), \\
C_1 &=& 3 d_1^2 n \left(-\frac{6\alpha}{n} + 2\beta - \gamma n + \delta n^2 \right).
\end{eqnarray}
We now calculate the expressions of the fractional energy densities of DM and DE in two different ways.\\
Using in the general definition of $\Omega_m$ given in Eq. (\ref{8}) the expression of $\rho_m$ given in Eq. (\ref{rhom4sol2}) along with the definition of $H$ given in Eq. (\ref{acca}), we obtain:
\begin{equation}
\Omega_m(t) = \frac{K + 9 d_1^2 n \left(-\frac{6\alpha}{n} + 2\beta - \gamma n + \delta n^2 \right) \ln t}{3 n^2}.\label{Omegam4sol2-1}
\end{equation}
Therefore, using the result of the Friedmann equation given in Eq. (\ref{11}), we can write:
\begin{equation}
\Omega_D(t) =1- \frac{K + 9 d_1^2 n \left(-\frac{6\alpha}{n} + 2\beta - \gamma n + \delta n^2 \right) \ln t}{3 n^2}.
\end{equation}
The evolutionary form of the fractional energy density of DE is given by:
\begin{equation}
\Omega_D'  
= - \frac{3 d_1^2 \left(-\frac{6\alpha}{n} + 2\beta - \gamma n + \delta n^2 \right)}{n^2},
\end{equation}
which is a constant depending on the parameters of the model and on the power law index of the scale factor we use.\\
 Instead, using in the general definition of $\Omega_m$ given in Eq.(\ref{8}) the expression of $\rho_m$ given in Eq. (\ref{rhom4sol2}) along with the definition of $H^2$ given in Eq. (\ref{acca4sol2}), we obtain:
\begin{eqnarray}
\Omega_m(t)=\frac{K+9d_1^2 n\left(-\dfrac{6\alpha}{n}+2\beta-\gamma n+\delta n^2\right)\ln t}
{K+3\left(-\dfrac{6\alpha}{n}+2\beta-\gamma n+\delta n^2\right)+9d_1^2 n\left(-\dfrac{6\alpha}{n}+2\beta-\gamma n+\delta n^2\right)\ln t}.\label{Omegam4sol2-2}
\end{eqnarray}
Therefore, using the result of the Friedmann equation given in Eq. (\ref{11}), we can write:
\begin{eqnarray}
\Omega_D(t)=1-\frac{K+9d_1^2 n\left(-\dfrac{6\alpha}{n}+2\beta-\gamma n+\delta n^2\right)\ln t}
{K+3\left(-\dfrac{6\alpha}{n}+2\beta-\gamma n+\delta n^2\right)+9d_1^2 n\left(-\dfrac{6\alpha}{n}+2\beta-\gamma n+\delta n^2\right)\ln t}.
\end{eqnarray}
The evolutionary form of the fractional energy density of DE is given by:
\begin{eqnarray}
\Omega_D'(t) \;=\; -\frac{27\,d_1^2\!\left(-\dfrac{6\alpha}{n}+2\beta-\gamma n+\delta n^2\right)^2}
{\left[K+3\left(-\dfrac{6\alpha}{n}+2\beta-\gamma n+\delta n^2\right)+9d_1^2 n\left(-\dfrac{6\alpha}{n}+2\beta-\gamma n+\delta n^2\right)\ln t\right]^2}.
\end{eqnarray}
We now want to obtain the expression of the pressure of DE $p_D$.\\ 
The general definition of $p_D$ is given by:
\begin{eqnarray}
    p_{D}= -\rho_{D} - \frac{\dot{\rho}_{D}}{3H} - \frac{Q_4}{3H}.
\end{eqnarray}
Using  the expression of $Q_4$ given in Eq. (\ref{Q4}), we can write:
\begin{eqnarray}
   \frac{Q_4}{3H} =  \left(d_1^2\rho_m + d_2^2\rho_D\right) .\label{Q4frac2}
\end{eqnarray}
Therefore, using the expressions of $\rho_{D}$ and $H$ we derived in Eqs. (\ref{cz3}) and (\ref{acca}) along with the result of Eq. (\ref{Q4frac2}), we can write:
\begin{eqnarray}
    p_{D}(t) 
&=&\left[ \frac{2}{n} -3-d_2^2 - 9d_1^4 n\ln t \right]\left(-\frac{6\alpha}{n} + 2\beta - \gamma n + \delta n^2\right)t^{-2}-d_1^2K\,t^{-2} .
\end{eqnarray}
We now want to calculate the expression of the EoS parameter of DE $\omega_D$. \\
In this case, the general definition of $\omega_D$ is given by:
\begin{eqnarray}
    \omega_{D} = -1 - \frac{\dot\rho_{D}}{3H\rho_{D}}- \frac{Q_4}{3H\rho_{D}}.
\end{eqnarray}
Using  the expression of $Q_4$ given in Eq. (\ref{Q4}), we can write:
\begin{eqnarray}
  \frac{Q_4}{3H\rho_{D}} = d_1^2 \left(\frac{\rho_m}{\rho_D}\right) + d_2^2= d_1^2 \left(\frac{\Omega_m}{\Omega_D}\right) + d_2^2 = d_1^2 \left(\frac{\Omega_m}{1-\Omega_m}\right) + d_2^2.
\end{eqnarray}
Therefore, using the expression of $\Omega_m$ given in Eq. (\ref{Omegam4sol2-1}), we obtain the following relation for $\omega_D$:
\begin{eqnarray}
    \omega_{D_{4,I-1,lim}} 
    &=& -1 + \frac{2  }{3n}- d_2^2- d_1^2\left[ \frac{K + 9 d_1^2 n \left(-\frac{6\alpha}{n} + 2\beta - \gamma n + \delta n^2 \right) \ln t}{3 n^2 - K - 9 d_1^2 n \left(-\frac{6\alpha}{n} + 2\beta - \gamma n + \delta n^2 \right) \ln t}  \right].    
\end{eqnarray}
Instead, using the expression of $\Omega_D$ given in Eq. (\ref{Omegam4sol2-2}), we obtain the following relation for $\omega_D$:
\begin{eqnarray}
    \omega_{D_{4I,-2,lim}} &=& -1 + \frac{2  }{3n}- d_1^2 \left[\frac{K + 9 d_1^2 n \left(-\dfrac{6\alpha}{n}+2\beta-\gamma n+\delta n^2\right) \ln t}{3 \left(-\dfrac{6\alpha}{n}+2\beta-\gamma n+\delta n^2\right)}\right]     - d_2^2.
\end{eqnarray}

\subsection{Fifth Interacting Case}
We now consider the fifth interaction term given by:
\begin{eqnarray}
    Q_5 = 3Hd^2\left( \frac{\rho_m\rho_D}{\rho_m+\rho_D}  \right).\label{Q5}
\end{eqnarray}
Inserting the expression of $Q_5$ along with the expression of $H$ given in Eq. (\ref{acca}) in the continuity equation for $\rho_m$ defined in Eq. (\ref{eq:DM_cons}), we obtain:
\begin{eqnarray}
\dot{\rho}_{m} +\left(\frac{3n}{t}\right)  \rho_{m} &=&\frac{3d^2n}{t}\left( \frac{\rho_m\rho_D}{\rho_m+\rho_D}  \right).
\end{eqnarray}
We now  want to make some consideration about the term $\frac{\rho_m\rho_D}{\rho_m+\rho_D}$.\\
We can write:
\begin{eqnarray}
    \frac{\rho_m\rho_D}{\rho_m+\rho_D}   &=& \frac{1}{\rho_{cr}}\left(\frac{\rho_m\rho_D}{\Omega_m+\Omega_D}  \right)\nonumber \\
    &=&\frac{\rho_m\rho_D}{\rho_{cr}}\nonumber\\
    &=& \frac{\rho_m\rho_D}{3H^2},
\end{eqnarray}
where we used the fact that $\Omega_m + \Omega_D =1$ and $\rho_{cr}=3H^2$.\\
Therefore, using the definitions of $H$ and $\rho_D$ given, respectively, in Eqs. (\ref{acca}) and (\ref{cz3}), we obtain:
\begin{eqnarray}
    Q_5 = \frac{3d^2}{nt} \left( -\frac{6\alpha}{n} + 2\beta - \gamma n + \delta n^2 \right)\rho_m.
\end{eqnarray}
Then, we can write the continuity equation for $\rho_m$ as:
\begin{eqnarray}
\dot{\rho}_{m} +\left(\frac{3n}{t}\right)  \rho_{m} &=&\frac{3d^2}{nt} \left( -\frac{6\alpha}{n} + 2\beta - \gamma n + \delta n^2 \right)\rho_m \label{rhom5gen}.
\end{eqnarray}
The solution of Eq. (\ref{rhom5gen}) is given by
\begin{eqnarray}
\rho_m(t) &= &\rho_{m0}\; t^{-3n + \frac{3d^2 C}{n}},\\
&=& 3H_0^2\Omega_{m0}\; t^{-3n + \frac{3d^2 C}{n}},\label{rhom5sol}
\end{eqnarray}
where $C$ is defined as:
\begin{eqnarray}
    C=-\frac{6\alpha}{n} + 2\beta - \gamma n + \delta n^2.
\end{eqnarray}
Using the expression of $\rho_m$ obtained in Eq. (\ref{rhom5sol}) along with the expression of $\rho_D$ obtained in Eq. (\ref{cz3}), we can obtain the expression of $H^2$ from the Friedmann equation given in Eq. (\ref{7}), which leads to
\begin{eqnarray}
 H^2(t)&=& H_0^2\left[ \Omega_{m0}\; t^{-3n + \frac{3d^2 C}{n}}  + \frac{1}{H_0^2}\left( -\frac{6\alpha}{n} + 2\beta - \gamma n + \delta n^2 \right)t^{-2}\right].\label{acca5sol}
\end{eqnarray}
 We now want to obtain the expression of the deceleration parameter  $q$ using the general definition given in Eq. (\ref{qgen}).\\
Using the expression of $H^2$ given in Eq. (\ref{acca5sol}), we obtain:
\begin{eqnarray}
q(t) &=& -1
- \frac{
H_0^2 \Omega_{m0}\!\left(-3n + \frac{3 d^2 C}{n}\right)
t^{-3n + \frac{3 d^2 C}{n} - 1}
- 2\!\left(-\frac{6\alpha}{n} + 2\beta - \gamma n + \delta n^2\right) t^{-3}
}{
2\!\left[
H_0^2 \Omega_{m0}\, t^{-3n + \frac{3 d^2 C}{n}}
+ \left(-\frac{6\alpha}{n} + 2\beta - \gamma n + \delta n^2\right)t^{-2}
\right]^{3/2}}\nonumber \\
&=&-1
- \frac{
H_0^2 \Omega_{m0}\!\left(-3n^2 +3 d^2 C\right)
t^{-3n + \frac{3 d^2 C}{n} - 1}
- 2n\left(-\frac{6\alpha}{n} + 2\beta - \gamma n + \delta n^2\right) t^{-3}
}{
2n\left[
H_0^2 \Omega_{m0}\, t^{-3n + \frac{3 d^2 C}{n}}
+ \left(-\frac{6\alpha}{n} + 2\beta - \gamma n + \delta n^2\right)t^{-2}
\right]^{3/2}}.
\end{eqnarray}
We now calculate the expressions of the fractional energy densities of DM and DE in two different ways.\\
Using in the general definition of $\Omega_m$ given in Eq. (\ref{8}) the expression of $\rho_m$ given in Eq. (\ref{rhom5sol}) along with the definition of $H$ given in Eq. (\ref{acca}), we obtain:
\begin{eqnarray}
\Omega_m(t) 
= \left(\frac{H_0^2 \, \Omega_{m0}}{n^2}\right) \, t^{-3n + \frac{3 d^2 C}{n} + 2}.\label{Omegam5sol1}
\end{eqnarray}
Therefore, using the result of the Friedmann equation given in Eq. (\ref{11}), we can write:
\begin{eqnarray}
\Omega_D(t) =1-
\left(\frac{H_0^2 \, \Omega_{m0}}{n^2}\right) \, t^{-3n + \frac{3 d^2 C}{n} + 2}.\label{Omegad5sol1}
\end{eqnarray}
The evolutionary form of the fractional energy density of DE is given by:
\begin{eqnarray}
\Omega_D'(t) = - \left(\frac{H_0^2 \, \Omega_{m0}}{n^3}\right) 
\left(-3n + \frac{3 d^2 C}{n} + 2\right) 
t^{-3n + \frac{3 d^2 C}{n} + 2}.
\end{eqnarray}
 Instead, using in the general definition of $\Omega_m$ given in Eq. (\ref{8}) the expression of $\rho_m$ given in Eq. (\ref{rhom5sol}) along with the definition of $H^2$ given in Eq. (\ref{acca5sol}), we obtain:
\begin{eqnarray}
\Omega_m(t) = 
\frac{
H_0^2 \Omega_{m0}\, t^{-3n + \frac{3 d^2 C}{n}}
}{
H_0^2 \Omega_{m0}\, t^{-3n + \frac{3 d^2 C}{n}}
+
\left(-\frac{6\alpha}{n} + 2\beta - \gamma n + \delta n^2\right)t^{-2}
}.\label{Omegam5sol2}
\end{eqnarray}
Therefore, using the result of the Friedmann equation given in Eq. (\ref{11}), we can write:
\begin{eqnarray}
\Omega_D(t) =1- 
\frac{
H_0^2 \Omega_{m0}\, t^{-3n + \frac{3 d^2 C}{n}}
}{
H_0^2 \Omega_{m0}\, t^{-3n + \frac{3 d^2 C}{n}}
+
\left(-\frac{6\alpha}{n} + 2\beta - \gamma n + \delta n^2\right)t^{-2}
}.\label{Omegad5sol2}
\end{eqnarray}
The evolutionary form of the fractional energy density of DE is given by:
\begin{eqnarray}
\Omega_D'(t)
&=&
-\frac{
\left(-\dfrac{6\alpha}{n} + 2\beta - \gamma n + \delta n^2\right)
\left(-3n + \dfrac{3 d^2 C}{n} + 2\right)
H_0^2 \Omega_{m0}\, t^{-3n + \frac{3 d^2 C}{n} - 2}
}{
n \left[
H_0^2 \Omega_{m0}\, t^{-3n + \frac{3 d^2 C}{n}}
+
\left(-\dfrac{6\alpha}{n} + 2\beta - \gamma n + \delta n^2\right)t^{-2}
\right]^2}\nonumber \\
&=& \frac{
\left(-\dfrac{6\alpha}{n} + 2\beta - \gamma n + \delta n^2\right)
\left[3n^2(1-C)-2(n+1)\right]
H_0^2 \Omega_{m0}\, t^{-3n + \frac{3 d^2 C}{n} - 2}
}{
n^2 \left[
H_0^2 \Omega_{m0}\, t^{-3n + \frac{3 d^2 C}{n}}
+
\left(-\dfrac{6\alpha}{n} + 2\beta - \gamma n + \delta n^2\right)t^{-2}
\right]^2}.
\end{eqnarray}
We now want to obtain the expression of the pressure of DE $p_D$.\\ 
The general expression of $p_D$ for this case is given by:
\begin{eqnarray}
    p_{D}= -\rho_{D} - \frac{\dot{\rho}_{D}}{3H} - \frac{Q_5}{3H}.
\end{eqnarray}
Using  the expression of $Q_5$ given in Eq. (\ref{Q5}), we can write:
\begin{eqnarray}
   \frac{Q_5}{3H} &=& d^2\left( \frac{\rho_m\rho_D}{\rho_m+\rho_D} \right)\nonumber \\
   &=& d^2\left[ \frac{\rho_m\rho_D}{\rho_{cr}(\Omega_m+\Omega_D)} \right]\nonumber \\
    &=&d^2\rho_m\Omega_D,
\end{eqnarray}
or alternatively 
\begin{eqnarray}
   \frac{Q_5}{3H} &=& d^2\left( \frac{\rho_m\rho_D}{\rho_m+\rho_D} \right)\nonumber \\
   &=& d^2\left[ \frac{\rho_m\rho_D}{\rho_{cr}(\Omega_m+\Omega_D)} \right]\nonumber \\
    &=&d^2\Omega_m\rho_D.
\end{eqnarray}
Using the expressions of $\rho_m$ given in Eq. (\ref{rhom5sol}) and the expression of $\Omega_D$ given in Eq. (\ref{Omegad5sol1}), we can write: 
\begin{eqnarray}
    p_{D}(t) 
    &=& \left( \frac{2}{n} -3 \right)\left[ -\frac{6\alpha}{n} + 2\beta - \gamma n + \delta n^2 \right]t^{-2}\nonumber \\
    &&-  3d^2H_0^2\Omega_{m0}\; t^{-3n + \frac{3d^2 C}{n}}\left[1-
\left(\frac{H_0^2 \, \Omega_{m0}}{n^2}\right)  t^{-3n + \frac{3 d^2 C}{n} + 2}    \right] .
\end{eqnarray}
Instead, using the expressions of $\rho_m$ given in Eq. (\ref{rhom5sol}) and the expression of $\Omega_D$ given in Eq. (\ref{Omegad5sol2}), we can write:
\begin{eqnarray}
    p_{D}(t) 
    &=& \left( \frac{2}{n} -3 \right)\left[ -\frac{6\alpha}{n} + 2\beta - \gamma n + \delta n^2 \right]t^{-2}\nonumber \\
    &&-  3d^2H_0^2\Omega_{m0} t^{-3n + \frac{3d^2 C}{n}}\times \nonumber \\
    &&\left[  1- 
\frac{
H_0^2 \Omega_{m0}\, t^{-3n + \frac{3 d^2 C}{n}}
}{
H_0^2 \Omega_{m0}\, t^{-3n + \frac{3 d^2 C}{n}}
+
\left(-\frac{6\alpha}{n} + 2\beta - \gamma n + \delta n^2\right)t^{-2}
}\right] . 
\end{eqnarray}
Using the expressions of $\rho_D$ given in Eq. (\ref{cz3}) and the expression of $\Omega_m$ given in Eq. (\ref{Omegam5sol1}), we can write:
\begin{eqnarray}
    p_{D}(t) 
    &=& \left( \frac{2}{n} -3 \right)\left[ -\frac{6\alpha}{n} + 2\beta - \gamma n + \delta n^2 \right]t^{-2} \nonumber \\
    &&-\frac{3d^2H_0^2 \, \Omega_{m0}}{n^2}  \left( -\frac{6\alpha}{n} + 2\beta - \gamma n + \delta n^2 \right)   \, t^{-3n + \frac{3 d^2 C}{n} }.
\end{eqnarray}
Instead, using the expressions of $\rho_D$ given in Eq. (\ref{cz3}) and the expression of $\Omega_m$ given in Eq. (\ref{Omegam5sol2}), we can write:
\begin{eqnarray}
    p_{D}(t) 
    &=& \left( \frac{2}{n} -3 \right)\left[ -\frac{6\alpha}{n} + 2\beta - \gamma n + \delta n^2 \right]t^{-2} \nonumber \\
    &&- \left( -\frac{6\alpha}{n} + 2\beta - \gamma n + \delta n^2 \right)\times \nonumber \\
    &&\left[\frac{3d^2H_0^2 \Omega_{m0}\, t^{-3n + \frac{3 d^2 C}{n}-2}
}{
H_0^2 \Omega_{m0}\, t^{-3n + \frac{3 d^2 C}{n}}
+
\left(-\frac{6\alpha}{n} + 2\beta - \gamma n + \delta n^2\right)t^{-2}
}\right].
\end{eqnarray}
We now want to calculate the expression of the EoS parameter of DE $\omega_D$. \\
The general expression of $\omega_D$ for this case is given by:
\begin{eqnarray}
    \omega_{D} = -1 - \frac{\dot\rho_{D}}{3H\rho_{D}}- \frac{Q_5}{3H\rho_{D}},
\end{eqnarray}

Using  the expression of $Q_5$ given in Eq. (\ref{}), we can write:
\begin{eqnarray}
   \frac{Q_5}{3H\rho_{D}} &=&    d^2\left( \frac{\rho_m}{\rho_m+\rho_D}  \right)\nonumber \\
   &=& d^2\left[ \frac{\rho_m}{\rho_{cr}(\Omega_m+\Omega_D)}  \right]\nonumber \\
   &=& d^2\left( \frac{\rho_m}{\rho_{cr}}  \right) = d^2\Omega_m
\end{eqnarray}
The final expression of $\omega_D$ is then given by:
\begin{eqnarray}
    \omega_{D} &=& -1 + \frac{2  }{3n }-   d^2\Omega_m
\end{eqnarray}
Therefore, using the expression of $\Omega_m$ given in Eq. (\ref{Omegam5sol1}), we obtain the following relation for $\omega_D$:
\begin{eqnarray}
    \omega_{D_{5,I-1}} &=& -1 + \frac{2  }{3n }-   
\left(\frac{d^2H_0^2 \, \Omega_{m0}}{n^2}\right) \, t^{-3n + \frac{3 d^2 C}{n} + 2}.
\end{eqnarray}
Instead, using the expression of $\Omega_m$ given in Eq. (\ref{Omegam5sol2}), we obtain the following relation for $\omega_D$:
\begin{eqnarray}
    \omega_{D_{5,I-2}} &=& -1 + \frac{2  }{3n }-   
\frac{
d^2H_0^2 \Omega_{m0}\, t^{-3n + \frac{3 d^2 C}{n}}
}{
H_0^2 \Omega_{m0}\, t^{-3n + \frac{3 d^2 C}{n}}
+
\left(-\frac{6\alpha}{n} + 2\beta - \gamma n + \delta n^2\right)t^{-2}
}.
\end{eqnarray}

\subsection{Sixth Interacting Case}
We now consider the sixth interaction term given by:
\begin{eqnarray}
    Q_6 = 3Hd^2\left( \frac{\rho_m^2}{\rho_m+\rho_D}  \right). \label{Q6}
\end{eqnarray}
We can now make some considerations about the quantity $\left( \frac{\rho_m^2}{\rho_m+\rho_D}  \right)$.\\
We have that:
\begin{eqnarray}
    \frac{\rho_m^2}{\rho_m+\rho_D}  &=& \frac{1}{\rho_{cr}}\left( \frac{\rho_m^2}{\Omega_m+\Omega_D}  \right)\nonumber \\
    &=&\frac{\rho_m^2}{\rho_{cr}},\label{marzia1}
\end{eqnarray}
where we used $\rho_m+\rho_D= \rho_{cr}\left(  \Omega_m+\Omega_D  \right)$ and $\Omega_m +\Omega_D=1$.\\
Then, using in the general expression of $Q_6$ the expression of $H$ given in Eq. (\ref{acca}) along with the result of Eq. (\ref{marzia1}), we can write :
\begin{eqnarray}
    Q_6 =  \left(\frac{d^2 t}{n}\right)\rho_m^2.
\end{eqnarray}
Therefore,  we obtain the following continuity equation for $\rho_m$:
\begin{eqnarray}
\dot{\rho}_{m} + \left(\frac{3n}{t}\right) \rho_{m} &=& \left(\frac{d^2t}{n}\right)\rho_m^2, \label{rho6gen}
\end{eqnarray}
Eq. (\ref{rho6gen}) is a Bernoulli-type differential equation with exponent $2$ and its general solution is given by:
\begin{equation}
\rho_{m}(t) = 
\frac{1}{\,C\,t^{3n} - \dfrac{d^{2}}{n(2-3n)}\,t^{2}}. \label{rhom6sol}
\end{equation}
We can relate the constant $C$ to the present day of $\rho_m$, indicated with $\rho_{m0}$, and to the other parameters of the model we consider.\\
At present time, i.e. for $t=t_0$ (with $t_0$ being the present day age of the Universe), we have that $\rho_m(t=t_0)=\rho_{m0}$, therefore from Eq. (\ref{rhom6sol}) we obtain:
\begin{equation}
C = \frac{1}{\rho_{m0}\,t_0^{3n}} 
+ \frac{d^{2}}{n(2-3n)}\,t_0^{2-3n}.
\end{equation}
Using the expression of $\rho_m$ obtained in Eq. (\ref{rhom6sol}) along with the expression of $\rho_D$ obtained in Eq. (\ref{cz3}), we can obtain the expression of $H^2$ from the Friedmann equation given in Eq. (\ref{7}), which leads to
\begin{eqnarray}
H^{2}(t)
=
\frac{1}{
3\!\left[
C\,t^{3n}
-
\dfrac{d^{2}}{n(2-3n)}\,t^{2}
\right]
}
+
\frac{1}{t^{2}}
\left(
-\frac{6\alpha}{n}
+2\beta
-\gamma n
+\delta n^{2}
\right). \label{acca6sol}
\end{eqnarray}
We can also relate the constant $K$ to the present day value of the Hubble parameter, i.e. $H_0$.\\
Considering $H(t=t_0)=H_0$ in Eq. (\ref{acca6sol}), we obtain:
\begin{eqnarray}
C = \frac{1}{t_0^{3n}}
\left[
\frac{1}{3\!\left(H_0^2 - \dfrac{1}{t_0^2}\left(-\frac{6\alpha}{n} + 2\beta - \gamma n + \delta n^2\right)\right)}
+ \frac{d^2}{n(2 - 3n)}\,t_0^2
\right].
\end{eqnarray}
We now want to obtain the expression of the deceleration parameter  $q$ using the general definition given in Eq. (\ref{qgen}).\\
Using the expression of $H^2$ given in Eq. (\ref{acca6sol}), we obtain:
\begin{eqnarray}
q(t)
&=& -1+
\left\{
3nC\,t^{3n-1}-\dfrac{2d^{2}t}{n(2-3n)}\right. \nonumber \\
&&\left.
+6\left(-\dfrac{6\alpha}{n}+2\beta-\gamma n+\delta n^{2}\right)\left[C\,t^{3n}-\dfrac{d^{2}t^{2}}{n(2-3n)}\right]^{2}t^{-3}
\right\}\times \nonumber \\
&&
\left\{
6\left[C\,t^{3n}-\dfrac{d^{2}t^{2}}{n(2-3n)}\right]^{2}\times \right. \nonumber \\
&&\left.\left\{
\frac{1}{3\left[C\,t^{3n}-\dfrac{d^{2}t^{2}}{n(2-3n)}\right]}
+\left(-\frac{6\alpha}{n}+2\beta-\gamma n+\delta n^{2}\right)t^{-2}
\right\}^{3/2}
\right\}^{-1}.
\end{eqnarray}
We now calculate the expressions of the fractional energy densities of DM and DE in two different ways.\\
Using in the general definition of $\Omega_m$ given in Eq. (\ref{8}) the expression of $\rho_m$ given in Eq. (\ref{rhom6sol}) along with the definition of $H$ given in Eq. (\ref{acca}), we obtain:
\begin{eqnarray}
\Omega_m(t) = 
\frac{1}{\,3 n^2 \left[ C\, t^{3n-2} - \dfrac{d^2}{n(2-3n)} \right] }.\label{Omegam6-1}
\end{eqnarray}
Therefore, using the result of the Friedmann equation given in Eq. (\ref{11}), we can write:
\begin{eqnarray}
\Omega_D(t) =1- 
\frac{1}{3 n^2 \left[ C\, t^{3n-2} - \dfrac{d^2}{n(2-3n)} \right]}.\label{Omegad6-1}
\end{eqnarray}
The evolutionary form of the fractional energy density of DE is given by:
\begin{eqnarray}
\Omega'_D(t) = 
\frac{3 n^2 C (3n-2)\, t^{3n-3}}{
\left\{ 3 n^2 \left[ C\, t^{3n-2} - \dfrac{d^2}{\,n(2-3n)} \right] \right\}^2
}.
\end{eqnarray}
 Instead, using in the general definition of $\Omega_m$ given in Eq. (\ref{8}) the expression of $\rho_m$ given in Eq. (\ref{}) along with the definition of $H^2$ given in Eq. (\ref{}), we obtain:
\begin{eqnarray}
\Omega_m(t) = 
\frac{1}{\,1 + 3 \left( -\frac{6\alpha}{n} + 2\beta - \gamma n + \delta n^2 \right) 
\left[ C\, t^{3n-2} - \frac{d^2}{\,n(2-3n)} \right] }.\label{Omegam6-2}
\end{eqnarray}
Therefore, using the result of the Friedmann equation given in Eq. (\ref{11}), we can write:
\begin{eqnarray}
\Omega_D(t) =1- 
\frac{1}{\,1 + 3 \left( -\frac{6\alpha}{n} + 2\beta - \gamma n + \delta n^2 \right) 
\left[ C\, t^{3n-2} - \frac{d^2}{\,n(2-3n)} \right] \,}.\label{Omegad6-2}
\end{eqnarray}
The evolutionary form of the fractional energy density of DE is given by:
\begin{eqnarray}
\Omega_D'(t) = 
\frac{3 \left(-\dfrac{6\alpha}{n} + 2\beta - \gamma n + \delta n^2\right) C (3n-2) t^{3n-3}}
{\left\{ 1 + 3 \left(-\dfrac{6\alpha}{n} + 2\beta - \gamma n + \delta n^2\right) 
\left[ C t^{3n-2} - \dfrac{d^2}{\,n(2-3n)} \right] \right\}^2}.
\end{eqnarray}
We now want to obtain the final expression of the pressure of DE $p_D$.\\
The general expression of $p_D$ is given by:
\begin{eqnarray}
    p_{D}= -\rho_{D} - \frac{\dot{\rho}_{D}}{3H} - \frac{Q_6}{3H}.
\end{eqnarray}
Using  the expression of $Q_6$ given in Eq. (\ref{Q6}), we can write:
\begin{eqnarray}
   \frac{Q_6}{3H} =  d^2\left( \frac{\rho_m^2}{\rho_m+\rho_D}  \right).
\end{eqnarray}
The general expression of $p_D$ is then given by:
\begin{eqnarray}
    p_{D}= -\rho_{D} - \frac{\dot{\rho}_{D}}{3H} - d^2\left( \frac{\rho_m^2}{\rho_m+\rho_D}\right). \label{piddi}
\end{eqnarray}
Therefore, using the expressions of $\rho_{D}$ and $H$ we derived in Eqs. (\ref{cz3}) and (\ref{acca}) along with the expression of $\rho_m$ obtained in Eq. (\ref{rhom6sol}), we can write:
\begin{eqnarray}
    p_{D}(t) 
    &=& \left( \frac{2}{n} -3 \right)\left( -\frac{6\alpha}{n} + 2\beta - \gamma n + \delta n^2 \right)t^{-2}\nonumber \\
    &&- d^2\left[C t^{3n} - \dfrac{d^2t^2}{n(2-3n)} \right]^{-1}\times \nonumber \\
    &&  \left\{ 1 + 3 \left(-\dfrac{6\alpha}{n} + 2\beta - \gamma n + \delta n^2 \right) \left[C t^{3n-2} - \dfrac{d^2}{n(2-3n)} \right] \right\}^{-1}         .
\end{eqnarray}
We now want to obtain the final expression of the EoS parameter of DE $\omega_D$.\\
The general expression of $\omega_D$ for this case is given by:
\begin{eqnarray}
    \omega_{D} = -1 - \frac{\dot\rho_{D}}{3H\rho_{D}}- \frac{Q_6}{3H\rho_{D}}.
\end{eqnarray}
Using  the expression of $Q_6$ given in Eq. (\ref{Q6}), we can write:
\begin{eqnarray}
    \frac{Q_6}{3H\rho_{D}} &=&  \frac{d^2}{\rho_{D}}\left( \frac{\rho_m^2}{\rho_m+\rho_D}  \right)\nonumber \\
     &=&  \frac{d^2}{\rho_{D}\rho_{cr}}\left( \frac{\rho_m^2}{\Omega_m+\Omega_D}  \right)\nonumber \\
     &=&  d^2\left( \frac{\rho_m^2}{\rho_{D}\rho_{cr}}  \right)\nonumber \\
   &=&  d^2\left( \frac{\rho_m}{\rho_{D} }   \right)\left( \frac{\rho_m}{\rho_{cr}}  \right)\nonumber \\
   &=&  d^2\left( \frac{\Omega_m}{\Omega_{D} }   \right)\Omega_m\nonumber \\
   &=&  d^2\left( \frac{\Omega_m^2}{\Omega_{D} }   \right) = d^2\left( \frac{\Omega_m^2}{1-\Omega_{m} }\right),
\end{eqnarray}
where we used the facts that $\rho_D + \rho_m = \rho_{cr}\left(\Omega_D+\Omega_m \right)$, $\Omega_D+\Omega_m=1$ and $\Omega_D = 1-\Omega_m$.\\
Therefore, we obtain the following relation for $\omega_D$
\begin{eqnarray}
    \omega_{D} &=& -1 + \frac{2  }{3n}-  d^2\left( \frac{\Omega_m^2}{\Omega_{D} } \right)\nonumber \\
    &=&  -1 + \frac{2  }{3n}-  d^2\left( \frac{\Omega_m^2}{1-\Omega_{m} } \right).
\end{eqnarray}
Using the expression of $\Omega_m$ given in Eq. (\ref{Omegam6-1}), we obtain the following relation for $\omega_D$:
\begin{eqnarray}
    \omega_{D_{6,I-1}} = -1 + \frac{2  }{3n}-  \frac{d^2}{ 
\left\{ 3 n^2 \left[ C\, t^{3n-2} - \dfrac{d^2}{\,n(2-3n)} \right] \right\}
\left\{ 3 n^2 \left[ C\, t^{3n-2} - \dfrac{d^2}{\,n(2-3n)} \right] - 1 \right\}
}.
\end{eqnarray}
Instead, using the expression of $\Omega_m$ given in Eq. (\ref{Omegam6-2}), we obtain the following relation for $\omega_D$:
\begin{eqnarray}
    \omega_{D_{6,I-2}} &=& -1 + \frac{2  }{3n}\nonumber\\
    &&-  
\left\{ 1 + 3 \left( -\frac{6\alpha}{n} + 2\beta - \gamma n + \delta n^2 \right) 
\left[C\, t^{3n-2} - \frac{d^2}{\,n(2-3n)} \right] \right\}^{-1}\times \nonumber \\
&&
\left\{ 3 \left( -\frac{6\alpha}{n} + 2\beta - \gamma n + \delta n^2 \right) 
\left[ C\, t^{3n-2} - \frac{d^2}{\,n(2-3n)} \right] \right\}^{-1}
.
\end{eqnarray}

\subsubsection{Limiting Case of  $n = 2/3$:}
We now consider the case corresponding to $3n = 2$, which is equivalent to $n = 2/3$.\\
In this case, the general solution of the continuity equation for $\rho_m$ is given by:
\begin{eqnarray}
\rho_{m}(t) &=& 
\frac{1}{\,t^{2}\!\left[C - \left(\dfrac{d^{2}}{n}\right)\ln t\right]}\nonumber \\
&=&\frac{n}{\,t^{2}\!\left(nC - d^{2}\ln t\right)}.\label{rhom6sol2}
\end{eqnarray}
Following the same procedure of the general case, we obtain:
\begin{eqnarray}
C &=& \frac{1}{\rho_{m0} t_0^2} + \left(\frac{d^2}{n}\right) \ln t_0.
\end{eqnarray}
Using the expression of $\rho_m$ obtained in Eq. (\ref{rhom6sol2}) along with the expression of $\rho_D$ obtained in Eq. (\ref{cz3}), we can obtain the expression of $H^2$ from the Friedmann equation given in Eq. (\ref{7}), which leads to
\begin{eqnarray}
H^{2}(t)
&=& \left[
\frac{1}{3\left[C-\left(\dfrac{d^{2}}{n}\right)\ln t\right]}+\left( -\frac{6\alpha}{n} + 2\beta - \gamma n + \delta n^2 \right)
\right]t^{-2}\nonumber \\
&=& \left[\frac{n}{3\left(nC - d^{2}\ln t\right)}
+\left( -\frac{6\alpha}{n} + 2\beta - \gamma n + \delta n^2 \right)
\right]t^{-2}.\label{acca6sol2}
\end{eqnarray}
Following the same procedure of the general case, we have:
\begin{eqnarray}
C &=& \frac{1}{3 \left[ H_0^2 t_0^2 - \left(-\frac{6\alpha}{n} + 2\beta - \gamma n + \delta n^2 \right) \right]} + \left(\frac{d^2}{n}\right) \ln t_0.
\end{eqnarray}
We now want to obtain the expression of the deceleration parameter  $q$ using the general definition given in Eq. (\ref{qgen}).\\
Using the expression of $H^2$ given in Eq. (\ref{acca6sol2}), we obtain:
\begin{eqnarray}
q(t)& &= -\,\frac{d^2}{6\,n \,\left(C - \frac{d^2}{n} \ln t\right)^2 \,
\left[ \frac{1}{3\left(C - \frac{d^2}{n}\ln t\right)} + \left(-\frac{6\alpha}{n} + 2\beta - \gamma n + \delta n^2\right) \right]}\nonumber \\
&=&-\,\frac{d^2}{2\,n\,\left(C - \frac{d^2}{n} \ln t\right)\,\Bigl[ 1 + 3 \left(-\frac{6\alpha}{n} + 2\beta - \gamma n + \delta n^2\right)\left(C - \frac{d^2}{n} \ln t\right) \Bigr]}.
\end{eqnarray}
We now calculate the expressions of the fractional energy densities of DM and DE in two different ways.\\
Using in the general definition of $\Omega_m$ given in Eq. (\ref{8}) the expression of $\rho_m$ given in Eq. (\ref{}) along with the definition of $H$ given in Eq. (\ref{acca}), we obtain:
\begin{eqnarray}
\Omega_m(t) &=& \frac{1}{\,3 n^2 \left[C -\left( \dfrac{d^{2}}{n} \right)\ln t \right]}\nonumber\\
 &=& \frac{1}{\,3 n \left(Cn - d^{2} \ln t \right)}.\label{Omegam6-3}
\end{eqnarray}

Therefore, using the result of the Friedmann equation given in Eq. (\ref{11}), we can write:
\begin{eqnarray}
\Omega_D(t) &=& 1-\frac{1}{\,3 n^2 \left[C - \left(\dfrac{d^{2}}{n}\right) \ln t \right]}\nonumber \\
&=&1- \frac{1}{\,3 n \left(Cn - d^{2} \ln t \right)}.\label{Omegad6-3}
\end{eqnarray}

The evolutionary form of the fractional energy density of DE is given by:
\begin{eqnarray}
\Omega_D'(t) &=& -\frac{3 n d^2}{\,t \left\{ 3 n^2 \left[C - \left(\dfrac{d^2}{n}\right) \ln t \right] \right\}^2}\nonumber \\
&=&-\frac{3 n d^2}{\,t \left[ 3 n \left(Cn - d^2 \ln t \right) \right]^2}.
\end{eqnarray}

 Instead, using in the general definition of $\Omega_m$ given in Eq. (\ref{8}) the expression of $\rho_m$ given in Eq. (\ref{}) along with the definition of $H^2$ given in Eq. (\ref{}), we obtain:
\begin{eqnarray}
\Omega_m(t) = \frac{1}{1 + 3 \left[C - \left(\frac{d^2}{n}\right) \ln t \right] \left(-\frac{6\alpha}{n} + 2\beta - \gamma n + \delta n^2\right)}.\label{Omegam6-4}
\end{eqnarray}
Therefore, using the result of the Friedmann equation given in Eq. (\ref{11}), we can write:
\begin{eqnarray}
\Omega_D(t) = 1-\frac{1}{1 + 3 \left[C - \left(\frac{d^2}{n}\right) \ln t \right] \left(-\frac{6\alpha}{n} + 2\beta - \gamma n + \delta n^2\right)}.\label{Omegad6-4}
\end{eqnarray}
The evolutionary form of the fractional energy density of DE is given by:
\begin{eqnarray}
\Omega_D'(t) = - \frac{3 \, d^2 \left(-\dfrac{6\alpha}{n} + 2\beta - \gamma n + \delta n^2\right)}
{\, t \left\{ 1 + 3 \left(-\dfrac{6\alpha}{n} + 2\beta - \gamma n + \delta n^2 \right) \left[ C - \left(\dfrac{d^2}{n}\right) \ln t \right] \right\}^2 }.
\end{eqnarray}

We now want to obtain the final expression of the pressure of DE $p_D$.\\
Therefore, using in Eq. (\ref{piddi}) the expressions of $\rho_{D}$ and $H$ we derived in Eqs. (\ref{cz3}) and (\ref{acca}) along with the result for $\rho_m$ obtained in Eq. (\ref{rhom6sol2}), we can write:
\begin{eqnarray}
    p_{D}(t) 
&=& \left( \frac{2}{n} -3 \right)\left[ -\frac{6\alpha}{n} + 2\beta - \gamma n + \delta n^2 \right]t^{-2}\nonumber \\
    &&-\frac{d^2n^2}
{\,t^2 \left(nC - d^2 \ln t\right)
\left[ n + 3 \left(-\frac{6\alpha}{n} + 2\beta - \gamma n + \delta n^2 \right) \left(nC - d^2 \ln t \right) \right] \,}.
\end{eqnarray}
We now want to calculate the expression of the EoS parameter of DE $\omega_D$. \\
Also for this case, we use the general relation:
\begin{eqnarray}
    \omega_{D} &=& -1 + \frac{2  }{3n}-  d^2\left( \frac{\Omega_m^2}{\Omega_{D} } \right)\nonumber \\
     &=&  -1 + \frac{2  }{3n}-  d^2\left( \frac{\Omega_m^2}{1-\Omega_{m} } \right).
\end{eqnarray}
Therefore, using the expression of $\Omega_m$ given in Eq. (\ref{Omegam6-3}), we obtain the following relation for $\omega_D$:
\begin{eqnarray}
    \omega_{D_{6,I-1,lim}} = -1 + \frac{2  }{3n}-  \frac{d^2}{\Big[ 3 n^2 \left(C - \frac{d^2}{n} \ln t \right) \Big] 
       \left[ 3 n^2 \left(C - \frac{d^2}{n} \ln t \right) - 1 \right]}.
\end{eqnarray}
Instead, using the expression of $\Omega_m$ given in Eq. (\ref{Omegam6-4}), we obtain the following relation for $\omega_D$:
\begin{eqnarray}
    \omega_{D_{6,I-2,lom}} &=& -1 + \frac{2  }{3n}- \nonumber \\
    &&\left\{ 1 + 3 \left[C - \left(\frac{d^2}{n}\right) \ln t \right] 
      \left(-\frac{6\alpha}{n} + 2\beta - \gamma n + \delta n^2\right) \right\}^{-1}\times \nonumber \\
      &&
      \left\{ 3 \left[C - \left(\frac{d^2}{n}\right) \ln t \right] 
      \left(-\frac{6\alpha}{n} + 2\beta - \gamma n + \delta n^2\right) \right\}^{-1}.
\end{eqnarray}

\subsection{Seventh Interacting Case}
We now consider the seventh interaction term given by:
\begin{eqnarray}
    Q_7 = 3Hd^2\left( \frac{\rho_D^2}{\rho_m+\rho_D}  \right).\label{Q7}
\end{eqnarray}
We now make some considerations about the term $\left( \frac{\rho_D^2}{\rho_m+\rho_D}  \right)$.\\
We have that:
\begin{eqnarray}
     \frac{\rho_D^2}{\rho_m+\rho_D}  &=& \frac{\rho_D^2}{\rho_{cr}(\Omega_m+\Omega_D)}\nonumber \\
     &=&\frac{\rho_D^2}{\rho_{cr}}\nonumber \\
      &=&\frac{\rho_D^2}{3H^2},
\end{eqnarray}
where we used the relations $\rho_m+\rho_D=\rho_{cr}(\Omega_m+\Omega_D)$, $\Omega_m+\Omega_D=1$ and $\rho_{cr}=3H^2$.\\
Using the expression of $H$ given in Eq. (\ref{acca}) along with the expression of $\rho_D$ given in Eq. (\ref{cz3}), we can write:
\begin{eqnarray}
\frac{\rho_D^2}{3H^2}=\frac{3}{n^2t^2} \left( -\frac{6\alpha}{n} + 2\beta - \gamma n + \delta n^2 \right)^2.
\end{eqnarray}
Therefore, the interaction term can be written as:
\begin{eqnarray}
    Q_7 = \frac{9d^2}{nt^3} \left( -\frac{6\alpha}{n} + 2\beta - \gamma n + \delta n^2 \right)^2,
\end{eqnarray}
and the continuity equation for $\rho_m$ as:
\begin{eqnarray}
\dot{\rho}_{m} + \left(\frac{3n}{t}\right)  \rho_{m} &=&\frac{9d^2}{nt^3} \left( -\frac{6\alpha}{n} + 2\beta - \gamma n + \delta n^2 \right)^2.\label{rho7gen}
\end{eqnarray}
The general solution of Eq. (\ref{rho7gen}) is given by:
\begin{equation}
\rho_m(t) = C\, t^{-3n}+ \frac{9 d^2}{n (3n - 2)} \left( -\frac{6\alpha}{n} + 2\beta - \gamma n + \delta n^2 \right)^2 t^{-2}. \label{rhom7sol}
\end{equation}
We can relate the constant $C$ to the present day of $\rho_m$, indicated with $\rho_{m0}$, and to the other parameters of the model we consider.\\
At present time, i.e. for $t=t_0$ (with $t_0$ being the present day age of the Universe), we have that $\rho_m(t=t_0)=\rho_{m0}$, therefore from Eq. (\ref{rhom7sol}) we obtain:
\begin{eqnarray}
C &=& t_0^{3n} \left[ \rho_{m0} - \frac{9 d^2}{n (3n - 2)} \left( -\frac{6\alpha}{n} + 2\beta - \gamma n + \delta n^2 \right)^2 t_0^{-2} \right].
\end{eqnarray}
Using the expression of $\rho_m$ obtained in Eq. (\ref{rhom7sol}) along with the expression of $\rho_D$ obtained in Eq. (\ref{cz3}), we can obtain the expression of $H^2$ from the Friedmann equation given in Eq. (\ref{7}), which leads to
\begin{eqnarray}
    H^2 &=&  \left( \frac{C}{3}\right)t^{-3n}   +      \frac{3 d^2}{n (3n - 2)} \left( -\frac{6\alpha}{n} + 2\beta - \gamma n + \delta n^2 \right)^2 t^{-2}  \nonumber \\
    &&\,\,\,\,\,\,\,\,\,\,\,\,\,\,\,\,\,\,\,\,\,\,\,\,\,\,\,\,\,\,\,\,\,\,\,\,\,\,\,\,\,\,\,\,\,\,\,\,\,\,\,\,\,\,+   \left( -\frac{6\alpha}{n} + 2\beta - \gamma n + \delta n^2 \right)t^{-2}.\label{acca7sol}
\end{eqnarray}
We can also relate the constant $C$ to the present day value of the Hubble parameter, i.e. $H_0$.\\
Considering $H(t=t_0)=H_0$ in Eq. (\ref{acca7sol}), we obtain:
\begin{eqnarray}
C &=& 3\, t_0^{3n} \Biggl\{ H_0^2 - \Biggl[ \frac{3 d^2}{n (3n - 2)} \left(-\frac{6\alpha}{n} + 2\beta - \gamma n + \delta n^2 \right)^2 \nonumber \\
&& \,\,\,\,\,\,\,\,\,\,\,\,\,+ \left(-\frac{6\alpha}{n} + 2\beta - \gamma n + \delta n^2 \right) \Biggr] t_0^{-2} \Biggr\}.
\end{eqnarray}
We now want to obtain the expression of the deceleration parameter  $q$ using the general definition given in Eq. (\ref{qgen}).\\
Using the expression of $H^2$ given in Eq. (\ref{acca7sol}), we obtain:
\begin{eqnarray}
    q(t) = -1 +
    \frac{\left(3nA + 2B\,t^{3n-2}\right)\,t^{\frac{3n-2}{2}}}
         {2\left(A + B\,t^{3n-2}\right)^{3/2}},
\end{eqnarray}
where we defined $A$ and $B$ as
\begin{eqnarray}
    A &=& \frac{C}{3}, \\
    B &=& \frac{3 d^2K^2}{n(3n - 2)}+ K,
\end{eqnarray}
and
\begin{eqnarray}
      K = -\frac{6\alpha}{n} + 2\beta - \gamma n + \delta n^2.
\end{eqnarray}
We now calculate the expressions of the fractional energy densities of DM and DE in two different ways.\\
Using in the general definition of $\Omega_m$ given in Eq. (\ref{8}) the expression of $\rho_m$ given in Eq. (\ref{rhom7sol}) along with the definition of $H$ given in Eq. (\ref{acca}), we obtain:
\begin{eqnarray}
\Omega_m(t) =\left(\frac{C}{3 n^2}\right) t^{2-3n} + \frac{3 d^2 K^2}{n^3 (3n-2)}.
\end{eqnarray}
Therefore, using the result of the Friedmann equation given in Eq. (\ref{11}), we can write:
\begin{eqnarray}
\Omega_D(t) =1-\left(\frac{C}{3 n^2}\right) t^{2-3n} - \frac{3 d^2 K^2}{n^3 (3n-2)}.\label{OmegaD1}
\end{eqnarray}
The evolutionary form of the fractional energy density of DE is given by:
\begin{eqnarray}
\Omega_D'(t)=- \left[\frac{C(2-3n)}{3 n^3}\right]  t^{2-3n}.
\end{eqnarray}
 Instead, using in the general definition of $\Omega_m$ given in Eq. (\ref{8}) the expression of $\rho_m$ given in Eq. (\ref{rhom7sol}) along with the definition of $H^2$ given in Eq. (\ref{acca7sol}), we obtain:
\begin{eqnarray}
\Omega_m(t) = \frac{C\, t^{2-3n} + \dfrac{9 d^2K^2}{n(3n-2)} }
                 {C\, t^{2-3n} + 3 \left[ \dfrac{3 d^2K^2}{n(3n-2)}  + K \right]}.
\end{eqnarray}

Therefore, using the result of the Friedmann equation given in Eq. (\ref{11}), we can write:
\begin{eqnarray}
\Omega_D(t) = 1-\frac{C\, t^{2-3n} + \dfrac{9 d^2K^2}{n(3n-2)} }
                 {C\, t^{2-3n} + 3 \left[ \dfrac{3 d^2K^2}{n(3n-2)}  + K \right]}\label{OmegaD2}.
\end{eqnarray}
The evolutionary form of the fractional energy density of DE is given by:
\begin{eqnarray}
\Omega_D' 
=  \frac{3 C K (2-3n) t^{2-3n}}{n \left\{ C t^{2-3n} + 3 \left[ \dfrac{3 d^2K^2}{n(3n-2)}  + K \right] \right\}^2 }.
\end{eqnarray}
We now want to find the final expression of the pressure of DE $p_D$.\\
The general definition for this case is given by:
\begin{eqnarray}
    p_{D}= -\rho_{D} - \frac{\dot{\rho}_{D}}{3H} - \frac{Q_7}{3H}.
\end{eqnarray}
Using  the expression of $Q_7$ given in Eq. (\ref{Q7}), we can write:
\begin{eqnarray}
   \frac{Q_7}{3H} = d^2\left( \frac{\rho_D^2}{\rho_m+\rho_D}  \right).
\end{eqnarray}
Therefore, we can write $p_D$ as:
\begin{eqnarray}
    p_{D}= -\rho_{D} - \frac{\dot{\rho}_{D}}{3H} - d^2\left( \frac{\rho_D^2}{\rho_m+\rho_D}  \right).
\end{eqnarray}
Using the expressions of $\rho_{D}$ and $H$ we derived in Eqs. (\ref{cz3}) and (\ref{acca}) along with the expression of $\rho_m$ given in Eq. (\ref{rhom7sol}), we can write:
\begin{eqnarray}
    p_{D}(t) 
    &=&  K \left[ \frac{2}{n} - 3 - \frac{9 d^2 K}{C t^{2-3n} + \dfrac{9 d^2 K^2}{n(3n-2)} + 3 K} \right]t^{-2}.
\end{eqnarray}
We now want to calculate the expression of the EoS parameter of DE $\omega_D$. \\
The general definition for this case is given by:
\begin{eqnarray}
    \omega_{D} = -1 - \frac{\dot\rho_{D}}{3H\rho_{D}}- \frac{Q_7}{3H\rho_{D}}.
\end{eqnarray}
Using  the expression of $Q_7$ given in Eq. (\ref{Q7}), we can write:
\begin{eqnarray}
   \frac{Q_7}{3H\rho_{D}} &=&  d^2  \left( \frac{\rho_D}{\rho_m+\rho_D}  \right)\nonumber \\
   &=& d^2  \left[ \frac{\rho_D}{\rho_{cr}(\Omega_m+\Omega_D)}  \right]\nonumber \\
    &=& d^2  \left( \frac{\rho_D}{\rho_{cr}}  \right) \nonumber \\
&=& d^2\Omega_D,
\end{eqnarray}
where we used the relations $\rho_m+\rho_D=\rho_{cr}\left( \Omega_m+\Omega_D\right)$, $\Omega_m+\Omega_D=1$ and the general definition of $\Omega_D$.\\
Therefore, using the expression of $\Omega_D$ given in Eq. (\ref{OmegaD1}), we obtain the following relation for $\omega_D$:
\begin{eqnarray}
    \omega_{D_{7,I-1}} &=& -1 + \frac{2   }{3n}\nonumber \\
    &&- d^2\left[ 1-\left(\frac{C}{3 n^2}\right) t^{2-3n} - \frac{3 d^2 K^2}{n^3 (3n-2)}  \right].
\end{eqnarray}
Instead, using the expression of $\Omega_D$ given in Eq. (\ref{OmegaD2}), we obtain the following relation for $\omega_D$:
\begin{eqnarray}
    \omega_{D_{7,I-2}} = -1 + \frac{2   }{3n}- d^2\left\{   1-\frac{C\, t^{2-3n} + \dfrac{9 d^2K^2}{n(3n-2)} }
                 {C\, t^{2-3n} + 3 \left[ \dfrac{3 d^2K^2}{n(3n-2)}  + K \right]}\right\}   .
\end{eqnarray}

\subsubsection{Limiting Case of  $n=2/3$}
We now consider the case corresponding to $3n-2=0$, which means $n=2/3$.\\
In this case, the solution of the continuity equation of DM is given by:
\begin{equation}\label{rhom7-2}
\rho_m(t) = Ct^{-2} + \frac{9 d^2}{n}\left(  -\frac{6\alpha}{n} + 2\beta - \gamma n + \delta n^2  \right)^2 t^{-2}\ln t.
\end{equation}
Following the same procedure of the general case, we obtain:
\begin{eqnarray}
C &=& \rho_{m0} t_0^2 - \frac{9 d^2}{n} \left(  -\frac{6\alpha}{n} + 2\beta - \gamma n + \delta n^2 \right)^2 \ln t_0.
\end{eqnarray}
Using the expression of $\rho_m$ obtained in Eq. (\ref{rhom7-2}) along with the expression of $\rho_D$ obtained in Eq. (\ref{cz3}), we can obtain the expression of $H^2$ from the Friedmann equation given in Eq. (\ref{7}), which leads to
\begin{eqnarray}
    H^2 &=&  \left(\frac{C}{3}\right) t^{-2} + \frac{3 d^2}{n}\left(  -\frac{6\alpha}{n} + 2\beta - \gamma n + \delta n^2  \right)^2 t^{-2}\ln t,   +   \left( -\frac{6\alpha}{n} + 2\beta - \gamma n + \delta n^2 \right)t^{-2}
    \nonumber \\
    &=&\left\{\frac{C}{3}+ K \left[1 + \left(\frac{3 d^2K}{n}\right) \ln t \right] \right\}t^{-2},\label{acca7sol2}
\end{eqnarray}
where we used the definition of $K$.\\
Following the same procedure of the general case, we have:
\begin{eqnarray}
C &=& 3 \Biggl[ H_0^2 t_0^2 - \left( -\frac{6\alpha}{n} + 2\beta - \gamma n + \delta n^2 \right) \nonumber \\
&& - \frac{3 d^2}{n}\left(  -\frac{6\alpha}{n} + 2\beta - \gamma n + \delta n^2  \right)^2 \ln t_0 \Biggr].
\end{eqnarray}
We now want to obtain the expression of the deceleration parameter  $q$ using the general definition given in Eq. (\ref{qgen}).\\
Using the expression of $H^2$ given in Eq. (\ref{acca7sol2}), we obtain:
\begin{eqnarray}
    q(t) = \frac{1}{2} - \frac{3 d^2 K^2}{4 n \left\{ \frac{C}{3} + K \left[ 1 + \left(\frac{3 d^2K}{n}\right)  \ln t \right] \right\} }.
\end{eqnarray}

We now calculate the expressions of the fractional energy densities of DM and DE in two different ways.\\
Using in the general definition of $\Omega_m$ given in Eq. (\ref{8}) the expression of $\rho_m$ given in Eq. (\ref{rhom7-2}) along with the definition of $H$ given in Eq. (\ref{acca}), we obtain:
\begin{eqnarray}
\Omega_m(t) =\frac{C}{3 n^2} + \left(\frac{3 d^2K^2}{n^3}\right)  \ln t.
\end{eqnarray}
Therefore, using the result of the Friedmann given in Eq. (\ref{7}), we can write:
\begin{eqnarray}
\Omega_D(t) =1-\frac{C}{3 n^2} - \left(\frac{3 d^2K^2}{n^3}\right)\ln t.\label{barnali1}
\end{eqnarray}
The evolutionary form of the fractional energy density of DE is given by:
\begin{eqnarray}
\Omega_D'(t)=- \frac{3 d^2 K^2}{n^4},
\end{eqnarray}
which is a constant depending on the parameters of the model and on the power law index of the scale factor we consider.\\
 Instead, using in the general definition of $\Omega_m$ given in Eq. (\ref{8}) the expression of $\rho_m$ given in Eq. (\ref{rhom7-2}) along with the definition of $H^2$ given in Eq. (\ref{acca7sol2}), we obtain:
\begin{eqnarray}
\Omega_m(t) =\frac{C + \left(\frac{9 d^2K^2}{n}\right)   \ln t}{C + 3 K \left[ 1 + \left(\frac{3 d^2K}{n}\right) \ln t \right]}.
\end{eqnarray}
Therefore, using the result of the Friedmann equation given in Eq. (\ref{11}), we can write:
\begin{eqnarray}
\Omega_D(t) = 1-\frac{C + \left(\frac{9 d^2K^2}{n}\right)  \ln t}{C + 3 K \left[ 1 + \left(\frac{3 d^2K}{n}\right) \ln t \right]}.
\end{eqnarray}\label{barnali2}
The evolutionary form of the fractional energy density of DE is given by:
\begin{eqnarray}
\Omega_D' 
=  - \frac{27 d^2 K^3}{n^2 \left\{ C + 3 K \left[ 1 + \left(\frac{3 d^2K}{n}\right) \ln t \right] \right\}^2}.
\end{eqnarray}
We now want to find the final expression of the pressure of DE $p_D$.\\
The general definition for this case is given by:
\begin{eqnarray}
    p_{D}= -\rho_{D} - \frac{\dot{\rho}_{D}}{3H} - \frac{Q_7}{3H}.
\end{eqnarray}
Using  the expression of $Q_7$ given in Eq. (\ref{Q7}), we can write:
\begin{eqnarray}
   \frac{Q_7}{3H} = d^2\left( \frac{\rho_D^2}{\rho_m+\rho_D}  \right).
\end{eqnarray}
Therefore, we can write $p_D$ as:
\begin{eqnarray}
    p_{D}= -\rho_{D} - \frac{\dot{\rho}_{D}}{3H} - d^2\left( \frac{\rho_D^2}{\rho_m+\rho_D}  \right).
\end{eqnarray}
Using the expressions of $\rho_{D}$ and $H$ we derived in Eqs. (\ref{cz3}) and (\ref{acca}) along with the expression of $\rho_m$ obtained in Eq. (\ref{rhom7-2}), we can write:
\begin{eqnarray}
    p_{D}(t) 
    &=&  K \left[ \frac{2}{n} - 3- \frac{9 d^2 K}{C + 3 K + \left(\frac{9 d^2K^2}{n}\right) \ln t} \right]t^{-2}.
\end{eqnarray}
We now want to calculate the expression of the EoS parameter of DE $\omega_D$. \\
The general definition for this case is given by:
\begin{eqnarray}
    \omega_{D} = -1 - \frac{\dot\rho_{D}}{3H\rho_{D}}- \frac{Q_7}{3H\rho_{D}}.
\end{eqnarray}
Using  the expression of $Q_7$ given in Eq. (\ref{Q7}), we can write:
\begin{eqnarray}
   \frac{Q_7}{3H\rho_{D}} &=&  d^2  \left( \frac{\rho_D}{\rho_m+\rho_D}  \right)\nonumber \\
   &=& d^2  \left[ \frac{\rho_D}{\rho_{cr}(\Omega_m+\Omega_D)}  \right]\nonumber \\
    &=& d^2  \left( \frac{\rho_D}{\rho_{cr}}  \right)\nonumber \\
    &=& d^2\Omega_D.
\end{eqnarray}
where we used the relations $\rho_m+\rho_D=\rho_{cr}\left( \Omega_m+\Omega_D\right)$, $\Omega_m+\Omega_D=1$ and the general definition of $\Omega_D$.\\
Therefore, using the expression of $\Omega_D$ given in Eq. (\ref{barnali1}), we obtain the following relation for $\omega_D$:
\begin{eqnarray}
    \omega_{D_{7,I-1,lim}} &=& -1 + \frac{2   }{3n}\nonumber \\
    &&- d^2\left[ 1-\frac{C}{3 n^2} - \left(\frac{3 d^2K^2}{n^3}\right)\ln t\right].
\end{eqnarray}
Instead, using the expression of $\Omega_D$ given in Eq. (\ref{barnali2}), we obtain the following relation for $\omega_D$:
\begin{eqnarray}
    \omega_{D_{7,I-2,lim}} = -1 + \frac{2   }{3n}- d^2\left\{  1-\frac{C + \left(\frac{9 d^2K^2}{n}\right)  \ln t}{C + 3 K \left[ 1 + \left(\frac{3 d^2K}{n}\right) \ln t \right]} \right\}. 
\end{eqnarray}

\subsection{Eighth Interacting Case}
We now consider the eighth interaction term given by:
\begin{eqnarray}
    Q_8 = 3Hd^2\left( \frac{\rho_D- \rho_m}{\rho_m+\rho_D}  \right).\label{Q8}
\end{eqnarray}

We now want to make some considerations about the term $\frac{\rho_D- \rho_m}{\rho_m+\rho_D} $.\\
We have that:
\begin{eqnarray}
 \left( \frac{\rho_D- \rho_m}{\rho_m+\rho_D}  \right) &=& \frac{1}{\rho_{cr}} \left( \frac{\rho_D- \rho_m}{\Omega_m+\Omega_D}  \right)\nonumber \\
 &=& \left( \frac{\rho_D- \rho_m}{\rho_{cr}}  \right)\nonumber\\
 &=& \left( \frac{\rho_D- \rho_m}{3H^2  }  \right),
\end{eqnarray}
where we used the fact that $\rho_D- \rho_m = \rho_{cr} \left( \Omega_m+\Omega_D  \right)$, $\Omega_m+\Omega_D=1$ and $\rho_{cr}=H^2$.\\
Then, the interaction term $Q_8$ can be written as:
\begin{eqnarray}
    Q_8 &=& d^2 \left( \frac{\rho_D- \rho_m}{H  }  \right) \nonumber \\
    &=&  \frac{3d^2}{n t} \left( -\frac{6\alpha}{n} + 2\beta - \gamma n + \delta n^2 \right)-d^2\left(\frac{t}{n}\right)\rho_m,
\end{eqnarray}
where we used the definitions of $H$ and $\rho_D$ given in Eqs. (\ref{acca}) and (\ref{cz3}).\\
The continuity equation for $\rho_m$ can be now written as:
\begin{eqnarray}
\dot{\rho}_{m} + \left(\frac{3n}{t}\right)  \rho_{m} &=& \frac{3d^2}{n t} \left( -\frac{6\alpha}{n} + 2\beta - \gamma n + \delta n^2 \right)-d^2\left(\frac{t}{n}\right)\rho_m,
\end{eqnarray}
and its general solutions is given by:
\begin{eqnarray}
\rho_m(t) =t^{-3 n} \, e^{- \frac{d^2 t^2}{2 n}} 
\left[ 
\frac{3 d^2 F}{n} \left(\frac{2 n}{d^2}\right)^{\frac{3 n - 2}{2}} \, 
\Gamma\left( \frac{3 n}{2}, -\frac{d^2 t^2}{2 n} \right) + C
\right],\label{rhom8sol}
\end{eqnarray}
where $C$ is an integration constant and  $\Gamma(a, x)$ represents the Gamma function. Moreover, $F$ is defined as:
\begin{equation}
F = -\frac{6 \alpha}{n} + 2 \beta - \gamma n + \delta n^2. 
\end{equation}
We can relate the constant $C$ to the present day of $\rho_m$, indicated with $\rho_{m0}$, and to the other parameters of the model we consider.\\
At present time, i.e. for $t=t_0$ (with $t_0$ being the present day age of the Universe), we have that $\rho_m(t=t_0)=\rho_{m0}$, therefore from Eq. (\ref{rhom8sol}) we obtain:
\begin{eqnarray}
C &=& e^{ \frac{d^2 t_0^2}{2 n} } t_0^{3 n} \, \rho_{m0} 
- \frac{3 d^2 F}{n} \left(\frac{2 n}{d^2}\right)^{\frac{3 n - 2}{2}} \, 
\Gamma\left( \frac{3 n}{2}, -\frac{d^2 t_0^2}{2 n} \right).
\end{eqnarray}
Using the expression of $\rho_m$ obtained in Eq. (\ref{rhom8sol}) along with the expression of $\rho_D$ obtained in Eq. (\ref{cz3}), we can obtain the expression of $H^2$ from the Friedmann equation given in Eq. (\ref{7}), which leads to
\begin{eqnarray}
    H^2 &=&  t^{-3 n} \, e^{- \frac{d^2 t^2}{2 n}} 
\left[ 
\frac{d^2 F}{n} \left(\frac{2 n}{d^2}\right)^{\frac{3 n - 2}{2}} \, 
\Gamma\left( \frac{3 n}{2}, -\frac{d^2 t^2}{2 n} \right)  + \frac{C}{3}
\right]+Ft^{-2}.\label{acca8sol}
\end{eqnarray}
We can also relate the constant $C$ to the present day value of the Hubble parameter, i.e. $H_0$.\\
Considering $H(t=t_0)=H_0$ in Eq. (\ref{acca8sol}), we obtain:
\begin{eqnarray}
C &=& 3\, t_0^{3 n} e^{\frac{d^2 t_0^2}{2 n}} 
\Biggl[ H_0^2 - F t_0^{-2} - t_0^{-3 n} e^{- \frac{d^2 t_0^2}{2 n}} \frac{d^2 F}{n} \left(\frac{2 n}{d^2}\right)^{\frac{3 n - 2}{2}} \Gamma\left( \frac{3 n}{2}, -\frac{d^2 t_0^2}{2 n} \right) \Biggr].
\end{eqnarray}
We now want to obtain the expression of the deceleration parameter  $q$ using the general definition given in Eq. (\ref{qgen}).\\
Using the expression of $H^2$ given in Eq. (\ref{acca8sol}), we obtain:
\begin{eqnarray}
q(t) &=& -1 - \frac{1}{2 H^2(t)} 
\left\{
t^{-3 n} e^{- \frac{d^2 t^2}{2 n}} 
\left(-\frac{3n}{t} - \frac{d^2 t}{n}\right)
\left[ \frac{d^2 F}{n} \left(\frac{2 n}{d^2}\right)^{\frac{3 n - 2}{2}} 
\Gamma\left( \frac{3 n}{2}, -\frac{d^2 t^2}{2 n} \right) + \frac{C}{3} \right]\right.\nonumber \\
&&\left.
\,\,\,\,\,\,\,\,\,\,\,\,\,\,\,\,\,\,\,\,\,\,\,\,\,\,\,\,\,\,\,\,\,\,\,\,\,\,+
\frac{d^2 F}{n} \left(\frac{2 n}{d^2}\right)^{\frac{3 n - 2}{2}} t^{-3 n} \frac{d^2 t}{n}
- 2 F t^{-3}
\right\},
\end{eqnarray}
where $H^2(t)$ is given in Eq. (\ref{acca8sol}).\\
We now calculate the expressions of the fractional energy densities of DM and DE in two different ways.\\
Using in the general definition of $\Omega_m$ given in Eq. (\ref{8}) the expression of $\rho_m$ given in Eq. (\ref{rhom8sol}) along with the definition of $H$ given in Eq. (\ref{acca}), we obtain:
\begin{eqnarray}
\Omega_m(t) = \frac{t^{2 - 3 n}}{3 n^2} \, 
e^{- \frac{d^2 t^2}{2 n}} 
\left[ 
\frac{3 d^2 F}{n} \left(\frac{2 n}{d^2}\right)^{\frac{3 n - 2}{2}} 
\Gamma\left( \frac{3 n}{2}, -\frac{d^2 t^2}{2 n} \right) + C 
\right].
\end{eqnarray}

Therefore, using the result of the Friedmann equation given in Eq. (\ref{11}), we can write:
\begin{eqnarray}
\Omega_{D_{8,I-1}}(t) = 1-\frac{t^{2 - 3 n}}{3 n^2}  
e^{- \frac{d^2 t^2}{2 n}} 
\left[ 
\frac{3 d^2 F}{n} \left(\frac{2 n}{d^2}\right)^{\frac{3 n - 2}{2}} 
\Gamma\left( \frac{3 n}{2}, -\frac{d^2 t^2}{2 n} \right) + C 
\right].\label{barbali3}
\end{eqnarray}
The evolutionary form of the fractional energy density of DE is given by:
\begin{eqnarray}
\Omega_D' &=& - \frac{t^{2-3 n}}{3 n^3} \, 
\left\{ 
\left( 2 - 3 n - \frac{d^2 t^2}{n} \right) 
\left[\frac{3 d^2 F}{n} \left(\frac{2 n}{d^2}\right)^{\frac{3 n - 2}{2}} 
\Gamma\left( \frac{3 n}{2}, -\frac{d^2 t^2}{2 n} \right) + C \right]\right.\nonumber \\
&&\left.
\,\,\,\,\,\,\,\,\,\,\,\,\,\,\,\,\,\,\,\,\,\,+ \frac{3 d^2 F}{n} \left(\frac{2 n}{d^2}\right)^{\frac{3 n - 2}{2}} \frac{d^2 t^2}{n}\right\}.
\end{eqnarray}
Instead, using in the general definition of $\Omega_m$ given in Eq. (\ref{8}) the expression of $\rho_m$ given in Eq. (\ref{rhom8sol}) along with the definition of $H^2$ given in Eq. (\ref{acca8sol}), we obtain:
\begin{eqnarray}
\Omega_m(t) = 
\frac{
\displaystyle \frac{3 d^2 F}{n} \left(\frac{2 n}{d^2}\right)^{\frac{3 n - 2}{2}} 
\Gamma\left(\frac{3 n}{2}, -\frac{d^2 t^2}{2 n}\right) + C
}{
\displaystyle 3 \left[
\frac{d^2 F}{n} \left(\frac{2 n}{d^2}\right)^{\frac{3 n - 2}{2}} 
\Gamma\left(\frac{3 n}{2}, -\frac{d^2 t^2}{2 n}\right) + \frac{C}{3} + F t^{3 n - 2} e^{\frac{d^2 t^2}{2 n}}
\right]
}.
\end{eqnarray}
Therefore, using the result of the Friedmann equation given in Eq. (\ref{11}), we can write:
\begin{eqnarray}
\Omega_{D_{8,I-2}}(t) =1- 
\frac{
\displaystyle \frac{3 d^2 F}{n} \left(\frac{2 n}{d^2}\right)^{\frac{3 n - 2}{2}} 
\Gamma\left(\frac{3 n}{2}, -\frac{d^2 t^2}{2 n}\right) + C
}{
\displaystyle 3 \left[
\frac{d^2 F}{n} \left(\frac{2 n}{d^2}\right)^{\frac{3 n - 2}{2}} 
\Gamma\left(\frac{3 n}{2}, -\frac{d^2 t^2}{2 n}\right) + \frac{C}{3} + F t^{3 n - 2} e^{\frac{d^2 t^2}{2 n}}
\right]
}.\label{barbali4}
\end{eqnarray}
The evolutionary form of the fractional energy density of DE is given by:
\begin{equation}
\Omega_D' = 
\frac{
3 F t^{3n-2} \, e^{\frac{d^2 t^2}{2 n}} 
\left[
(A \, \Gamma + C) \left( 3n - 2 + \frac{d^2 t^2}{n} \right) 
- A \frac{d^2 t^2}{n} \left( \frac{d^2 t^2}{2 n} \right)^{\frac{3n}{2}-1} e^{\frac{d^2 t^2}{2 n}}
\right]
}
{
n\left[ A \, \Gamma + C + 3 F t^{3 n - 2} e^{\frac{d^2 t^2}{2 n}} \right]^2
},
\end{equation}
\noindent
where $A$ and $\Gamma$ are defined as:
\begin{eqnarray}
A &=& \frac{3 d^2 F}{n} \left(\frac{2 n}{d^2}\right)^{\frac{3 n - 2}{2}}, \\
\Gamma &=& \Gamma\left(\frac{3 n}{2}, -\frac{d^2 t^2}{2 n}\right).
\end{eqnarray}
We now want to find the final expression of the pressure of DE $p_D$.\\
The general definition of $p_D$ is given by:
\begin{eqnarray}
    p_{D}= -\rho_{D} - \frac{\dot{\rho}_{D}}{3H} - \frac{Q_8}{3H}.
\end{eqnarray}
Using  the expression of $Q_8$ given in Eq. (\ref{Q8}), we can write:
\begin{eqnarray}
   \frac{Q_8}{3H} =d^2\left( \frac{\rho_D- \rho_m}{\rho_m+\rho_D}  \right). \label{fracq8}
\end{eqnarray}
Therefore, using the expressions of $\rho_{D}$ and $H$ we derived in Eqs. (\ref{cz3}) and (\ref{acca}) along with the result of Eq. (\ref{fracq8}), we can write:
\begin{eqnarray}
    p_{D}(t) 
    &=& \left( \frac{2}{n} -3 \right)\left[ -\frac{6\alpha}{n} + 2\beta - \gamma n + \delta n^2 \right]t^{-2}\nonumber \\
    &&-d^2\left[\frac{3 K - t^{-3 n + 2} e^{- \frac{d^2 t^2}{2 n}} G}{3 K + t^{-3 n + 2} e^{- \frac{d^2 t^2}{2 n}} G}.  \right],
\end{eqnarray}
where $G$ is defined as:
\begin{eqnarray}
    G= \frac{3 d^2 F}{n} \left(\frac{2 n}{d^2}\right)^{\frac{3 n - 2}{2}} 
\Gamma\left( \frac{3 n}{2}, -\frac{d^2 t^2}{2 n} \right) + C.
\end{eqnarray}

We now want to calculate the expression of the EoS parameter of DE $\omega_D$. \\
The general expression for this case is given by:
\begin{eqnarray}
    \omega_{D} = -1 - \frac{\dot\rho_{D}}{3H\rho_{D}}- \frac{Q_8}{3H\rho_{D}}.
\end{eqnarray}
Using  the expression of $Q_8$ given in Eq. (\ref{Q8}), we can write:
\begin{eqnarray}
 \frac{Q_8}{3H\rho_{D}} &=&  \frac{d^2}{\rho_{D}}  \left( \frac{\rho_D- \rho_m}{\rho_m+\rho_D}  \right)\nonumber \\
  &=&  \frac{d^2}{\rho_{D}\rho_{cr}}  \left( \frac{\rho_D- \rho_m}{\Omega_m+\Omega_D}  \right)\nonumber \\
  &=&  d^2  \left( \frac{\rho_D- \rho_m}{\rho_{D}\rho_{cr}}  \right)\nonumber \\
    &=&  d^2 \left(\frac{2 \Omega_D - 1}{\Omega_D\rho_{cr}}\right).
\end{eqnarray}
We obtain then:
\begin{eqnarray}
     \frac{Q_8}{3H\rho_{D}} &=& \frac{d^2t^2}{3n^2}  \left(\frac{2 \Omega_D - 1}{\Omega_D}\right)
\end{eqnarray}
Therefore, using the expression of $\Omega_D$ given in Eq. (\ref{barbali3}), we obtain the following relation for $\omega_D$:
\begin{eqnarray}
    \omega_{D_{8,I-1}} &=& -1 +\frac{2}{3n}-  \frac{d^2t^2}{3n^2}\left(\frac{2 \Omega_{D_{8,I-1}} - 1}{\Omega_{D_{8,I-1}}}\right)
\end{eqnarray}
Instead, using the expression of $\Omega_D$ given in Eq. (\ref{barbali4}), we obtain the following relation for $\omega_D$:
\begin{eqnarray}
    \omega_{D_{8,I-2}} &=& -1 +\frac{2}{3n}-  \frac{d^2t^2}{3n^2}\left(\frac{2 \Omega_{D_{8,I-2}} - 1}{\Omega_{D_{8,I-2}}}\right)
\end{eqnarray}

\subsection{Ninth Interacting Case}
We now consider the ninth interaction term given by:
\begin{eqnarray}
    Q_9 &=& 3Hd^2\left( \frac{\rho_m- \rho_D}{\rho_m+\rho_D}  \right).\label{Q9}
\end{eqnarray}
Following the same procedure made for the case with $Q_8$, we obtain the following expression for the continuity equation of $\rho_m$:
\begin{eqnarray}
\dot{\rho}_{m} + \left(\frac{3n}{t}\right)  \rho_{m} &=&- \frac{3d^2}{n t} \left( -\frac{6\alpha}{n} + 2\beta - \gamma n + \delta n^2 \right)+d^2\left(\frac{t}{n}\right)\rho_m.\label{rho9gen}
\end{eqnarray}
The general solution of Eq. (\ref{rho9gen}) is given by:
\begin{equation}
\rho_m(t) = t^{-3n}\,e^{\frac{d^2 t^2}{2n}}
\left[
3K\left(\frac{2n}{d^2}\right)^{\frac{3n-2}{2}}
\Gamma\!\left(\frac{3n}{2},\frac{d^2 t^2}{2n}\right)
+ C_1
\right],\label{rhom9sol}
\end{equation}
where $C_1$ is an integration constant, $K$ is defined as
\begin{eqnarray}
    K = -\frac{6\alpha}{n} + 2\beta - \gamma n + \delta n^2.
\end{eqnarray}
Moreover, $\Gamma(a,x)$ is the Gamma function.\\
We can relate the constant $C$ to the present day of $\rho_m$, indicated with $\rho_{m0}$, and to the other parameters of the model we consider.\\
At present time, i.e. for $t=t_0$ (with $t_0$ being the present day age of the Universe), we have that $\rho_m(t=t_0)=\rho_{m0}$, therefore from Eq. (\ref{rhom8sol}) we obtain:
\begin{eqnarray}
C_1 &=& \rho_{m0}\, t_0^{3n} \, e^{-\frac{d^2 t_0^2}{2n}} 
- 3K\left(\frac{2n}{d^2}\right)^{\frac{3n-2}{2}}
\Gamma\left(\frac{3n}{2},\frac{d^2 t_0^2}{2n}\right).
\end{eqnarray}
Using the expression of $\rho_m$ obtained in Eq. (\ref{rhom9sol}) along with the expression of $\rho_D$ obtained in Eq. (\ref{cz3}), we can obtain the expression of $H^2$ from the Friedmann equation given in Eq. (\ref{7}), which leads to
\begin{eqnarray}
    H^2= t^{-3n}\,e^{\frac{d^2 t^2}{2n}}
\left[
K\left(\frac{2n}{d^2}\right)^{\frac{3n-2}{2}}
\Gamma\!\left(\frac{3n}{2},\frac{d^2 t^2}{2n}\right)
+ \frac{C_1}{3}\right]+ Kt^{-2}.\label{acca9sol}
\end{eqnarray}
We can also relate the constant $C_1$ to the present day value of the Hubble parameter, i.e. $H_0$.\\
Considering $H(t=t_0)=H_0$ in Eq. (\ref{acca9sol}), we obtain:
\begin{eqnarray}
C_1 &=& 3 t_0^{3 n} e^{- \frac{d^2 t_0^2}{2n}} \Biggl[ H_0^2 - K t_0^{-2} - t_0^{-3 n} e^{\frac{d^2 t_0^2}{2n}} K\left(\frac{2n}{d^2}\right)^{\frac{3n-2}{2}}
\Gamma\!\left(\frac{3n}{2},\frac{d^2 t_0^2}{2n}\right) \Biggr].
\end{eqnarray}
We now want to obtain the expression of the deceleration parameter  $q$ using the general definition given in Eq. (\ref{qgen}).\\
Using the expression of $H^2$ given in Eq. (\ref{acca9sol}), we obtain:
\begin{eqnarray}
q(t) &=& -1 - \left\{
t^{-3n-1} e^{\frac{d^2 t^2}{2n}} \left( \frac{d^2 t^2}{n} - 3 n \right) \left[ K \left(\frac{2n}{d^2}\right)^{\frac{3n-2}{2}} \Gamma\left(\frac{3n}{2}, \frac{d^2 t^2}{2n}\right) + \frac{C_1}{3} \right] 
\right. \nonumber \\
&&\left.\,\,\,\,\,\,\,\,\,\,\,\,\,\,\,\,\,\,\,\,- K \left(\frac{2n}{d^2}\right)^{\frac{3n-2}{2}} \frac{d^2 t}{n} t^{-3n} \left( \frac{d^2 t^2}{2n} \right)^{\frac{3n}{2}-1} - 2 K t^{-3}
\right\}\times \nonumber \\
&&\left\{2 \left\{ t^{-3n} e^{\frac{d^2 t^2}{2n}} \left[ K \left(\frac{2n}{d^2}\right)^{\frac{3n-2}{2}} \Gamma\left(\frac{3n}{2}, \frac{d^2 t^2}{2n}\right) + \frac{C_1}{3} \right] + K t^{-2} \right\}^{3/2}\right\}^{-1}.
\end{eqnarray}
We now calculate the expressions of the fractional energy densities of DM and DE in two different ways.\\
Using in the general definition of $\Omega_m$ given in Eq. (\ref{8}) the expression of $\rho_m$ given in Eq. (\ref{rhom9sol}) along with the definition of $H$ given in Eq. (\ref{acca}), we obtain:
\begin{equation}
\Omega_m(t) = \left(\frac{ t^{2-3n}  }{3 n^2}\right) e^{\frac{d^2 t^2}{2n}}
\left[ 3 K \left(\frac{2n}{d^2}\right)^{\frac{3n-2}{2}} 
\Gamma\!\left(\frac{3n}{2}, \frac{d^2 t^2}{2n}\right) + C_1 \right].
\end{equation}
Therefore, using the result of the Friedmann equation given in Eq. (\ref{11}), we can write:
\begin{equation}
\Omega_{D_{9-1}}(t) = 1-\left(\frac{ t^{2-3n} \,  }{3 n^2}\right)e^{\frac{d^2 t^2}{2n}} 
\left[ 3 K \left(\frac{2n}{d^2}\right)^{\frac{3n-2}{2}} 
\Gamma\!\left(\frac{3n}{2}, \frac{d^2 t^2}{2n}\right) + C_1 \right].\label{barbali5}
\end{equation}
The evolutionary form of the fractional energy density of DE is given by:
\begin{eqnarray}
\Omega_{D_{9-1}}' &=& - \frac{1}{3 n^3} \left\{
t^{2-3n} e^{\frac{d^2 t^2}{2n}} \left( \frac{d^2 t^2}{n} + 2 - 3n \right)
\left[ 3 K A \, \Gamma\!\left(\frac{3n}{2}, \frac{d^2 t^2}{2n}\right) + C_1 \right] 
\right. \nonumber \\
&&\left.- 3 K A \frac{d^2 t}{n} t^{3-3n} \left(\frac{d^2 t^2}{2n}\right)^{\frac{3n}{2}-1} 
\right\}, 
\end{eqnarray}
where $A$ is defined as:
\begin{eqnarray}
    A = \left(\frac{2n}{d^2}\right)^{\frac{3n-2}{2}}.
\end{eqnarray}
 Instead, using in the general definition of $\Omega_m$ given in Eq. (\ref{8}) the expression of $\rho_m$ given in Eq. (\ref{rhom9sol}) along with the definition of $H^2$ given in Eq. (\ref{acca9sol}), we obtain:
\begin{equation}
\Omega_m(t) = \frac{
3 K \left(\frac{2n}{d^2}\right)^{\frac{3n-2}{2}} \Gamma\!\left(\frac{3n}{2}, \frac{d^2 t^2}{2n}\right) + C_1
}{
3 \left[ K \left(\frac{2n}{d^2}\right)^{\frac{3n-2}{2}} \Gamma\!\left(\frac{3n}{2}, \frac{d^2 t^2}{2n}\right) + \frac{C_1}{3} + K t^{3n-2} e^{-d^2 t^2/(2n)} \right]
}.
\end{equation}
Therefore, using the result of the Friedmann equation given in Eq. (\ref{11}), we can write:
\begin{equation}
\Omega_{D_{9-2}}(t) =1- \frac{
3 K \left(\frac{2n}{d^2}\right)^{\frac{3n-2}{2}} \Gamma\!\left(\frac{3n}{2}, \frac{d^2 t^2}{2n}\right) + C_1
}{
3 \left[ K \left(\frac{2n}{d^2}\right)^{\frac{3n-2}{2}} \Gamma\!\left(\frac{3n}{2}, \frac{d^2 t^2}{2n}\right) + \frac{C_1}{3} + K t^{3n-2} e^{-d^2 t^2/(2n)} \right]
}.\label{barnali6}
\end{equation}
The evolutionary form of the fractional energy density of DE is given by:
\begin{eqnarray}
\Omega_{D_{9-2}}' = - \frac{t}{n} \frac{N' D - N D'}{D^2}, 
\end{eqnarray}
where we defined:
\begin{eqnarray}
A &=& \left(\frac{2n}{d^2}\right)^{\frac{3n-2}{2}},\\
\Gamma &=& \Gamma\left(\frac{3n}{2}, \frac{d^2 t^2}{2n}\right),\\
E &=& e^{-d^2 t^2/(2n)},\\
N &=& 3 K A \Gamma + C_1, \\
D &=& 3 \left( K A \Gamma + C_1/3 + K t^{3n-2} E \right), \\
N' &=& - \frac{3 K A d^2 t}{n} \left(\frac{d^2 t^2}{2n}\right)^{3n/2-1} E, \\
D' &=& 3 \Big[ N' + K t^{3n-3} E \left( 3n-2 - \frac{d^2 t^2}{n} \right) \Big].
\end{eqnarray}
We now want to derive the final expression of the pressure of DE $p_D$.\\
The general definition for this case is given by:
\begin{eqnarray}
    p_{D}= -\rho_{D} - \frac{\dot{\rho}_{D}}{3H} - \frac{Q_9}{3H}.
\end{eqnarray}
Using the expressions of $\rho_{D}$, $\dot{\rho}_{D}$ and $H$ we derived before, we can write: 
\begin{eqnarray}
    p_{D}(t) 
    &=& \left( \frac{2}{n} -3 \right)\left[ -\frac{6\alpha}{n} + 2\beta - \gamma n + \delta n^2 \right]t^{-2}.
\end{eqnarray}
Using  the expression of $Q_9$ given in Eq. (\ref{Q9}), we can write:
\begin{eqnarray}
   \frac{Q_9}{3H} = d^2 \left( \frac{\rho_m- \rho_D}{\rho_m+\rho_D}\right).
\end{eqnarray}
The final expression of $p_D$ is then given by:
\begin{eqnarray}
    p_{D}(t) 
    &=& \left( \frac{2}{n} -3 \right)\left[ -\frac{6\alpha}{n} + 2\beta - \gamma n + \delta n^2 \right]t^{-2}.\nonumber \\
    &&-d^2\left[ \frac{ t^{-3n+2} e^{\frac{d^2 t^2}{2n}} G - 3 K }{ t^{-3n+2} e^{\frac{d^2 t^2}{2n}} G + 3 K }\right],
\end{eqnarray}
where $G$ is defined as
\begin{eqnarray}
 G= 3 K \left(\frac{2n}{d^2}\right)^{\frac{3n-2}{2}} \Gamma\left(\frac{3n}{2}, \frac{d^2 t^2}{2n}\right) + C_1.
\end{eqnarray}
We now want to calculate the expression of the EoS parameter of DE $\omega_D$. \\
The general expression for this case is given by:
\begin{eqnarray}
    \omega_{D} = -1 - \frac{\dot\rho_{D}}{3H\rho_{D}}- \frac{Q_9}{3H\rho_{D}},
\end{eqnarray}
Using  the expression of $Q_9$ given in Eq. (\ref{Q9}), we can write:
\begin{eqnarray}
 \frac{Q_9}{3H\rho_{D}} &=&  \frac{d^2}{\rho_{D}}  \left( \frac{\rho_m- \rho_D}{\rho_m+\rho_D}  \right)\nonumber \\
  &=&  \frac{d^2}{\rho_{D}\rho_{cr}}  \left( \frac{\rho_m- \rho_D}{\Omega_m+\Omega_D}  \right)\nonumber \\
  &=&  d^2  \left( \frac{\rho_m- \rho_D}{\rho_{D}\rho_{cr}}  \right)\nonumber \\
    &=&  d^2\left( \frac{1 - 2 \, \Omega_D}{\Omega_D \, \rho_{cr}}\right).
\end{eqnarray}
Therefore, using the expression of $\Omega_D$ given in Eq. (\ref{barbali5}), we obtain the following relation for $\omega_D$:
\begin{eqnarray}
    \omega_{D_{9,I-1}} &=& -1 +\frac{2}{3n}-  \frac{d^2t^2}{3n^2}\left(\frac{1 - 2 \Omega_{D_{9-1}}}{\Omega_{D_{9-1}}}  \right).
\end{eqnarray}
Instead, using the expression of $\Omega_{D_{9-2}}$ given in Eq. (\ref{barnali6}), we obtain the following relation for $\omega_D$:
\begin{eqnarray}
    \omega_{D_{9,I-2}} &=& -1 +\frac{2}{3n}-  \frac{d^2t^2}{3n^2}\left( \frac{1 - 2 \Omega_{D_{9-2}}}{\Omega_{D_{9-2}}}\right).
\end{eqnarray}

\section{Reconstruction of Scalar Field Models}
In this Section, we establish a correspondence between the DE model studied in this work and a set of widely used scalar field frameworks. In particular, we focus on the tachyon, the k-essence, the quintessence, the Yang–Mills (YM) and the Non-Linear Electrodynamics (NLED) scalar field models.

\subsection{The Tachyon Scalar Field Model}
We begin by considering the first scalar field model, namely the tachyon field.\\
In recent years, there has been significant interest in inflationary models involving the tachyon, which is regarded as a potential candidate for driving DE \cite{ref33,ref33-2}. The tachyon is an unstable scalar field naturally emerging in string theory, particularly through its appearance in the Dirac-Born-Infeld (DBI) action (the DBI model will be discussed later), which describes the dynamics of D-branes \cite{ref34,ref34-1,ref34-2,ref34-3}. It has been suggested that tachyon condensation near the maximum of the effective scalar potential could have triggered cosmological inflation in the early Universe. A rolling tachyon exhibits a distinctive equation of state (EoS), with its parameter varying smoothly between $-1$ and $0$, thereby interpolating between a cosmological constant and pressureless dust. This feature has motivated models in which dark energy is treated as a dynamical quantity, inspiring the concept of a variable cosmological constant and encouraging further studies of tachyon-driven inflationary scenarios.

The tachyon scalar field possesses an effective Lagrangian derived from open string field theory~\cite{sen}, making it a compelling candidate to explain cosmic acceleration. 
Its dynamics are governed by the following Lagrangian~\cite{sen1}:
\begin{eqnarray}
L=-V(\phi)\sqrt{1-g^{\mu \nu}\partial _{\mu}\phi \partial_{\nu}\phi}.\label{39}
\end{eqnarray}
Here, $V(\phi)$ denotes the tachyon potential, while $g^{\mu \nu}$ represents the metric tensor. \\
The energy density $\rho_{\phi}$ and pressure $p_{\phi}$ associated with the tachyon field can be derived from the effective Lagrangian and are given by:
\begin{eqnarray}
    \rho_{\phi}&=&\frac{V(\phi)}{\sqrt{1-\dot{\phi}^2}},\label{40}\\
p_{\phi}&=& -V(\phi)\sqrt{1-\dot{\phi}^2}.\label{41}
\end{eqnarray}
Furthermore, the equation of state (EoS) parameter $\omega_{\phi}$ corresponding to the tachyon scalar field is expressed as:
\begin{eqnarray}
\omega_{\phi}=\frac{p_{\phi}}{\rho_{\phi}}=\dot{\phi}^2-1\rightarrow  \dot{\phi}^2 = 1+\omega_{\phi}.\label{42}
\end{eqnarray}
For the tachyon field, the energy density $\rho_{\phi}$ is real only if the field velocity satisfies $-1 < \dot{\phi} < 1$. 
As a result, from Eq.~(\ref{42}), the equation of state (EoS) parameter $\omega_{\phi}$ is constrained to the interval $-1 < \omega_{\phi} < 0$. 
This implies that, while the tachyon field can drive the accelerated expansion of the Universe, it cannot enter the phantom regime characterized by $\omega_{\Lambda} < -1$.\\
We now aim to establish a correspondence between the tachyon scalar field model and the dark energy models considered in this work.\\
By equating the dark energy density $\rho_D$ with that in Eq.~(\ref{40}), we obtain the following expression for the tachyon potential $V(\phi)$:
\begin{eqnarray}
    V(\phi)=\rho_{\Lambda} \sqrt{1-\dot{\phi}^2} =\rho_{\Lambda} \sqrt{-\omega_{\phi}}.\label{43}
\end{eqnarray}
Instead, using the expression of the EoS parameter  $\omega_{D_{NI}}$ for the non interacting case given in Eq.~(\ref{42}), we obtain the following expression for the $\dot{\phi}^2$:
\begin{eqnarray}
\dot{\phi}^2_{1}&=& 1 + \omega_{D_{1}}= \frac{2}{3n}\rightarrow \phi_1 = \sqrt{\frac{2}{3n}}\cdot t,
\label{44}\\
    V( \phi  )_{1} &=&\rho_D\sqrt{-\omega_{D_{1}}}\nonumber \\
    &=& \sqrt{1-\frac{2}{3n}}\cdot\rho_D\nonumber \\
    &=& 3\sqrt{1-\frac{2}{3n}}\cdot  \left( -\frac{6\alpha}{n} + 2\beta - \gamma n + \delta n^2 \right)t^{-2}    ,\label{45}
\end{eqnarray}
Therefore, we can write $V$ as function of $\phi$ as follows:
\begin{eqnarray}
V(\phi)_1 = \frac{2 }{n} \sqrt{1 - \frac{2}{3n}}
\left(-\frac{6\alpha}{n} + 2\beta - \gamma n + \delta n^2\right) \cdot \phi_1^{-2}.
\end{eqnarray}
    For the first interacting case, we obtain:
\begin{eqnarray}
\dot{\phi}^2_{1,I}&=& 1 + \omega_{D_{1,I}},
\label{}\\
    V( \phi  )_{1,I} &=&\rho_D\sqrt{-\omega_{D_{1,I}}}.
\end{eqnarray}
For the second interacting case, we obtain:
\begin{eqnarray}
\dot{\phi}^2_{2,I}&=& 1 + \omega_{D_{1,I}} = \frac{2}{3n}-d^2\rightarrow \phi_{2,I} = \sqrt{\frac{2}{3n}-d^2}\cdot t,
\label{}\\
    V( \phi  )_{2,I} &=&\rho_D\sqrt{-\omega_{D_{1,I}}}\nonumber \\&=& \sqrt{1-\frac{2}{3n}+d^2}\cdot\rho_D\nonumber \\
    &=& 3\sqrt{1-\frac{2}{3n}+d^2}\cdot  \left( -\frac{6\alpha}{n} + 2\beta - \gamma n + \delta n^2 \right)t^{-2}  .
\end{eqnarray}
Therefore, we can write:
\begin{eqnarray}
V(\phi)_{2,I} = 3 \left(\frac{2}{3n} - d^2\right) 
\sqrt{1 - \frac{2}{3n} + d^2} 
\left(-\frac{6\alpha}{n} + 2\beta - \gamma n + \delta n^2 \right) \cdot\phi_{2,I}^{-2}.
\end{eqnarray}
In the limiting case of $d^2=0$, we recover the same result of the non interacting case, as it is expected.\\
For the third interacting case, we obtain:
\begin{eqnarray}
\dot{\phi}^2_{3,I-1}&=& 1 + \omega_{D_{3,I-1}},
\\
    V( \phi  )_{3,I-1} &=&\rho_D\sqrt{-\omega_{D_{3,I-1}}},
\\
\dot{\phi}^2_{3,I-2}&=& 1 + \omega_{D_{3,I-2}},
\\
    V( \phi  )_{3,I-2} &=&\rho_D\sqrt{-\omega_{D_{3,I-2}}},
\end{eqnarray}
and
 \begin{eqnarray}
\dot{\phi}^2_{3,I-1,lim}&=& 1 + \omega_{D_{3,I-1,lim}},
\\
    V( \phi  )_{3,I-1,lim} &=&\rho_D\sqrt{-\omega_{D_{3,I-1,lim}}},
\\
\dot{\phi}^2_{3,I-2,lim}&=& 1 + \omega_{D_{3,I-2,lim}},
\\
    V( \phi  )_{3,I-2,lim} &=&\rho_D\sqrt{-\omega_{D_{3,I-2,lim}}}.
\end{eqnarray}
For the fouth interacting case, we obtain:
\begin{eqnarray}
\dot{\phi}^2_{4,I-1}&=& 1 + \omega_{D_{4,I-1}},
\\
    V( \phi  )_{4,I-1} &=&\rho_D\sqrt{-\omega_{D_{4,I-1}}},
\\
\dot{\phi}^2_{4,I-2}&=& 1 + \omega_{D_{4,I-2}},
\\
    V( \phi  )_{4,I-2} &=&\rho_D\sqrt{-\omega_{D_{4,I-2}}},
\end{eqnarray}
and
 \begin{eqnarray}
\dot{\phi}^2_{4,I-1,lim}&=& 1 + \omega_{D_{4,I-1,lim}},
\\
    V( \phi  )_{4,I-1,lim} &=&\rho_D\sqrt{-\omega_{D_{4,I-1,lim}}},
\\
\dot{\phi}^2_{4,I-2,lim}&=& 1 + \omega_{D_{4,I-2,lim}},
\\
    V( \phi  )_{4,I-2,lim} &=&\rho_D\sqrt{-\omega_{D_{4,I-2,lim}}}.
\end{eqnarray}
For the fifth interacting case, we obtain:
\begin{eqnarray}
\dot{\phi}^2_{5,I-1}&=& 1 + \omega_{D_{5,I-1}},
\\
    V( \phi  )_{5,I-1} &=&\rho_D\sqrt{-\omega_{D_{5,I-1}}},
\\
\dot{\phi}^2_{5,I-2}&=& 1 + \omega_{D_{5,I-2}},
\\
    V( \phi  )_{5,I-2} &=&\rho_D\sqrt{-\omega_{D_{5,I-2}}}.
\end{eqnarray}
For the sixth interacting case, we obtain:
\begin{eqnarray}
\dot{\phi}^2_{6,I-1}&=& 1 + \omega_{D_{6,I-1}},
\\
    V( \phi  )_{6,I-1} &=&\rho_D\sqrt{-\omega_{D_{6,I-1}}},
\\
\dot{\phi}^2_{6,I-2}&=& 1 + \omega_{D_{6,I-2}},
\\
    V( \phi  )_{6,I-2} &=&\rho_D\sqrt{-\omega_{D_{6,I-2}}},
\end{eqnarray}
and
 \begin{eqnarray}
\dot{\phi}^2_{6,I-1,lim}&=& 1 + \omega_{D_{6,I-1,lim}},
\\
    V( \phi  )_{6,I-1,lim} &=&\rho_D\sqrt{-\omega_{D_{6,I-1,lim}}},
\\
\dot{\phi}^2_{6,I-2,lim}&=& 1 + \omega_{D_{6,I-2,lim}},
\\
    V( \phi  )_{6,I-2,lim} &=&\rho_D\sqrt{-\omega_{D_{6,I-2,lim}}}.
\end{eqnarray}
For the seventh interacting case, we obtain:
\begin{eqnarray}
\dot{\phi}^2_{7,I-1}&=& 1 + \omega_{D_{7,I-1}},\\
    V( \phi  )_{7,I-1} &=&\rho_D\sqrt{-\omega_{D_{7,I-1}}},\\
\dot{\phi}^2_{7,I-2}&=& 1 + \omega_{D_{7,I-2}},\\
    V( \phi  )_{7,I-2} &=&\rho_D\sqrt{-\omega_{D_{7,I-2}}},
\end{eqnarray}
and
 \begin{eqnarray}
\dot{\phi}^2_{7,I-1,lim}&=& 1 + \omega_{D_{7,I-1,lim}},\\
    V( \phi  )_{7,I-1,lim} &=&\rho_D\sqrt{-\omega_{D_{7,I-1,lim}}},\\
\dot{\phi}^2_{7,I-2,lim}&=& 1 + \omega_{D_{7,I-2,lim}},
\\
    V( \phi  )_{7,I-2,lim} &=&\rho_D\sqrt{-\omega_{D_{7,I-2,lim}}}.
\end{eqnarray}
For the  eighth interacting case, we obtain:
\begin{eqnarray}
\dot{\phi}^2_{8,I-1}&=& 1 + \omega_{D_{8,I-1}},\\
    V( \phi  )_{8,I-1} &=&\rho_D\sqrt{-\omega_{D_{8,I-1}}},\\
\dot{\phi}^2_{8,I-2}&=& 1 + \omega_{D_{8,I-2}},\\
    V( \phi  )_{8,I-2} &=&\rho_D\sqrt{-\omega_{D_{8,I-2}}}.
\end{eqnarray}
For the ninth interacting case, we obtain:
\begin{eqnarray}
\dot{\phi}^2_{9,I-1}&=& 1 + \omega_{D_{9,I-1}},\\
    V( \phi  )_{9,I-1} &=&\rho_D\sqrt{-\omega_{D_{9,I-1}}},\\
\dot{\phi}^2_{9,I-2}&=& 1 + \omega_{D_{9,I-2}},\\
    V( \phi  )_{9,I-2} &=&\rho_D\sqrt{-\omega_{D_{9,I-2}}}.
\end{eqnarray}

\subsection{The K-essence Scalar Field Model}
Next, we focus on the second scalar field model, namely K-essence.\\
K-essence models are characterized by a scalar field whose kinetic term enters the Lagrangian in a non-canonical way. This class of models, originally inspired by the Born–Infeld action in string theory, has been widely employed to describe the late-time accelerated expansion of the Universe~\cite{35zim}. 
The general action $S$ for the K-essence scalar field depends on the field $\phi$ and the kinetic term $\chi = \frac{1}{2}\dot{\phi}^2$, and can be written as~\cite{36zim-1,36zim-2}:
\begin{eqnarray}
S = \int d^4x \, \sqrt{-g} \, p(\phi, \chi). \label{51}
\end{eqnarray}
Here, $p(\phi, \chi)$ denotes the Lagrangian density, which coincides with the pressure of the scalar field, and $g$ is the determinant of the metric tensor $g^{\mu \nu}$. 
Assuming a separable form $p(\phi, \chi) = f(\phi) g(\chi)$, the pressure and energy density of the scalar field $\phi$ can be written as:
\begin{eqnarray}
    p(\phi, \chi )&=&f(\phi)( -\chi+\chi ^2   ), \label{52}\\
        \rho(\phi, \chi )&=&f(\phi)(-\chi+3\chi ^2).\label{53}
\end{eqnarray}
The function $f(\phi)$ is an arbitrary function, chosen based on physical or mathematical considerations, which determines the explicit dependence of the pressure and energy density on the scalar field $\phi$. 
It can be interpreted as a multiplicative potential that modifies the kinetic contribution of the field. When $f(\phi) > 0$, it scales both the pressure and energy density positively. 
In this sense, $f(\phi)$ can be regarded as an effective potential or interaction for the scalar field. \\
In K-essence cosmology, specific forms of $f(\phi)$ are selected to model the evolution of the scalar field. For instance, one can take $f(\phi) = V(\phi)$ as a scalar potential, or adopt more complex forms to achieve particular dynamical features, such as cosmic acceleration, tracking behavior, or transitions of the equation of state parameter $\omega$. \\
Consequently, the equation of state (EoS) parameter $\omega_K$ for the K-essence scalar field can be expressed as:
\begin{eqnarray}
    \omega _K= \frac{p(\phi, \chi )}{\rho(\phi, \chi )}=\frac{\chi-1}{3\chi -1}.\label{54}
\end{eqnarray}
From the expression of $\omega_K$ given above, it follows that the K-essence scalar field exhibits phantom behavior, i.e., $\omega_K < -1$, when the kinetic term satisfies $1/3 < \chi < 1/2$. \\
We now establish a correspondence between the K-essence EoS parameter, $\omega_K$, and the dark energy models considered in this work. \\
For the non interacting case, the value of $\chi$ corresponding to the model under study can be obtained by equating the expressions of $\omega_{D_{NI}}$  with Eq.~(\ref{54}), yielding:
For the non interacting case, we obtain:
\begin{eqnarray}
    \chi_{NI} &=& \frac{\omega_{D_{NI}}-1}{3\omega_{D_{NI}}-1}\nonumber \\ 
    &=&  \frac{1 - 3n}{3(1 - 2n)}.
\end{eqnarray}
Moreover, using the expression of $\rho_{D}$  in Eq. (\ref{53}), we obtain:
For the non interacting case, we obtain:
\begin{eqnarray}
    f_{NI}(\phi )&=&\frac{\rho_{D}}{\chi_{NI}(3\chi_{NI}-1)}\nonumber \\
    &=& - \frac{3 (1 - 2n)^2}{n (1 - 3n)} \, \rho_D \nonumber \\
    &=&  - \frac{9 (1 - 2n)^2}{n (1 - 3n)} \left( -\frac{6\alpha}{n} + 2\beta - \gamma n + \delta n^2 \right)t^{-2}.
\end{eqnarray}
Furthermore, using the expression of $\chi_{NI}$, we have: 
\begin{eqnarray}
\chi_{NI} (3 \chi_{NI} - 1) &=& \frac{2(1 - \omega_{D_{NI}})}{(3 \omega_{D_{NI}} - 1)^2} \nonumber \\
&=&- \frac{n (1 - 3n)}{3 (1 - 2n)^2}.
\end{eqnarray}
For the first interacting case, we obtain:
\begin{eqnarray}
    \chi_{1,I-1} &=& \frac{\omega_{D_{1,I-1}}-1}{3\omega_{D_{1,I-1}}-1},
\\
    f_{1,I-1}(\phi )&=&\frac{\rho_{D}}{\chi_{1,I-1}(3\chi_{1,I-1}-1)} ,
\\
\chi_{1,I-1} (3 \chi_{1,I-1} - 1) &=& \frac{2(1 - \omega_{D_{1,I-1}})}{(3 \omega_{D_{1,I-1}} - 1)^2},
\\
    \chi_{1,I-2} &=& \frac{\omega_{D_{1,I-2}}-1}{3\omega_{D_{1,I-2}}-1},
\\
    f_{1,I-2}(\phi )&=&\frac{\rho_{D}}{\chi_{1,I-2}(3\chi_{1,I-2}-1)} ,
\\
\chi_{1,I-2} (3 \chi_{1,I-2} - 1) &=& \frac{2(1 - \omega_{D_{1,I-2}})}{(3 \omega_{D_{1,I-2}} - 1)^2}.
\end{eqnarray}
For the second interacting case, we obtain:
\begin{eqnarray}
    \chi_{2,I} &=& \frac{\omega_{D_{2,I}}-1}{3\omega_{D_{2,I}}-1}, \nonumber \\
    &=&\frac{2 - 6 n - 3 n d^2}{3 \left( 2 - 4 n - 3 n d^2 \right)}\nonumber \\
    &=&\frac{2 - 3n(2+d^2)}{3 \left[ 2 - n(4 + 3  d^2) \right]},
\\
    f_{2,I}(\phi )&=&\frac{\rho_{D}}{\chi_{2,I}(3\chi_{2,I}-1)}\nonumber \\
    &=& -\frac{3 \left[ 2 - n (4 + 3 d^2) \right]^2}{2 n \left[ 2 - 3 n (2 + d^2) \right]}\rho_D\nonumber \\
    &=&-\frac{9 \left[ 2 - n (4 + 3 d^2) \right]^2}{2 n \left[ 2 - 3 n (2 + d^2) \right]} \left( -\frac{6\alpha}{n} + 2\beta - \gamma n + \delta n^2 \right)t^{-2},\\
\chi_{2,I} (3 \chi_{2,I} - 1) &=& \frac{2(1 - \omega_{D_{2,I}})}{(3 \omega_{D_{2,I}} - 1)^2} \nonumber \\
&=& - \frac{2 n \left[ 2 - 3 n (2 + d^2) \right]}{3 \left[ 2 - n (4 + 3 d^2) \right]^2}.
\end{eqnarray}

For the third interacting case, we obtain:
\begin{eqnarray}
    \chi_{3,I-1} &=& \frac{\omega_{D_{3,I-1}}-1}{3\omega_{D_{3,I-1}}-1},
\\
    f_{3,I-1}(\phi )&=&\frac{\rho_{D}}{\chi_{3,I-1}(3\chi_{3,I-1}-1)} ,
\\
\chi_{3,I-1} (3 \chi_{3,I-1} - 1) &=& \frac{2(1 - \omega_{D_{3,I-1}})}{(3 \omega_{D_{3,I-1}} - 1)^2},
\\
    \chi_{3,I-2} &=& \frac{\omega_{D_{3,I-2}}-1}{3\omega_{D_{3,I-2}}-1},
\\
    f_{3,I-2}(\phi )&=&\frac{\rho_{D}}{\chi_{3,I-2}(3\chi_{3,I-2}-1)} ,
\\
\chi_{3,I-2} (3 \chi_{3,I-2} - 1) &=& \frac{2(1 - \omega_{D_{3,I-2}})}{(3 \omega_{D_{3,I-2}} - 1)^2},
\end{eqnarray}

and
\begin{eqnarray}
    \chi_{3,I-1,lim} &=& \frac{\omega_{D_{3,I-1,lim}}-1}{3\omega_{D_{3,I-1,lim}}-1},    
\\
    f_{3,I-1,lim}(\phi )&=&\frac{\rho_{D}}{\chi_{3,I-1}(3\chi_{3,I-1,lim}-1)} ,
\\
\chi_{3,I-1,lim} (3 \chi_{3,I-1,lim} - 1) &=& \frac{2(1 - \omega_{D_{3,I-1,lim}})}{(3 \omega_{D_{3,I-1,lim}} - 1)^2},
\\
    \chi_{3,I-2,lim} &=& \frac{\omega_{D_{3,I-2,lim}}-1}{3\omega_{D_{3,I-2}}-1},    
\\
    f_{3,I-2,lim}(\phi )&=&\frac{\rho_{D}}{\chi_{3,I-2,lim}(3\chi_{3,I-2}-1)} ,
\\
\chi_{3,I-2,lim} (3 \chi_{3,I-2,lim} - 1) &=& \frac{2(1 - \omega_{D_{3,I-2}})}{(3 \omega_{D_{3,I-2,lim}} - 1)^2}.
\end{eqnarray}
For the fourth interacting case, we obtain:
\begin{eqnarray}
    \chi_{4,I-1} &=& \frac{\omega_{D_{4,I-1}}-1}{3\omega_{D_{4,I-1}}-1},    
\\
    f_{4,I-1}(\phi )&=&\frac{\rho_{D}}{\chi_{4,I-1}(3\chi_{4,I-1}-1)} ,
\\
\chi_{4,I-1} (3 \chi_{4,I-1} - 1) &=& \frac{2(1 - \omega_{D_{4,I-1}})}{(3 \omega_{D_{3,I-1}} - 1)^2},
\\
    \chi_{4,I-2} &=& \frac{\omega_{D_{4,I-2}}-1}{3\omega_{D_{4,I-2}}-1},    
\\
    f_{4,I-2}(\phi )&=&\frac{\rho_{D}}{\chi_{4,I-2}(3\chi_{4,I-2}-1)} ,
\\
\chi_{4,I-2} (3 \chi_{4,I-2} - 1) &=& \frac{2(1 - \omega_{D_{4,I-2}})}{(3 \omega_{D_{3,I-2}} - 1)^2},
\end{eqnarray}
and
\begin{eqnarray}
     \chi_{4,I-1,lim} &=& \frac{\omega_{D_{4,I-1,lim}}-1}{3\omega_{D_{4,I-1,lim}}-1}, \\
    f_{4,I-1,lim}(\phi )&=&\frac{\rho_{D}}{\chi_{4,I-1}(3\chi_{4,I-1,lim}-1)} ,
\\
\chi_{4,I-1,lim} (3 \chi_{4,I-1,lim} - 1) &=& \frac{2(1 - \omega_{D_{4,I-1,lim}})}{(3 \omega_{D_{4,I-1,lim}} - 1)^2},
\\
    \chi_{4,I-2,lim} &=& \frac{\omega_{D_{4,I-2,lim}}-1}{3\omega_{D_{4,I-2,lim}}-1},    
\\
    f_{4,I-2,lim}(\phi )&=&\frac{\rho_{D}}{\chi_{4,I-2,lim}(3\chi_{4,I-2,lim}-1)} ,
\\
\chi_{4,I-2,lim} (3 \chi_{4,I-2,lim} - 1) &=& \frac{2(1 - \omega_{D_{4,I-2,lim}})}{(3 \omega_{D_{4,I-2,lim}} - 1)^2}.
\end{eqnarray}
For the fifth interacting case, we obtain:
\begin{eqnarray}
    \chi_{5,I-1} &=& \frac{\omega_{D_{5,I-1}}-1}{3\omega_{D_{5,I-1}}-1},    
\\
    f_{5,I-1}(\phi )&=&\frac{\rho_{D}}{\chi_{5,I-1}(3\chi_{5,I-1}-1)},
\\
\chi_{5,I-1} (3 \chi_{5,I-1} - 1) &=& \frac{2(1 - \omega_{D_{5,I-1}})}{(3 \omega_{D_{5,I-1}} - 1)^2},
\\
    \chi_{5,I-2} &=& \frac{\omega_{D_{5,I-2}}-1}{3\omega_{D_{5,I-2}}-1},    
\\
    f_{5,I-2}(\phi )&=&\frac{\rho_{D}}{\chi_{5,I-2}(3\chi_{5,I-2}-1)} ,
\\
\chi_{5,I-2} (3 \chi_{5,I-2} - 1) &=& \frac{2(1 - \omega_{D_{5,I-2}})}{(3 \omega_{D_{5,I-2}} - 1)^2}.
\end{eqnarray}
For the sixth interacting case, we obtain:
\begin{eqnarray}
    \chi_{6,I-1} &=& \frac{\omega_{D_{6,I-1}}-1}{3\omega_{D_{6,I-1}}-1},    
\\
    f_{6,I-1}(\phi )&=&\frac{\rho_{D}}{\chi_{6,I-1}(3\chi_{6,I-1}-1)} ,
\\
\chi_{6,I-1} (3 \chi_{6,I-1} - 1) &=& \frac{2(1 - \omega_{D_{6,I-1}})}{(3 \omega_{D_{6,I-1}} - 1)^2},
\\
    \chi_{6,I-2} &=& \frac{\omega_{D_{6,I-2}}-1}{3\omega_{D_{6,I-2}}-1},    
\\
    f_{6,I-2}(\phi )&=&\frac{\rho_{D}}{\chi_{6,I-2}(3\chi_{6,I-2})} ,
\\
\chi_{6,I-2} (3 \chi_{6,I-2} - 1) &=& \frac{2(1 - \omega_{D_{6,I-2}})}{(3 \omega_{D_{6,I-2}} - 1)^2},
\end{eqnarray}

and
\begin{eqnarray}
    \chi_{6,I-1,lim} &=& \frac{\omega_{D_{6,I-1,lim}}-1}{3\omega_{D_{6,I-1,lim}}-1},    
\\
    f_{6,I-1,lim}(\phi )&=&\frac{\rho_{D}}{\chi_{6,I-1}(3\chi_{6,I-1,lim}-1)} ,
\\
\chi_{6,I-1,lim} (3 \chi_{6,I-1,lim} - 1) &=& \frac{2(1 - \omega_{D_{6,I-1,lim}})}{(3 \omega_{D_{6,I-1,lim}} - 1)^2},
\\
    \chi_{6,I-2,lim} &=& \frac{\omega_{D_{6,I-2,lim}}-1}{3\omega_{D_{6,I-2,lim}}-1},    
\\
    f_{6,I-2,lim}(\phi )&=&\frac{\rho_{D}}{\chi_{6,I-2,lim}(3\chi_{6,I-2,lim}-1)} ,
\\
\chi_{6,I-2,lim} (3 \chi_{6,I-2,lim} - 1) &=& \frac{2(1 - \omega_{D_{6,I-2,lim}})}{(3 \omega_{D_{6,I-2,lim}} - 1)^2}.
\end{eqnarray}
For the seventh interacting case, we obtain:
\begin{eqnarray}
    \chi_{7,I-1} &=& \frac{\omega_{D_{7,I-1}}-1}{3\omega_{D_{7,I-1}}-1},    
\\
    f_{7,I-1}(\phi )&=&\frac{\rho_{D}}{\chi_{7,I-1}(3\chi_{7,I-1}-1)} ,
\\
\chi_{7,I-1} (3 \chi_{7,I-1} - 1) &=& \frac{2(1 - \omega_{D_{7,I-1}})}{(3 \omega_{D_{7,I-1}} - 1)^2},
\\
    \chi_{7,I-2} &=& \frac{\omega_{D_{7,I-2}}-1}{3\omega_{D_{7,I-2}}-1},    
\\
    f_{7,I-2}(\phi )&=&\frac{\rho_{D}}{\chi_{7,I-2}(3\chi_{7,I-2}-1)} ,
\\
\chi_{7,I-2} (3 \chi_{7,I-2} - 1) &=& \frac{2(1 - \omega_{D_{7,I-2}})}{(3 \omega_{D_{7,I-2}} - 1)^2},
\end{eqnarray}
and
\begin{eqnarray}
    \chi_{7,I-1,lim} &=& \frac{\omega_{D_{7,I-1,lim}}-1}{3\omega_{D_{7,I-1,lim}}-1},    
\\
    f_{7,I-1,lim}(\phi )&=&\frac{\rho_{D}}{\chi_{7,I-1}(3\chi_{7,I-1,lim}-1)} ,
\\
\chi_{7,I-1,lim} (3 \chi_{7,I-1,lim} - 1) &=& \frac{2(1 - \omega_{D_{7,I-1,lim}})}{(3 \omega_{D_{7,I-1,lim}} - 1)^2},
\\
    \chi_{7,I-2,lim} &=& \frac{\omega_{D_{7,I-2,lim}}-1}{3\omega_{D_{7,I-2}}-1},    
\\
    f_{7,I-2,lim}(\phi )&=&\frac{\rho_{D}}{\chi_{7,I-2,lim}(3\chi_{7,I-2}-1)} ,
\\
\chi_{7,I-2,lim} (3 \chi_{7,I-2,lim} - 1) &=& \frac{2(1 - \omega_{D_{7,I-2}})}{(3 \omega_{D_{7,I-2,lim}} - 1)^2}.
\end{eqnarray}
For the eigth interacting case, we obtain:
\begin{eqnarray}
    \chi_{8,I-1} &=& \frac{\omega_{D_{8,I-1}}-1}{3\omega_{D_{8,I-1}}-1},    
\\
    f_{8,I-1}(\phi )&=&\frac{\rho_{D}}{\chi_{8,I-1}(3\chi_{8,I-1}-1)},
\\
\chi_{8,I-1} (3 \chi_{8,I-1} - 1) &=& \frac{2(1 - \omega_{D_{8,I-1}})}{(3 \omega_{D_{8,I-1}} - 1)^2},
\\
    \chi_{8,I-2} &=& \frac{\omega_{D_{8,I-2}}-1}{3\omega_{D_{8,I-2}}-1},    
\\
    f_{8,I-2}(\phi )&=&\frac{\rho_{D}}{\chi_{8,I-2}(3\chi_{8,I-2}-1)} ,
\\
\chi_{8,I-2} (3 \chi_{8,I-2} - 1) &=& \frac{2(1 - \omega_{D_{8,I-2}})}{(3 \omega_{D_{8,I-2}} - 1)^2}.
\end{eqnarray}
For the ninth interacting case, we obtain:
\begin{eqnarray}
    \chi_{9,I-1} &=& \frac{\omega_{D_{9,I-1}}-1}{3\omega_{D_{9,I-1}}-1},    
\\
    f_{9,I-1}(\phi )&=&\frac{\rho_{D}}{\chi_{9,I-1}(3\chi_{9,I-1}-1)} ,
\\
\chi_{9,I-1} (3 \chi_{9,I-1} - 1) &=& \frac{2(1 - \omega_{D_{9,I-1}})}{(3 \omega_{D_{9,I-1}} - 1)^2},
\\
    \chi_{9,I-2} &=& \frac{\omega_{D_{9,I-2}}-1}{3\omega_{D_{9,I-2}}-1},    
\\
    f_{9,I-2}(\phi )&=&\frac{\rho_{D}}{\chi_{9,I-2}(3\chi_{9,I-2}-1)}, 
\\
\chi_{9,I-2} (3 \chi_{9,I-2} - 1) &=& \frac{2(1 - \omega_{D_{9,I-2}})}{(3 \omega_{D_{9,I-2}} - 1)^2}.
\end{eqnarray}

\subsection{The Quintessence Scalar Field Model}
We now turn our attention to the quintessence scalar field model.\\
In this framework, quintessence is represented by a homogeneous scalar field $\phi$ evolving in time and minimally coupled to gravity. Its dynamics are governed by a potential $V(\phi)$, which drives the accelerated expansion of the Universe. The action $S_Q$ describing the quintessence scalar field is given by
\cite{mou1}:
\begin{eqnarray}
    S_Q=\int d^4x \sqrt{-g}\,\Big[-\frac{1}{2}g^{\mu \nu} \partial _{\mu} \phi   \partial _{\nu} \phi - V( \phi )  \Big].\label{69}
\end{eqnarray}
The energy-momentum tensor $T_{\mu \nu}$ associated with the quintessence scalar field can be obtained by varying the action $S_Q$, given in Eq.~(\ref{69}), with respect to the metric tensor $g^{\mu \nu}$:
\begin{eqnarray}
T_{\mu \nu}=\frac{2}{\sqrt{-g}} \frac{\delta S}{\delta g^{\mu \nu}},\label{70}
\end{eqnarray}
which leads to the following expression:
\begin{eqnarray}
    T_{\mu \nu}=\partial _{\mu} \phi   \partial _{\nu} \phi - g_{\mu \nu}\Big[\frac{1}{2}g^{\alpha \beta} \partial _{\alpha} \phi   \partial _{\beta} \phi + V( \phi )  \Big].\label{71}
\end{eqnarray}
In a FLRW background, the pressure $p_Q$ and the energy density $\rho_Q$ of the quintessence scalar field model are given, respectively, by:
\begin{eqnarray}
p_Q&=&T_i^i=\frac{1}{2}\dot{\phi}^2-V(\phi),\label{73}\\
    \rho_Q&=&-T_0^0=\frac{1}{2}\dot{\phi}^2+V(\phi).\label{72}
\end{eqnarray}
Therefore, using the results obtained in Eqs. (\ref{73}) and (\ref{72}), we derive that the Equation of State (EoS) parameter $\omega_Q$ for the quintessence scalar field model can be expressed as:
\begin{eqnarray}    \omega_Q=\frac{p_Q}{\rho_Q}=\frac{\dot{\phi}^2-2V(\phi)}{\dot{\phi}^2+2V(\phi)}.\label{74}
\end{eqnarray}
We find from Eq.~(\ref{74}) that, when $\omega_Q < -1/3$, the Universe undergoes accelerated expansion if $\dot{\phi}^2 < V(\phi)$.\\
Varying the quintessence action $S_Q$, as defined in Eq.~(\ref{69}), with respect to the scalar field $\phi$ yields the following equation of motion:
\begin{equation}
\ddot{\phi} + 3 H \dot{\phi} + \frac{dV}{d\phi} = 0,
\end{equation}
where $H$ is the Hubble parameter.\\
We now establish a correspondence between the interacting scenario and the quintessence DE model. Making the correspondences $\omega_Q=\omega_D$ and  $\rho_Q=\rho_D$, we derive the following expressions for $\dot{\phi}^2$ and $ V( \phi )$:
\begin{eqnarray}
    \dot{\phi}^2&=&(1+\omega_{\Lambda})\rho_{\Lambda}, \label{75}\\
    V( \phi ) &=& \frac{1}{2}(1-\omega_{\Lambda})\rho_{\Lambda}.\label{76}
\end{eqnarray}
For the non interacting case, substituting $\omega_{D_{NI}}$  in Eqs. (\ref{75}) and (\ref{76}), the kinetic energy term $\dot{\phi}^2$ and the quintessence potential energy $V( \phi )$ can be expressed as follows:
\begin{eqnarray}
    \dot{\phi}_{NI}^2&=&(1+\omega_{D})\rho_D\nonumber \\
    &=&\left( \frac{2}{3n}\right)\rho_D\nonumber \\
    &=&\left( \frac{2}{n}\right)  \left( -\frac{6\alpha}{n} + 2\beta - \gamma n + \delta n^2 \right)    t^{-2},\label{77-77}\\
    V_{NI}( \phi ) &=&\frac{\rho_{D}}{2}(1-\omega_D)\nonumber \\
    &=&\left( \frac{3n-1}{n}  \right)\left( -\frac{6\alpha}{n} + 2\beta - \gamma n + \delta n^2 \right)    t^{-2}. \label{78}
\end{eqnarray}
From Eq. (\ref{77-77}), we obtain:
\begin{eqnarray}
    \phi_{NI} = \sqrt{\left( \frac{2}{n}\right)  \left( -\frac{6\alpha}{n} + 2\beta - \gamma n + \delta n^2 \right) }\cdot \ln t.
\end{eqnarray}
We can also write
\begin{eqnarray}
     V_{NI}(\phi) = \left(\frac{3n - 1}{n}\right)  A  \exp\Bigg[- 2  \phi_{NI} \sqrt{\frac{n}{2 A}} \Bigg] ,
\end{eqnarray}
where $A$ is defined as:
\begin{eqnarray}
    A = -\frac{6\alpha}{n} + 2\beta - \gamma n + \delta n^2.
\end{eqnarray}
For the first interacting case, we obtain:
\begin{eqnarray}
    \dot{\phi}_{1,I-1}^2&=&(1+\omega_{D_{1,I-1}})\rho_D,\label{77}\\
    \dot{\phi}_{1,I-2}^2&=&(1+\omega_{D_{1,I-2}})\rho_{D },\label{77-2}\\
    V_{1,I-1}( \phi ) &=&\frac{\rho_{D}}{2}(1-\omega_{D_{1,I-1}}), \label{78}\\
    V_{1,I-2}( \phi ) &=&\frac{\rho_{D}}{2}(1-\omega_{D_{1,I-2}})\label{78-2}.
\end{eqnarray}
For the second interacting case, we obtain:
\begin{eqnarray}
    \dot{\phi}_{2,I}^2&=&(1+\omega_{D_{2,I}})\rho_D\nonumber \\
    &=&\left( \frac{2}{3n} -d^2\right)\rho_D \nonumber \\
    &=& \left( \frac{2}{n} -3d^2\right)\left( -\frac{6\alpha}{n} + 2\beta - \gamma n + \delta n^2 \right)    t^{-2},\label{77-77-}\\
    V_{2,I}( \phi ) &=&\frac{\rho_{D}}{2}(1-\omega_{D_{2,I}})\nonumber \\
    &=&\left( \frac{3n-1}{n} +\frac{3d^2}{2} \right)\left( -\frac{6\alpha}{n} + 2\beta - \gamma n + \delta n^2 \right)    t^{-2}. \label{78}
\end{eqnarray}
From Eq. (\ref{77-77-}), we obtain:
\begin{eqnarray}
    \phi_{2,I} = \sqrt{\left( \frac{2}{n}-3d^2\right)  \left( -\frac{6\alpha}{n} + 2\beta - \gamma n + \delta n^2 \right) }\cdot \ln t.
\end{eqnarray}
We can also write:
\begin{eqnarray}
    \left( \frac{3n-1}{n} + \frac{3 d^2}{2} \right) A \, \exp\left[ - 2 \frac{\phi}{\sqrt{B A}} \right],
\end{eqnarray}
where $A$ and $B$ are defined as:
\begin{eqnarray}
    A &=& -\frac{6\alpha}{n} + 2\beta - \gamma n + \delta n^2,\\
    B&=& \frac{2}{n} - 3 d^2.
\end{eqnarray}
In the limiting case of $d^2=0$, we recover the same results for the non interacting case, as it is expected.\\
For the third interacting case, we obtain:
\begin{eqnarray}
    \dot{\phi}_{3,I-1}^2&=&(1+\omega_{D_{3,I-1}})\rho_D,\label{77}\\
    \dot{\phi}_{3,I-2}^2&=&(1+\omega_{D_{3,I-2}})\rho_{D },\label{77-2}\\
    V_{3,I-1}( \phi ) &=&\frac{\rho_{D}}{2}(1-\omega_{D_{3,I-1}}), \label{78}\\
    V_{3,I-2}( \phi ) &=&\frac{\rho_{D}}{2}(1-\omega_{D_{3,I-2}}),\label{78-2}
\end{eqnarray}
and
\begin{eqnarray}
    \dot{\phi}_{3,I-1,lim}^2&=&(1+\omega_{D_{3,I-1,lim}})\rho_D,\label{77}\\
    \dot{\phi}_{3,I-2,lim}^2&=&(1+\omega_{D_{3,I-2,lim}})\rho_{D },\label{77-2}\\
    V_{3,I-1,lim}( \phi ) &=&\frac{\rho_{D}}{2}(1-\omega_{D_{3,I-1,lim}}), \label{78}\\
    V_{3,I-2,lim}( \phi ) &=&\frac{\rho_{D}}{2}(1-\omega_{D_{3,I-2,lim}}).\label{78-2}
\end{eqnarray}

For the fourth interacting case, we obtain:
\begin{eqnarray}
    \dot{\phi}_{4,I-1}^2&=&(1+\omega_{D_{4,I-1}})\rho_D,\label{77}\\
    \dot{\phi}_{4,I-2}^2&=&(1+\omega_{D_{4,I-2}})\rho_{D },\label{77-2}\\
    V_{4,I-1}( \phi ) &=&\frac{\rho_{D}}{2}(1-\omega_{D_{4,I-1}}), \label{78}\\
    V_{4,I-2}( \phi ) &=&\frac{\rho_{D}}{2}(1-\omega_{D_{4,I-2}}),\label{78-2}
\end{eqnarray}
and
\begin{eqnarray}
    \dot{\phi}_{4,I-1,lim}^2&=&(1+\omega_{D_{4,I-1,lim}})\rho_D,\label{77}\\
    \dot{\phi}_{4,I-2,lim}^2&=&(1+\omega_{D_{4,I-2,lim}})\rho_{D },\label{77-2}\\
    V_{4,I-1,lim}( \phi ) &=&\frac{\rho_{D}}{2}(1-\omega_{D_{4,I-1,lim}}), \label{78}\\
    V_{4,I-2,lim}( \phi ) &=&\frac{\rho_{D}}{2}(1-\omega_{D_{4,I-2,lim}}).\label{78-2}
\end{eqnarray}

For the fifth interacting case, we obtain:
\begin{eqnarray}
    \dot{\phi}_{5,I-1}^2&=&(1+\omega_{D_{5,I-1}})\rho_D,\label{77}\\
    \dot{\phi}_{5,I-2}^2&=&(1+\omega_{D_{5,I-2}})\rho_{D },\label{77-2}\\
    V_{5,I-1}( \phi ) &=&\frac{\rho_{D}}{2}(1-\omega_{D_{5,I-1}}), \label{78}\\
    V_{5,I-2}( \phi ) &=&\frac{\rho_{D}}{2}(1-\omega_{D_{5,I-2}})\label{78-2}.
\end{eqnarray}

For the sixth interacting case, we obtain:
\begin{eqnarray}
    \dot{\phi}_{6,I-1}^2&=&(1+\omega_{D_{6,I-1}})\rho_D,\label{77}\\
    \dot{\phi}_{6,I-2}^2&=&(1+\omega_{D_{6,I-2}})\rho_{D },\label{77-2}\\
    V_{6,I-1}( \phi ) &=&\frac{\rho_{D}}{2}(1-\omega_{D_{6,I-1}}), \label{78}\\
    V_{6,I-2}( \phi ) &=&\frac{\rho_{D}}{2}(1-\omega_{D_{6,I-2}}),\label{78-2}
\end{eqnarray}
and
\begin{eqnarray}
    \dot{\phi}_{6,I-1,lim}^2&=&(1+\omega_{D_{6,I-1,lim}})\rho_D,\label{77}\\
    \dot{\phi}_{6,I-2,lim}^2&=&(1+\omega_{D_{6,I-2,lim}})\rho_{D },\label{77-2}\\
    V_{6,I-1,lim}( \phi ) &=&\frac{\rho_{D}}{2}(1-\omega_{D_{6,I-1,lim}}), \label{78}\\
    V_{6,I-2,lim}( \phi ) &=&\frac{\rho_{D}}{2}(1-\omega_{D_{6,I-2,lim}}).\label{78-2}
\end{eqnarray}

For the seventh interacting case, we obtain:
\begin{eqnarray}
    \dot{\phi}_{7,I-1}^2&=&(1+\omega_{D_{7,I-1}})\rho_D,\label{77}\\
    \dot{\phi}_{7,I-2}^2&=&(1+\omega_{D_{7,I-2}})\rho_{D },\label{77-2}\\
    V_{7,I-1}( \phi ) &=&\frac{\rho_{D}}{2}(1-\omega_{D_{7,I-1}}), \label{78}\\
    V_{7,I-2}( \phi ) &=&\frac{\rho_{D}}{2}(1-\omega_{D_{7,I-2}}),\label{78-2}
\end{eqnarray}
and
\begin{eqnarray}
    \dot{\phi}_{7,I-1,lim}^2&=&(1+\omega_{D_{7,I-1,lim}})\rho_D,\label{77}\\
    \dot{\phi}_{7,I-2,lim}^2&=&(1+\omega_{D_{7,I-2,lim}})\rho_{D },\label{77-2}\\
    V_{7,I-1,lim}( \phi ) &=&\frac{\rho_{D}}{2}(1-\omega_{D_{7,I-1,lim}}), \label{78}\\
    V_{7,I-2,lim}( \phi ) &=&\frac{\rho_{D}}{2}(1-\omega_{D_{7,I-2,lim}}).\label{78-2}
\end{eqnarray}

For the eighth interacting case, we obtain:
\begin{eqnarray}
    \dot{\phi}_{8,I-1}^2&=&(1+\omega_{D_{8,I-1}})\rho_D,\label{77}\\
    \dot{\phi}_{8,I-2}^2&=&(1+\omega_{D_{8,I-2}})\rho_{D },\label{77-2}\\
    V_{8,I-1}( \phi ) &=&\frac{\rho_{D}}{2}(1-\omega_{D_{8,I-1}}), \label{78}\\
    V_{8,I-2}( \phi ) &=&\frac{\rho_{D}}{2}(1-\omega_{D_{8,I-2}}).\label{78-2}
\end{eqnarray}

For the ninth interacting case, we obtain:
\begin{eqnarray}
    \dot{\phi}_{9,I-1}^2&=&(1+\omega_{D_{9,I-1}})\rho_D,\label{77}\\
    \dot{\phi}_{9,I-2}^2&=&(1+\omega_{D_{9,I-2}})\rho_{D },\label{77-2}\\
    V_{9,I-1}( \phi ) &=&\frac{\rho_{D}}{2}(1-\omega_{D_{9,I-1}}), \label{78}\\
    V_{9,I-2}( \phi ) &=&\frac{\rho_{D}}{2}(1-\omega_{D_{9,I-2}}).\label{78-2}
\end{eqnarray}

\subsection{The Yang-Mills (YM) Scalar Field Model}
We now consider the Yang-Mills (YM) scalar field model.\\
This model, rooted in non-Abelian gauge theory, has been studied as a possible source of dark energy (DE). Below we derive the expressions for the energy density and pressure of the YM field and investigate how it can effectively describe the dark-energy dynamics in our cosmological framework.

Recent works indicate that the Yang-Mills condensate (YMC) can be a viable DE candidate~\cite{ym1,ym2,ym3,ym9-2,ym9-3,ym9-4,ym9-5}. Two principal motivations support this choice. First, unlike ad hoc scalar fields, the YM field has a firm foundation in particle physics: it arises naturally from the gauge structure of the Standard Model and its extensions. Second, the YM field can realize scenarios where the weak energy condition is violated, allowing a richer cosmological dynamics that can include accelerated expansion.

The effective YMC model exhibits features that make it attractive for modeling DE. Being connected to gauge bosons, it can be embedded within unified interaction theories, and its effective equation-of-state parameter differs markedly from that of canonical scalar fields. In particular, the YMC can produce a variety of behaviors, from quintessence-like regimes with $-1 < \omega < 0$ to phantom-like regimes with $\omega < -1$, thus accommodating different accelerating scenarios.

In what follows we present the effective Lagrangian and derive the corresponding energy density and pressure for the YM condensate, then discuss the resulting cosmological implications. The effective Yang--Mills DE Lagrangian density is given by:
\begin{eqnarray}
L_{YMC} = \frac{bF}{2} \left( \ln \left| \frac{F}{\kappa^2} \right| - 1 \right), \label{murano117}
\end{eqnarray}
where the constant $\kappa$ is a renormalization scale with dimensions of mass squared,  while the term $F$ plays the role of the order parameter of the Yang-Mills condensate and it is given by:
\begin{eqnarray}
F = -\frac{1}{2} F_{\mu \nu}^\alpha F^{\alpha \mu \nu} = E^2 - B^2, \label{murano118}
\end{eqnarray}
where the quantity $F_{\mu \nu}^\alpha$ indicates the Yang-Mills field strength tensor, while $E$ and $B$ are the effective electric and magnetic components of the field, respectively.\\
We can easily observe that the  pure electric case is recovered in the limiting case corresponding to $B = 0$: in this case, we have that $F = E^2 $.\\
Moreover, the parameter $b$ denotes the Callan-Symanzik coefficient~\cite{ym18,ym18-1}, 
which for the gauge group $SU(N)$ is given by:
\begin{eqnarray}
b = \frac{11N - 2N_f}{24\pi^2}, \label{murano119}
\end{eqnarray}
where the quantity $N_f$ denotes the number of quark flavors. 

Considering the gauge group $SU(2)$, the Callan--Symanzik \begin{eqnarray}
b = 2 \times \frac{11}{24\pi^2},
\end{eqnarray}
when fermionic contributions are neglected, while it assumes the \begin{eqnarray}
b = 2 \times \frac{5}{24\pi^2},
\end{eqnarray}
when the number of quark flavors is $N_f = 6$. 

In the case of the $SU(3)$ gauge group, the effective Lagrangian presented in Eq.~\eqref{murano117} 
offers a phenomenological framework to describe the phenomenon of asymptotic freedom exhibited by 
quarks confined within hadrons~\cite{ym21,ym21-1}.

It is crucial to clarify that the $SU(2)$ Yang-Mills scalar field model considered in this work 
as a possible candidate for dark energy is fundamentally distinct from the QCD gluon fields or 
the electroweak gauge bosons such as $Z^0$ and $W^{\pm}$. The Yang--Mills condensate (YMC) relevant 
for cosmology operates at an energy scale characterized by $\kappa^{1/2} \approx 10^{-3}\,\text{eV}$, 
which is many orders of magnitude smaller than the energy scales typical of quantum chromodynamics 
and electroweak processes. This vast difference in scale highlights the phenomenological independence 
of the YMC dark energy model from standard particle physics gauge fields.

The form of the effective YM Lagrangian $L_{\text{YMC}}$ described in Eq.~\eqref{murano117} can be understood 
as a one-loop quantum corrected Lagrangian~\cite{ym21,ym21-1}. The classical $SU(N)$ Yang--Mills Lagrangian 
can be written as
\begin{equation}
L = \frac{1}{2g_0^2} F, 
\label{murano120}
\end{equation}
where the quantity $g_0$ is the bare coupling constant. 

Including one-loop quantum corrections leads to a running coupling constant $g$ that replaces 
the bare coupling, as
\begin{equation}
g_0^2 \rightarrow g^2 = 
\frac{4 \times 12 \pi^2}{11N \ln\!\left(\frac{k}{k_0^2}\right)} 
= \frac{2}{b \ln\!\left(\frac{k}{k_0^2}\right)}, 
\label{murano121}
\end{equation}
where the quantity $k$ represents the momentum transfer, while $k_0$ denotes the corresponding energy scale.

To construct an effective theory, the momentum scale $k^2$ is replaced by the field strength $F$ via the substitution
\begin{equation}
\ln\!\left(\frac{k}{k_0^2}\right) 
\rightarrow 
2 \ln\!\left|\frac{F}{\kappa^2 e}\right| 
= 
2 \ln\!\left|\frac{F}{\kappa^2} - 1\right|.
\label{murano122}
\end{equation}

This replacement restores the effective Lagrangian form shown in Eq.~\eqref{murano117}. 
The effective Yang-Mills (YM) action exhibits several notable theoretical properties, 
including the correct trace anomaly, asymptotic freedom, gauge invariance, and Lorentz invariance~\cite{ym16}. 

The logarithmic dependence of the field strength in the YMC Lagrangian closely resembles 
the Coleman--Weinberg scalar effective potential~\cite{ym19} 
and shares structural similarities with the Parker-Raval effective gravity Lagrangian~\cite{ym20}. 
These analogies underline the quantum corrections that give rise to the model and 
emphasize its potential significance in cosmology.

It is worth stressing that the renormalization scale $\kappa$ 
is the only free parameter in this effective Yang--Mills model. 
This feature distinguishes it from scalar-field dark energy models, 
where the potential function is typically chosen \emph{ad hoc} to match observations. 
In contrast, the YMC framework determines the Lagrangian form strictly from one-loop quantum corrections 
derived from the underlying gauge theory. 
As a result, the model is theoretically more robust and predictive, 
since its dynamics are governed by quantum field theory principles rather than phenomenological assumptions. 
This makes the Yang--Mills condensate a compelling and physically motivated candidate for describing dark energy.

Starting from the effective Lagrangian in Eq.~\eqref{murano117}, 
the corresponding expressions for the Yang-Mills condensate (YMC) energy density $\rho_{YMC}$ 
and pressure $p_{YMC}$ can be written as
\begin{align}
\rho_{YMC} &= \frac{\epsilon E^2}{2} + \frac{b E^2}{2}, \label{murano123} \\[4pt]
p_{YMC} &= \frac{\epsilon E^2}{6} - \frac{b E^2}{2}. \label{murano124}
\end{align}

Here, $\epsilon$ denotes the dielectric constant associated with the Yang-Mills field, 
which plays a crucial role in determining the effective coupling strength within the condensate. 
It is defined as
\begin{align}
\epsilon = 2 \frac{\partial L_{\text{eff}}}{\partial F} 
= b \ln \left| \frac{F}{\kappa^2} \right|.
\label{murano125}
\end{align}

These quantities characterize the macroscopic cosmological behavior of the Yang--Mills condensate (YMC) 
and are derived through the standard procedure of varying the Lagrangian with respect to the metric tensor 
to obtain the components of the energy--momentum tensor. 
The resulting expressions provide the foundation for studying the dynamical properties of the dark energy component described by the YMC.

Equations~\eqref{murano123} and~\eqref{murano124} can also be expressed as
\begin{eqnarray}
\rho_y &=& \frac{1}{2} b \kappa^2 (y+1) e^{y}, \label{murano126}\\
p_y &=& \frac{1}{6} b \kappa^2 (y-3) e^{y}, \label{murano127}
\end{eqnarray}
or in the following equivalent form:
\begin{eqnarray}
\rho_y &=& \frac{1}{2} (y+1) b E^2, \label{murano128} \\
p_y &=& \frac{1}{6} (y-3) b E^2, \label{murano129}
\end{eqnarray}
where the dimensionless parameter $y$ is introduced and defined as
\begin{eqnarray}
y = \frac{\epsilon}{b} = \ln \left| \frac{F}{\kappa^2} \right| = \ln \left| \frac{E^2}{\kappa^2} \right|. \label{defiy}
\end{eqnarray}
Using the expressions for $\rho_y$ and $p_y$ in Eqs.~\eqref{murano126} and~\eqref{murano127} 
(or equivalently in Eqs.~\eqref{murano128} and~\eqref{murano129}), 
the equation of state (EoS) parameter $\omega_y$ of the YMC model can be expressed as
\begin{align}
\omega_y = \frac{p_y}{\rho_y} = \frac{y - 3}{3(y + 1)}.
\label{omegay}
\end{align}

At the critical point where $\epsilon = 0$ (i.e., $y = 0$), we obtain $\omega_y = -1$,  corresponding to a de~Sitter phase of cosmic expansion.  In the neighborhood of this point, the sign of the dielectric constant $\epsilon$  plays a decisive role in determining the behavior of the EoS parameter $\omega_y$. 
When $\epsilon < 0$, the EoS parameter drops below the phantom divide ($\omega_y < -1$),  indicating a phantom-like regime characterized by super-accelerated expansion. 
Conversely, for $\epsilon > 0$, one finds $\omega_y > -1$, 
corresponding to a quintessence-like behavior in which the dark energy density  evolves more moderately. 
This sensitivity to the sign of $\epsilon$ underscores its importance in  governing the dynamical properties of the Yang--Mills condensate as a viable dark energy candidate.  

Hence, as previously discussed, the YMC model naturally accommodates both regimes,  $0 > \omega_y > -1$ and $\omega_y < -1$.  The expression for $\omega_y$ in Eq.~\eqref{omegay} can be inverted to yield
\begin{align}
y = -\,\frac{3(\omega_y + 1)}{3\omega_y - 1}.
\label{murano130}
\end{align}

To ensure the physical viability of the model, the energy density of the Yang--Mills condensate must remain positive, which imposes a lower bound on the dimensionless parameter $y$:
\begin{align}
y > 1.
\end{align}
This requirement guarantees that the YMC satisfies the basic energy conditions of cosmology and leads to a corresponding lower limit on the condensate strength:
\begin{align}
F > \frac{\kappa^2}{e} \simeq 0.368\,\kappa^2.
\end{align}

Before exploring specific cosmological scenarios, it is useful to analyze the behavior of the equation of state parameter $\omega_y$ as a function of $F$. In the regime where the dielectric constant is large, $\epsilon \gg b$ (or equivalently $F \gg \kappa^2$), the YMC behaves like a radiation fluid, with
\begin{align}
p_y &= \frac{1}{3} \rho_y, \\
\omega_y &= \frac{1}{3}.
\end{align}

At the opposite limit, corresponding to the critical point $\epsilon = 0$ (i.e., $F = \kappa^2$), the YMC mimics a cosmological constant:
\begin{align}
\omega_y &= -1, \\
p_y &= -\rho_y,
\end{align}
with the associated energy density
\begin{align}
\rho_y = \frac{1}{2} b \kappa^2,
\end{align}
defining the critical energy density \cite{ym1,ym3}.

A remarkable aspect of the YMC model is the smooth transition of the equation of state from the radiation-like regime at high energies ($\omega_y \simeq 1/3$ for $F \gg \kappa^2$) to a cosmological constant–like regime at low energies ($\omega_y = -1$ for $F = \kappa^2$). This feature naturally allows for scaling solutions of the dark energy component \cite{ym10,ym10-1}. 
Moreover, $\omega_y$ is a continuous function of the dimensionless variable $y$, ensuring a smooth evolution across the entire range $(-1, \infty)$.

We can further investigate the possibility for the EoS parameter to cross the phantom divide at $\omega_y = -1$. According to Eq.~\eqref{omegay}, $\omega_y$ is fully determined by the condensate strength $F$. 
From a theoretical perspective, the phantom regime, $\omega_y < -1$, is achieved when
\begin{eqnarray}
F < \kappa^2,
\end{eqnarray}
and, in this case, the crossing is also smooth with respect to the parameter $F$.\\
When incorporated into a cosmological setting as a dark energy component alongside matter and radiation, the condensate strength $F$ in the YMC model is no longer a free parameter but evolves dynamically with cosmic time $t$.\\
In the absence of interactions with matter or radiation, the equation of state parameter $\omega_y$ gradually approaches $-1$ from above, without ever crossing this value. In this non-interacting scenario, the YMC effectively behaves like a cosmological constant at late times. By contrast, if the YMC exchanges energy with matter and/or radiation via a coupling or interaction term, the evolution changes significantly. In such interacting cases, $\omega_y$ may drop below the phantom divide ($\omega_y < -1$), entering a phantom-like regime. The asymptotic value of $\omega_y$ depends on the strength of the interaction and, for moderate couplings, can stabilize at a value notably different from $-1$.\\
A key advantage of the phantom regime in the YMC model is that all relevant physical quantities—including the energy density $\rho_y$, pressure $p_y$, and the EoS parameter $\omega_y$—remain smooth and well-defined throughout cosmic evolution. This feature contrasts with many scalar field dark energy models, where entering the phantom regime can induce instabilities or finite-time singularities such as Big Rip scenarios. In the YMC framework, the phantom-like behavior emerges naturally from the underlying gauge field dynamics, ensuring both theoretical consistency and cosmological viability.\\

By setting the equation of state (EoS) parameter of the Yang-Mills condensate, $\omega_y$, equal to the EoS parameter of the DE model under consideration, the dimensionless variable $y$ can be expressed as:
\begin{eqnarray}
y = - \frac{3\left(\omega_D+1\right)}{3\omega_D-1}.\label{murano131}
\end{eqnarray}
We can easily from Eq. (\ref{murano131}) that the condition  $y>1$ is obtained when:
\begin{eqnarray}
    -\frac{1}{3}< \omega_{D} <\frac{1}{3}
\end{eqnarray}
We now consider the results obtained in this work.\\
For the non interacting case, we obtain:
\begin{eqnarray}
y_{NI} &=& - \frac{3\left(\omega_D+1\right)}{3\omega_D-1} \nonumber \\
&=&\frac{1}{2n-1}. \label{yni}
\end{eqnarray}
Therefore, we must have that $n\neq 1/2$.\\
Moreover, we obtain from Eq. (\ref{yni}) that $y>1$ when $\frac{1}{2}<n<1$.\\
For the first interacting case, we obtain:
\begin{eqnarray}
y_{1,I} &=& - \frac{3\left(\omega_{D_{1,I}}+1\right)}{3\omega_{D_{1,I}}-1}.
\end{eqnarray}
For the second interacting case, we obtain:
\begin{eqnarray}
y_{2,I} &=& - \frac{3\left(\omega_{D_{2,I}}+1\right)}{3\omega_{D_{2,I}}-1}\nonumber \\
&=&\frac{2 - 3n d^2}{n\left(4+3d^2  \right) - 2 }. \label{y2I}
\end{eqnarray}
Therefore, we must have $n\neq \frac{2}{4+3d^2}$. In the limiting case of $d^2=0$ we recover the same results of the non interacting case obtained in Eq. (\ref{}), as it is expected.\\
Moreover, we find from Eq. (\ref{y2I}) that $y>1$ when $\frac{2}{4 + 3 d^2} < n < \frac{2}{2 + 3 d^2}$.\\
For the third interacting case, we obtain:
\begin{eqnarray}
y_{3,I-1} &=& - \frac{3\left(\omega_{D_{3,I-1}}+1\right)}{3\omega_{D_{3,I-I}}-1},\label{}\\
y_{3,I-2} &=& - \frac{3\left(\omega_{D_{3,I-2}}+1\right)}{3\omega_{D_{3,I-2}}-1},\label{}\\
y_{3,I-1,lim} &=& - \frac{3\left(\omega_{D_{3,I-1\,lim}}+1\right)}{3\omega_{D_{3,I-1,lim}}-1},\label{}\\
y_{3,I-2,lim} &=& - \frac{3\left(\omega_{D_{3,I-2,lim}}+1\right)}{3\omega_{D_{3,I-2\lim}}-1}.\label{}
\end{eqnarray}

For the fourth interacting case, we obtain:
\begin{eqnarray}
y_{4,I-1} &=& - \frac{3\left(\omega_{D_{4,I-1}}+1\right)}{3\omega_{D_{4,I-I}}-1},\label{}\\
y_{4,I-2} &=& - \frac{3\left(\omega_{D_{4,I-2}}+1\right)}{3\omega_{D_{4,I-2}}-1},\label{}\\
y_{4,I-1,lim} &=& - \frac{3\left(\omega_{D_{4,I-1\,lim}}+1\right)}{3\omega_{D_{4,I-1,lim}}-1},\label{}\\
y_{4,I-2,lim} &=& - \frac{3\left(\omega_{D_{4,I-2,lim}}+1\right)}{3\omega_{D_{4,I-2\lim}}-1}.\label{}
\end{eqnarray}

For the fifth interacting case, we obtain:
\begin{eqnarray}
y_{5,I-1} &=& - \frac{3\left(\omega_{D_{5,I-1}}+1\right)}{3\omega_{D_{5,I-I}}-1},\label{}\\
y_{5,I-2} &=& - \frac{3\left(\omega_{D_{5,I-2}}+1\right)}{3\omega_{D_{5,I-2}}-1}.\label{}
\end{eqnarray}

For the sixth interacting case, we obtain:
\begin{eqnarray}
y_{6,I-1} &=& - \frac{3\left(\omega_{D_{6,I-1}}+1\right)}{3\omega_{D_{6,I-I}}-1},\label{}\\
y_{6,I-2} &=& - \frac{3\left(\omega_{D_{6,I-2}}+1\right)}{3\omega_{D_{6,I-2}}-1},\label{}\\
y_{6,I-1,lim} &=& - \frac{3\left(\omega_{D_{6,I-1\,lim}}+1\right)}{3\omega_{D_{6,I-1,lim}}-1},\label{}\\
y_{6,I-2,lim} &=& - \frac{3\left(\omega_{D_{6,I-2,lim}}+1\right)}{3\omega_{D_{6,I-2\lim}}-1}.\label{}
\end{eqnarray}

For the seventh interacting case, we obtain:
\begin{eqnarray}
y_{7,I-1} &=& - \frac{3\left(\omega_{D_{7,I-1}}+1\right)}{3\omega_{D_{7,I-I}}-1},\label{}\\
y_{7,I-2} &=& - \frac{3\left(\omega_{D_{7,I-2}}+1\right)}{3\omega_{D_{7,I-2}}-1},\label{}\\
y_{7,I-1,lim} &=& - \frac{3\left(\omega_{D_{7,I-1\,lim}}+1\right)}{3\omega_{D_{7,I-1,lim}}-1},\label{}\\
y_{7,I-2,lim} &=& - \frac{3\left(\omega_{D_{7,I-2,lim}}+1\right)}{3\omega_{D_{7,I-2\lim}}-1}.\label{}
\end{eqnarray}

For the eighth interacting case, we obtain:
\begin{eqnarray}
y_{8,I-1} &=& - \frac{3\left(\omega_{D_{8,I-1}}+1\right)}{3\omega_{D_{8,I-I}}-1},\label{}\\
y_{8,I-2} &=& - \frac{3\left(\omega_{D_{8,I-2}}+1\right)}{3\omega_{D_{8,I-2}}-1}.\label{}\
\end{eqnarray}

For the ninth interacting case, we obtain:
\begin{eqnarray}
y_{9,I-1} &=& - \frac{3\left(\omega_{D_{9,I-1}}+1\right)}{3\omega_{D_{9,I-I}}-1},\label{}\\
y_{9,I-2} &=& - \frac{3\left(\omega_{D_{9,I-2}}+1\right)}{3\omega_{D_{9,I-2}}-1}.\label{}\
\end{eqnarray}

\subsection{The Non-Linear Electro-Dynamics (NLED) Scalar Field Model}

We now turn our attention to the final model under consideration, namely the Nonlinear Electrodynamics (NLED) scalar field model.\\
Recently, NLED has been proposed as a promising framework to address cosmic singularities by extending classical Maxwell electrodynamics with nonlinear corrections. Within this approach, exact solutions of Einstein's field equations coupled to NLED demonstrate that these nonlinear terms become relevant in regimes of strong gravitational and electromagnetic fields. Such modifications can smooth out singularities and yield physically consistent solutions. Moreover, the interaction of General Relativity (GR) with NLED has been shown to naturally generate an inflationary phase in the early Universe, providing a viable classical mechanism for primordial inflation without invoking quantum corrections.\\
In classical electrodynamics, the evolution of the free electromagnetic field is determined by the Maxwell Lagrangian density, $L_M$, which is given by \cite{ele1,ele2}:
\begin{eqnarray}
L_M = - \frac{F^{\mu \nu} F_{\mu \nu}}{4\mu}, \label{murano135}
\end{eqnarray}
where $F^{\mu \nu}$ is the electromagnetic field strength tensor, defined as $F^{\mu\nu} = \partial^\mu A^\nu - \partial^\nu A^\mu$, 
and $\mu$ represents the magnetic permeability of the medium, which must be included when restoring physical units in the field equations.

We now turn to a nonlinear extension of the classical Maxwell Lagrangian, incorporating second-order corrections in the electromagnetic field invariants. The modified Lagrangian density is expressed as:
\begin{eqnarray}
L = -\frac{F}{4\mu_0} + \omega F^2 + \eta F^{*2}, \label{murano136}
\end{eqnarray}
where $\omega$ and $\eta$ are free parameters quantifying the strength of the nonlinear contributions, and $F^*$ represents the electromagnetic pseudo-invariant, defined by
\begin{eqnarray}
F^* = F_{\mu\nu} \tilde{F}^{\mu\nu} = \frac{1}{2} \epsilon^{\mu\nu\rho\sigma} F_{\mu\nu} F_{\rho\sigma}.
\end{eqnarray}
Here, $\tilde{F}^{\mu\nu}$ is the dual field strength tensor and $\epsilon^{\mu\nu\rho\sigma}$ denotes the totally antisymmetric Levi-Civita symbol.\\

We focus on the regime in which the homogeneous electric field $E$ in the plasma is rapidly suppressed due to backreaction from the induced electric currents. In this case, the electric contribution is negligible ($E^2 \approx 0$), and the magnetic field dominates the dynamics, allowing the simplification
\begin{eqnarray}
F \simeq 2 B^2,
\end{eqnarray}
so that $F$ depends solely on the squared magnetic field $B^2$.\\

The pressure $p_{NLED}$ and energy density $\rho_{NLED}$ of the NLED field can be obtained from the corresponding energy-momentum tensor in a spatially homogeneous and isotropic Universe. They read:
\begin{align}
p_{NLED} &= \frac{B^2}{6\mu}\left(1 - 40 \mu \omega B^2 \right), \label{murano138}\\
\rho_{NLED} &= \frac{B^2}{2\mu}\left(1 - 8 \mu \omega B^2 \right). \label{murano139}
\end{align}

The weak energy condition, $\rho_{NLED} > 0$, requires
\begin{eqnarray}
B < \frac{1}{2\sqrt{2 \mu \omega}}.
\end{eqnarray}
Meanwhile, the pressure becomes negative when
\begin{eqnarray}
B > \frac{1}{2\sqrt{10 \mu \omega}}.
\end{eqnarray}

The magnetic field can act as an effective source of dark energy if the strong energy condition is violated, i.e., when
\begin{eqnarray}
\rho_B + 3 p_B < 0,
\end{eqnarray}
which occurs for magnetic field strengths satisfying
\begin{eqnarray}
B > \frac{1}{2\sqrt{6 \mu \omega}}.\label{bcond}
\end{eqnarray}

The equation of state (EoS) parameter $\omega_{NLED}$ for the Nonlinear Electrodynamics (NLED) field is defined as
\begin{eqnarray}
\omega_{NLED} = \frac{p_{NLED}}{\rho_{NLED}} = \frac{1 - 40 \mu \omega B^2}{3 \left(1 - 8 \mu \omega B^2 \right)}, \label{murano140}
\end{eqnarray}
which can be inverted to express the magnetic field intensity $B^2$ in terms of $\omega_{NLED}$ as
\begin{eqnarray}
B^2 = \frac{1 - 3 \omega_{NLED}}{8 \mu \omega \left(5 - 3 \omega_{NLED}\right)}. \label{murano141}
\end{eqnarray}
By establishing a correspondence between the equation of state parameter of the NLED model and that of the dark energy model under consideration, we obtain
\begin{eqnarray}
B^2 = \frac{1 - 3 \omega_D}{8 \mu \omega \left(5 - 3 \omega_D\right)}. \label{murano142}
\end{eqnarray}
We now consider the results obtained in this work.\\
For the non interacting case, we obtain:
\begin{eqnarray}
    B_{NI}^2 &=& \frac{1 - 3 \omega_D }{8 \mu \omega \left(5 - 3 \omega_{D  }\right)}=\frac{2n - 1}{8\,\mu\,\omega\,(4n - 1)}. \label{bni} 
\end{eqnarray}
Therefore, in order to avoid singularities in Eq. (\ref{bni}), we must have $n\neq 1/4$. \\
Moreover, Eq. (\ref{bni}) satisfies the condition given in Eq. (\ref{bcond}) when $n>1$.\\
For the first interacting case, we obtain:
\begin{eqnarray}
    B_{1,I}^2 &=& \frac{1 - 3 \omega_{D_{1,I}  } }{8 \mu \omega \left(5 - 3 \omega_{D_{1,I}  }\right)}.
\end{eqnarray}
For the second interacting case, we obtain:
\begin{eqnarray}
    B_{2,I}^2 &=& \frac{1 - 3 \omega_{D_{2,I}  } }{8 \mu \omega \left(5 - 3 \omega_{D_{2,I}  }\right)}\nonumber\\
    &=&\dfrac{4n-2+3n d^2}{8\,\mu\,\omega\,(8n-2+3n d^2)}\nonumber \\
    &=&\dfrac{n\left( 4+3d^2  \right)-2}{8\,\mu\,\omega\,\left[n\left(8+3d^2  \right)-2\right]}. \label{bni2}
\end{eqnarray}
In order to avoid singularities in Eq. (\ref{bni2}), we must have $n\neq \frac{2}{8+3d^2}$.\\
Moreover, Eq. (\ref{bni2}) satisfies the condition given in Eq. (\ref{bcond}) when $n>\frac{2}{2+3d^2}$.\\
Therefore, we must have that $n\neq \frac{2}{8+3d^2 }$ (in the limiting case of $d^2=0$ we recover the same result of the non interacting case given in Eq. (\ref{}) as it is expected).\\
For the third interacting case, we obtain:
\begin{eqnarray}
B_{3,I-1}^2 &=& \frac{1 - 3 \omega_{D_{3,I-1}}}{8 \mu \omega \left(5 - 3 \omega_{D_{3,I-1}}\right)},\\
B_{3,I-2}^2 &=& \frac{1 - 3 \omega_{D_{3,I-2}}}{8 \mu \omega \left(5 - 3 \omega_{D_{3,I-2}}\right)},\\
B_{3,I-1,lim}^2 &=& \frac{1 - 3 \omega_{D_{3,I-1,lim}}}{8 \mu \omega \left(5 - 3 \omega_{D_{3,I-1,lim}}\right)},\\
B_{3,I-2,lim}^2 &=& \frac{1 - 3 \omega_{D_{3,I-2,lim}}}{8 \mu \omega \left(5 - 3 \omega_{D_{3,I-2}}\right)}.
\end{eqnarray}

For the fourth interacting case, we obtain:
\begin{eqnarray}
B_{4,I-1}^2 &=& \frac{1 - 3 \omega_{D_{4,I-1}}}{8 \mu \omega \left(5 - 3 \omega_{D_{4,I-1}}\right)},\\
B_{4,I-2}^2 &=& \frac{1 - 3 \omega_{D_{4,I-2}}}{8 \mu \omega \left(5 - 3 \omega_{D_{4,I-2}}\right)},\\
B_{4,I-1,lim}^2 &=& \frac{1 - 3 \omega_{D_{4,I-1,lim}}}{8 \mu \omega \left(5 - 3 \omega_{D_{4,I-1,lim}}\right)},\\
B_{4,I-2,lim}^2 &=& \frac{1 - 3 \omega_{D_{4,I-2,lim}}}{8 \mu \omega \left(5 - 3 \omega_{D_{4,I-2}}\right)}.
\end{eqnarray}

For the fifth interacting case, we obtain:
\begin{eqnarray}
B_{5,I-1}^2 &=& \frac{1 - 3 \omega_{D_{5,I-1}}}{8 \mu \omega \left(5 - 3 \omega_{D_{5,I-1}}\right)},\\
B_{5,I-2}^2 &=& \frac{1 - 3 \omega_{D_{5,I-2}}}{8 \mu \omega \left(5 - 3 \omega_{D_{5,I-2}}\right)}.
\end{eqnarray}

For the sixth interacting case, we obtain:
\begin{eqnarray}
B_{6,I-1}^2 &=& \frac{1 - 3 \omega_{D_{6,I-1}}}{8 \mu \omega \left(5 - 3 \omega_{D_{6,I-1}}\right)},\\
B_{6,I-2}^2 &=& \frac{1 - 3 \omega_{D_{6,I-2}}}{8 \mu \omega \left(5 - 3 \omega_{D_{6,I-2}}\right)},\\
B_{6,I-1,lim}^2 &=& \frac{1 - 3 \omega_{D_{6,I-1,lim}}}{8 \mu \omega \left(5 - 3 \omega_{D_{6,I-1,lim}}\right)},\\
B_{6,I-2,lim}^2 &=& \frac{1 - 3 \omega_{D_{6,I-2,lim}}}{8 \mu \omega \left(5 - 3 \omega_{D_{6,I-2}}\right)}.
\end{eqnarray}

For the seventh interacting case, we obtain:
\begin{eqnarray}
B_{7,I-1}^2 &=& \frac{1 - 3 \omega_{D_{7,I-1}}}{8 \mu \omega \left(5 - 3 \omega_{D_{7,I-1}}\right)},\\
B_{7,I-2}^2 &=& \frac{1 - 3 \omega_{D_{7,I-2}}}{8 \mu \omega \left(5 - 3 \omega_{D_{7,I-2}}\right)},\\
B_{7,I-1,lim}^2 &=& \frac{1 - 3 \omega_{D_{7,I-1,lim}}}{8 \mu \omega \left(5 - 3 \omega_{D_{7,I-1,lim}}\right)},\\
B_{7,I-2,lim}^2 &=& \frac{1 - 3 \omega_{D_{7,I-2,lim}}}{8 \mu \omega \left(5 - 3 \omega_{D_{7,I-2}}\right)}.
\end{eqnarray}

For the eighth interacting case, we obtain:
\begin{eqnarray}
B_{8,I-1}^2 &=& \frac{1 - 3 \omega_{D_{8,I-1}}}{8 \mu \omega \left(5 - 3 \omega_{D_{8,I-1}},\right)}\\
B_{8,I-2}^2 &=& \frac{1 - 3 \omega_{D_{8,I-2}}}{8 \mu \omega \left(5 - 3 \omega_{D_{8,I-2}}\right)}.
\end{eqnarray}

For the ninth interacting case, we obtain:
\begin{eqnarray}
B_{9,I-1}^2 &=& \frac{1 - 3 \omega_{D_{9,I-1}}}{8 \mu \omega \left(5 - 3 \omega_{D_{9,I-1}}\right)},\\
B_{9,I-2}^2 &=& \frac{1 - 3 \omega_{D_{9,I-2}}}{8 \mu \omega \left(5 - 3 \omega_{D_{9,I-2}}\right)}.
\end{eqnarray}

\section{Conclusions}
In this work, we investigated a Dark Energy (DE) density model that incorporates the Hubble parameter and its higher-order time derivatives, namely $H^2 $, $\dot{H} $, $\ddot{H} $, and $\dddot{H} $. By assuming a power-law scale factor $a(t) \propto t^n $, we derived analytical expressions for a wide range of cosmological quantities, including the energy densities of matter ($\rho_m $) and DE ($\rho_D $), their corresponding fractional densities ($\Omega_m $ and $\Omega_D $), the Hubble function $H^2 $, the deceleration parameter $q $, the evolutionary form of the DE fractional energy density $\Omega'_D $, the DE pressure $p_D $ and the DE Equation of State (EoS) $\omega_D$.\\ 
For the interacting scenario, we analyzed nine different coupling forms of the interaction term $Q $, each depending on $H $, $\rho_m $ and/or $\rho_D $. These interactions affect, in different ways, the dynamical evolution of the quantities derived in this work.\\
Finally, by establishing correspondences with scalar field models, including tachyon, k-essence, quintessence, Yang-Mills (YM) and Nonlinear Electrodynamics (NLED), we have provided a deeper theoretical foundation for the proposed DE model.

\end{document}